\documentclass[a4paper,11pt]{article}
\pdfoutput=1

\usepackage{jheppub} 

\usepackage[T1]{fontenc} 

\usepackage{amssymb}\usepackage[normalem]{ulem} 
\usepackage{ulem}

\usepackage{verbatim}
\usepackage{graphicx} 
\usepackage{graphics}
\usepackage{epsfig} 
\usepackage{color}
\usepackage{amssymb}
\usepackage{amsmath}
\usepackage{doi}
\usepackage{dsfont}
\usepackage{setspace}
\usepackage{float}
\usepackage{caption}
\usepackage{subcaption}

\usepackage{slashed}
\usepackage{booktabs}
\usepackage{multirow}
\usepackage{soul}

\title{Four top final states with NLO accuracy in perturbative QCD: 3 lepton  channel}

\author{Nikolaos Dimitrakopoulos}
\author{and Malgorzata Worek\,}

\affiliation{Institute for Theoretical Particle Physics
and Cosmology, RWTH Aachen University, \\D-52056 Aachen, Germany}

\emailAdd{ndimitrak@physik.rwth-aachen.de}
\emailAdd{worek@physik.rwth-aachen.de}

\abstract{We examine the effect of higher-order QCD corrections on the four top-quark production cross section in the $3$ lepton decay channel.  Top-quark and $W$ gauge-boson decays are included at next-to-leading order in perturbative QCD. The narrow-width approximation, that preserves spin correlations, is used to perform the calculation for this complex  $2 \to 12 \,(13)$ process.  We present differential cross-section distributions of top-quark decay products for the LHC Run III energy of  $\sqrt{s}=13.6$ TeV, using realistic selection cuts. Our results show that QCD corrections and jet radiation in top-quark and $W$ decays can lead to significant changes in shapes of basic kinematic distributions and, therefore, need to be included for the accurate description of $pp \to t\bar{t}t\bar{t}+X$ process in the $3\ell$ decay channel. The NLO QCD results obtained in this work are important in the context of the recent observation of the $pp\to t\bar{t}t\bar{t}+X$ process by the ATLAS and CMS collaborations and are of great significance for the High-Luminosity phase of the LHC.}

\dedicated{\rm TTK-24-32, P3H-24-062}


\begin{document} 

\maketitle
\flushbottom

%
\section{Introduction}
%

The production of four top quarks $(t\bar{t}t\bar{t})$ represents one of the rarest and most intriguing processes studied at the Large Hadron Collider (LHC). As the heaviest known elementary particle, the top quark plays a crucial role in the Standard Model (SM) of particle physics, particularly in the electroweak symmetry-breaking mechanism and the coupling to the Higgs boson. The simultaneous production of four top quarks in high-energy proton-proton collisions offers a unique opportunity to probe the SM at the highest energy scales and to search for potential new physics beyond it. In the SM, the production of four top quarks is predicted to occur with a low cross section \cite{Bevilacqua:2012em}, making it a challenging process to observe. However, its study is vital for several reasons. Firstly, the production of four top quarks provides a rigorous test of SM predictions, especially in the context of QCD and the complex interaction of top quarks. Measuring the  $pp \to t\bar{t}t\bar{t}+X$ process and comparing it with theoretical predictions allows physicists to validate or refine existing Monte Carlo tools and approximations they employ. Secondly, the large mass of the top quark provides a substantial coupling to the Higgs boson $(Y_t=\sqrt{2} \, m_t/v \approx 1)$, making the top quark a key player in the understanding of the Higgs mechanism. The production of four top quarks is sensitive to the top-Higgs interaction and the Higgs-boson width, offering insight into the nature of electroweak symmetry breaking \cite{Cao:2016wib,Cao:2019ygh}. Thirdly, many theories beyond the SM, such as supersymmetry, extra dimensions, and models involving new heavy particles, predict an increase in the production rate of the four top quarks \cite{Guchait:2007jd,Pomarol:2008bh,Plehn:2008ae,Jung:2010ms,Gregoire:2011ka,Calvet:2012rk,Arina:2016cqj,Alvarez:2016nrz,Baer:2016wkz,Baer:2017pba,Alvarez:2019uxp,Anisha:2023xmh}. For example, some new particles or interactions may lead not only to an increased cross section, but also to new signatures in the four top events that would affect various dimensionful and dimensionless differential cross-section distributions. Any deviation from SM predictions will, of course, signal the presence of new physics beyond the SM (BSM). However, for all of these BSM studies, a good theoretical control of the SM $pp\to t\bar{t}t\bar{t}+X$ process is an essential condition for the correct interpretation of possible signals of new physics that may appear in this channel.  Fourthly, four top-quark production at hadron colliders can also be examined in the Standard Model Effective Field Theory.  Although the $pp\to t\bar{t}$ process seems to be the leading process for exploring new physics in the top-quark sector, there are situations where dimension-six operators are suppressed in this process. In such situations, a much better way to investigate these contributions can be through the direct production of four top quarks \cite{Degrande:2010kt,Zhang:2017mls,Banelli:2020iau,Aoude:2022deh}.  Fifthly, the decay of four top quarks results in various and complex final states with multiple leptons, light- and $b$-jets as well as missing transverse momentum from escaping neutrinos, providing a rich environment for testing advanced particle reconstruction methods and analysis techniques. These final states also allow for the exploration of rare decay modes and interactions that are otherwise difficult to study. Finally, observing four top-quark production at the LHC is a significant experimental challenge due to the rarity of the process and the complexity of the resulting events. Sophisticated analysis methods, including machine learning algorithms, have to be employed to distinguish signal events from the large background processes. Despite these challenges, the continued study of four top-quark production holds great potential for advancing our understanding of fundamental interactions within the SM  and possibly uncovering new phenomena that could reshape our understanding of the universe.  

The ATLAS and CMS collaborations at the CERN LHC performed searches for $t\bar{t}t\bar{t}$ production using the $pp$ collision data recorded between 2015 and 2018 at a center-of-mass energy of 13 TeV, covering the decay channels with zero to four leptons.  These searches found evidence for $t\bar{t}t\bar{t}$ production with a significance of more than $3$ standard deviations from the background-only hypothesis, however, the observation level of $5$ standard deviations was not reached \cite{ATLAS:2020hpj,CMS:2023zdh}. Finally, in 2023 with the help of the updated identification techniques for charged leptons and $b$-jets, and by employing a revised multivariate analysis strategy to distinguish the signal process from the main backgrounds, the four top-quark production was observed with a significance larger than $5$ standard deviations by both the ATLAS and CMS collaborations \cite{ATLAS:2023ajo,CMS:2023ftu}.  In both measurements, events with two same-sign, three, and four charged leptons (electrons and muons) and additional jets have been analysed. Although the observation of the $pp \to t\bar{t}t\bar{t}+X$ process marked a milestone in the top-quark physics program at the LHC, detailed measurements of $pp \to t\bar{t}t\bar{t}+X$ in all final states, i.e. in the  $4\ell$, $3\ell$, $2\ell$ (including two same-sign as well as two opposite-sign charged leptons), $1\ell$ and  $0\ell$ decay channels, along with various differential cross-section distributions of interest to SM and beyond SM physics, are still pending.  

As far as theory is concerned, numerous theoretical predictions regarding the $pp\to t\bar{t}t\bar{t}$ process can be found in the literature. Briefly, they can be divided into two categories, predictions involving the stable top quark, which are important for the overall normalization of the process, and predictions incorporating top-quark decays, which are needed when studying various differential cross-section distributions in various fiducial phase-space regions that are of interest to the ATLAS and CMS collaborations. For the case of stable top quark, the first NLO QCD predictions were calculated in Ref. \cite{Bevilacqua:2012em} and later recomputed in Refs. \cite{Alwall:2014hca,Maltoni:2015ena}. The so-called complete NLO corrections,  that include all leading and subleading contributions together with their corresponding higher-order corrections at perturbative orders from ${\cal O}(\alpha_s^5)$ to ${\cal O}(\alpha^5)$, were provided in Ref. \cite{Frederix:2017wme}, where it was shown in detail that accidental cancellations occurred between different terms in $\alpha_s$ and $\alpha$ for the $pp \to t\bar{t}t\bar{t}+X$ process. Consequently, at the level of the total cross section, the NLO effects for this process are dominated by $\alpha_s$ corrections to the leading QCD process at ${\cal O}(\alpha_s^4)$. It should be noted, however, that the impact of the various subleading NLO corrections can be enhanced  at the differential cross-section level when the cuts on the final states are applied. Finally, the calculation  of  higher-order QCD corrections from soft gluon emission at the next-to-leading logarithmic accuracy was also performed for the $pp \to t\bar{t}t\bar{t}+X$ process \cite{vanBeekveld:2022hty}. These theoretical predictions also took into account constant ${\cal O}(\alpha_s)$ non-logarithmic contributions that do not vanish at the absolute production threshold at $4m_t$. 
To provide theoretical predictions in the fiducial phase-space regions, it is necessary to study the final state with $12$ particles, which greatly complicates the calculation of higher-order corrections for this process. Depending on the decay mode considered, the corresponding final state consists of four $b$-jets, additional light jets, and/or charged leptons together with (significant) missing transverse momentum from neutrinos escaping detection. To model top-quark decays,  one can straightforwardly incorporate leading order $t\to W^+b \to \ell^+ \nu_\ell \, b$ or $t\to W^+b\to  q \bar{q}^{\,\prime}\, b$ decays and employ parton shower approximations for additional collinear parton splitting or soft gluon emission contributions. This method has already been utilised in the case of the $pp\to t\bar{t}t\bar{t}+X$ process in the $1\ell$ decay channel \cite{Jezo:2021smh}. Another approach is to calculate higher-order QCD corrections not only to the production of four top quarks, but also in top-quark decays already at the level of matrix elements, thereby preserving spin correlations at the NLO level in QCD. For this purpose, in  Ref. \cite{Dimitrakopoulos:2024qib} the narrow-width approximation (NWA) was utilized to obtain higher-order predictions in the $4\ell$ decay channel.  Unfortunately, results for other decay channels are not yet available in the literature. 

The purpose of this article is to remedy this situation and to provide theoretical predictions for the $pp \to t\bar{t}t\bar{t}+X$ process in the $3\ell$ decay channel,  which include higher-order QCD corrections to the production and decays of four top quarks. Similarly to the case of the $4\ell$ decay channel, the NWA will also be used here. Such calculations would not only include all spin correlations for $2 \to 12$ processes at the NLO level in QCD, but would also allow us to use arbitrary cuts for the observed final states. In this way, the magnitude of higher-order corrections and the size of the theoretical uncertainties due to scale dependence can be verified in the realistic environment studied by the ATLAS and CMS collaborations. 

The article is organized as follows. In Section \ref{sec:calculation}, we briefly describe the calculations we perform. Our computational setup is provided in Section \ref{sec:setup}. Our LO and NLO QCD integrated (fiducial) cross-section results are discussed in Section \ref{sec:integrated}. On the other hand, various differential cross-section distributions are provided and analyzed in Section \ref{sec:differential}. Section \ref{sec:summary}  closes the text and discusses some obvious further directions of research.

%
\section{Description of the calculation}
\label{sec:calculation}
%

The production of four top quarks, while rare, can occur in high-energy $pp$ collisions when the center-of-mass energy of the colliding partons exceeds the $4m_t$ threshold. The unstable top quarks decay further into various final states, including those with $4\ell$, $3\ell$, $2\ell$, $1\ell$ or $0\ell$. In this study, our focus lies on the $3\ell$ decay channel, where one $W^\pm$ gauge boson decays hadronically while the remaining three $W$ gauge bosons decay leptonically. Therefore, the processes that we consider  are as follows:
\begin{align}\label{process}
\begin{split}
& pp \to t\bar{t}t\bar{t} +X\to W^+W^-W^+W^-b\bar{b} b\bar{b} +X \to\ell^+ \nu_\ell  \, \ell^- \bar{\nu}_\ell \,  \ell^+ \nu_\ell \, q \bar{q}^{\, \prime} \, b\bar{b}  b\bar{b} +X \,,\\
& pp \to t\bar{t}t\bar{t} +X\to W^+W^-W^+W^-b\bar{b} b\bar{b} +X\to\ell^+ \nu_\ell\,   \ell^- \bar{\nu}_\ell \,  \ell^- \bar{\nu}_\ell  \,q\bar{q}^{\, \prime} b\bar{b}  b\bar{b} +X\,,
\end{split}
\end{align}
where $\ell^{\pm} = e^{\pm},\mu^{\pm}$ as well as $q\bar{q}^{\, \prime} = d\bar{u}, s\bar{c}$ or $q\bar{q}^{\, \prime} =u\bar{d}, c\bar{s}$ for $W^-$ and $W^+$ respectively. At LO we only consider contributions of the order of $\mathcal{O}(\alpha_s^4 \alpha^8)$ as these are the dominant ones, see e.g. Ref. \cite{Frederix:2017wme}. In this case, the production of four top quarks occurs exclusively through QCD interactions and the electroweak couplings become relevant only during the decay phase. In addition, we calculate NLO QCD corrections of the order of $\mathcal{O}(\alpha_s^5 \alpha^8)$ to the aforementioned Born contributions. The unstable top quarks and $W$ bosons are treated in the NWA that allows us to separate the production and the decay stages of the four top quarks. In this approximation, by taking the limits $\Gamma_t / m_t, \; \Gamma_W / m_W \to 0 $ we force the intermediate resonances to be on-shell.  This means that only diagrams with four intermediate top-quark resonances are considered in the calculation. Off-shell effects related to triple, double, single and non-resonant top-quark contributions as well as finite top-quark and $W$ width effects are simply neglected throughout the calculation. The neglected contributions are suppressed by the $\Gamma/m$ ratio for sufficiently inclusive observables \cite{Fadin:1993kt}. They can, however, be enhanced in specific phase-space regions, like for example in the high $p_T$ tail of various dimensionful observables or in the vicinity of kinematical thresholds/edges, see e.g. Refs.  \cite{Bevilacqua:2019quz,Bevilacqua:2020pzy,Stremmer:2021bnk,Bevilacqua:2022nrm,Hermann:2022vit}. There is no doubt that the full off-shell prediction should be used if available, as it provides the most realistic description of the studied processes. However, obtaining NLO QCD results with full off-shell effects for the $2\to 12$ process, especially when hadronic decays of the $W^\pm$ gauge boson are involved, is not only computationally demanding but also very cumbersome, see Ref. \cite{Denner:2017kzu} for the simpler case of $t\bar{t}$ production. Consequently,  for the $pp \to t\bar{t}t\bar{t}+X$ process in the $3\ell$ decay channel, we focus on the on-shell approximation only. On the other hand, the separation of production and decays inherited in the NWA allows us to analyze the importance of higher-order effects at these two stages separately. Indeed, we can examine scenarios in which QCD corrections are applied both at the production and the decay stages, labeled as $\rm NLO_{full}$, as well as scenarios in which QCD corrections are solely applied during the production stage, denoted as $\rm NLO_{LO_{dec}}$. In this case, top-quark decays are considered with the LO accuracy only. In addition, we can assess the impact of expanding the top-quark width in the calculation, referred to as $\rm NLO_{exp}$. All the aforementioned approaches have been introduced in Ref. \cite{Dimitrakopoulos:2024qib}.

At LO, two types of subprocesses contribute to the processes listed in Eq.~\eqref{process}, namely $gg$ and $q\bar{q}/\bar{q}q$, where $q = u,d,c,s,b$. The former contribution is the dominant one with $72$ Feynman diagrams, while the latter consists of $14$ Feynman diagrams only and accounts for almost $12\%$ of the LO (fiducial) cross section. At NLO, we compute both virtual and real emission contributions. The virtual corrections are obtained by taking the interference of the sum of all one-loop diagrams with the Born amplitudes. For the real emission part, we can have an additional parton in the final state, thus,  the following subprocesses contribute: $gg, q\bar{q}/\bar{q}q, qg/gq, \bar{q}g/g\bar{q}$, where $q = u,d,c,s,b$. For the $gg, q\bar{q}/\bar{q}q$ initiated subprocesses the extra parton is always a gluon, that can be emitted during both the production and decay stages. For the $ qg/gq, \bar{q}g/g\bar{q}$ subprocesses the extra light- or $b$-quark is emitted only at the production stage of the process. Our NLO calculations are performed within the  \textsc{Helac-NLO} framework \cite{Bevilacqua:2011xh}  that consists of \textsc{Helac-1Loop} \cite{vanHameren:2009dr} and \textsc{Helac-Dipoles} \cite{Czakon:2009ss}. The \textsc{Helac-1Loop} Monte Carlo library comprises \textsc{CutTools} \cite{Ossola:2007ax} for the numerical evaluation of one-loop amplitudes using the OPP reduction method \cite{Ossola:2008zzb, Ossola:2008xq, Draggiotis:2009yb}, and \textsc{OneLoop} \cite{vanHameren:2010cp} for the evaluation of the scalar integrals. The real emission part is computed within the \textsc{Helac-Dipoles} framework, where both the Catani-Seymour \cite{Catani:1996vz, Catani:2002hc} and Nagy-Soper \cite{Bevilacqua:2013iha} subtraction schemes have been implemented for the extraction of soft and collinear singularities. The advantage of the latter approach lies in the fact
that in each case, there are fewer terms to be evaluated compared to the Catani-Seymour approach. The difference corresponds to the total number of possible spectators in the process under scrutiny, which are relevant only in the Catani-Seymour case. For this reason, the Nagy-Soper subtraction scheme is our default method throughout the computations. Finally, the phase-space integration is performed and optimized with \textsc{Parni} \cite{vanHameren:2007pt} and \textsc{Kaleu} \cite{vanHameren:2010gg}. 

Similar to our previous NLO QCD calculations for the $pp\to t \bar{t}t\bar{t}$ process in the  $4 \ell$ decay channel, also in this case we have conducted several cross-checks to ensure the correctness and reliability of our theoretical results. To begin with, we have used two different subtraction schemes for extracting the soft and collinear infrared singularities in the real emission part of the NLO calculation. We have reproduced our NLO QCD results with the help of the Catani-Seymour subtraction scheme and found excellent agreement with the results obtained using the default method. 
In the case of both subtraction schemes we have also verified the independence of our (complete) real emission results on an additional phase-space restriction imposed by the unphysical parameter $\alpha_{max}$, see e.g. Refs. \cite{Nagy:1998bb,Nagy:2003tz,Bevilacqua:2009zn,Czakon:2015cla} for the definition of this parameter. Additionally, the cancellation of the infrared $1/\epsilon$ and $1/\epsilon^2$ poles between virtual and real corrections, as provided by the ${\cal I}$-operator, has been checked numerically for several phase-space points for both the $gg$ and $q\bar{q}$ subprocesses.  We also monitor the numerical stability by checking Ward identities at every phase-space point. The events which do not pass this check have not been discarded from the calculation of the finite part, but rather recalculated with quadruple precision. On the other hand, for the $q\bar{q}$ subprocess we use the so-called scale test \cite{Badger:2010nx}, which is based on momentum rescaling. To further confirm the correctness of our results, we have performed various cross-checks using the \textsc{MadGraph}${}_{-}$\textsc{aMC@NLO} \cite{Alwall:2014hca} and \textsc{Recola} \cite{Actis:2016mpe} Monte Carlo programs.  Specifically, with the help of the \textsc{MadGraph}${}_{-}$\textsc{aMC@NLO}  program we have cross-checked the $1/\epsilon$ and $1/\epsilon^2$ poles, separately for the production and decays of the four top quarks. In the latter case, we explicitly calculated the infrared $\epsilon$ poles  for  $t \to W^+ b \to q\bar{q}^{\,\prime} b$ and  $t \to W^+ b \to l^+ \nu_l b$. On the other hand, using the \textsc{Recola} program we have calculated the aforementioned poles for the whole process, i.e. for the production and decays simultaneously. Finally, the finite part of the 1-loop amplitude has also been cross-checked with \textsc{Recola} for a few phase-space points for both $gg$ and $q\bar{q}$ subprocesses.

All our LO and NLO results are stored in modified \textsc{Les Houches Files} \cite{Alwall:2006yp} and then converted into  \textsc{Root Ntuple Files} \cite{Antcheva:2009zz, Bern:2013zja}. This allows us to efficiently change renormalization and factorization scale settings, use different PDF sets, fairly quickly estimate the internal PDF uncertainties from PDF error sets, produce various (infrared-safe) differential cross-section distributions as well as change their binning or apply more stringent cuts, all without rerunning the entire process from scratch. All these tasks are carried out with the help of the \textsc{HEPlot} program \cite{Bevilacqua:HEPlot}.

%
\section{Computational setup}
\label{sec:setup}
%

We consider the $pp \to t\bar{t}t\bar{t}+X$ process in the $3\ell$ decay channel as defined in Eq.~\eqref{process}.  We provide results at LO and NLO in QCD for the LHC Run III using a center-of-mass-energy of $\sqrt{s} = 13.6$ TeV. Since at the LHC the $\tau$ leptons are often analyzed separately due to their short lifetime and complex decay pattern, we will not consider them in our analysis. We work in the five-flavor scheme and keep the Cabibbo-Kobayashi-Maskawa mixing matrix diagonal. The numerical values of the  SM input parameters that enter our calculations are as follows:
\begin{equation}
\begin{array}{lll}
 G_{ \mu}= 1.166 3787 \cdot 10^{-5} ~{\rm GeV}^{-2}\,,  
& \quad \quad \quad &  m_{t}=172.5 ~{\rm GeV} \,,
\vspace{0.2cm}\\
 m_{W}= 80.379 ~{\rm GeV} \,, 
&&\Gamma_{W}^{\rm NLO} = 2.0972 ~{\rm GeV}\,, 
\vspace{0.2cm}\\
  m_{Z}=91.1876  ~{\rm GeV} \,, 
&& m_b = 0 ~{\rm GeV}\,.
\end{array}
\end{equation}
Given that three of the four $W$ bosons decay leptonically and do not receive QCD corrections, we account for higher-order effects in their decays by employing the NLO QCD value for $\Gamma_W$, evaluated for $\mu_R = m_W$, in both LO and NLO calculations. The electromagnetic coupling $\alpha$ is evaluated in the $G_\mu$-scheme and it is derived via the following relationship
\begin{equation}
\alpha_{G_\mu}=\frac{\sqrt{2}}{\pi} 
\,G_F \,m_W^2  \left( 1-\frac{m_W^2}{m_Z^2} \right)\,.\\ \vspace{0.2cm}
\end{equation}
We use the LO (NLO) top-quark width for our LO (NLO) calculations. They are obtained using the formulas in Ref. \cite{Denner:2012yc}, where $\alpha_s(\mu_R=m_t)$. The values for  $\Gamma_t^{\rm  LO}$  and  $\Gamma_t^{\rm NLO}$ read 
\begin{equation}
\label{top_widths}
\begin{array}{lll}
 \Gamma_{t}^{\rm LO} = 1.4806842 ~{\rm GeV}\,, &
 \quad \quad\quad \quad &
 \Gamma_{t}^{\rm NLO} = 1.3535983  ~{\rm GeV}\,.
\end{array}
\end{equation}
For the LO and NLO QCD results we follow the PDF4LHC working group's recommendation for SM processes \cite{Butterworth:2015oua} and employ the following three PDF sets: MSHT20 \cite{Bailey:2020ooq}, NNPDF3.1 \cite{NNPDF:2017mvq} and CT18 \cite{Hou:2019efy}. We consider the MSHT20  PDF sets as our default sets. Since CT18 lacks a LO PDF set, we replace it with CT14. Specifically, we use the CT14llo PDF set with $\alpha_s (m_Z) = 0.130$ \cite{Dulat:2015mca} for our LO calculations. The running of the strong coupling is provided with two-loop (one-loop) accuracy at NLO (LO) via the LHAPDF library \cite{Buckley:2014ana} where we assume $N_f=5$. We compute the cross section for two different  choices of the renormalization $(\mu_R)$ and factorization scale $(\mu_F)$ 
\begin{equation}
\begin{split}
\mu_R=\mu_F =\mu_0 & = 2 m_t \,,\\
\mu_R=\mu_F  =\mu_0 & =  \frac{1}{4} E_T \,,
\end{split}
\end{equation}
with $E_T$ given by
\begin{equation}
E_T= \sqrt{m_t^2 + p_T^2(t_1)} + \sqrt{m_t^2 + p_T^2(t_2)}+ \sqrt{m_t^2 + p_T^2(\,\bar{t}_1\,)}+ \sqrt{m_t^2 + p_T^2(\,\bar{t}_2\,)}\,,
\end{equation}
where $m_t$ is the nominal mass of the top quark. Finally, top and anti-top momenta needed in the definition of $E_T$ are reconstructed from their decay products. We estimate the uncertainty from missing higher orders by performing a conventional $7$-point scale variation around the central $\mu_{R,F}$ values by factors of $2$ subject to the constraint $0.5 \le \mu_R/\mu_F \le 2$. In addition, we calculate the internal PDF errors for the three PDF sets, using prescriptions provided by each collaboration. 

Light- and $b$-jets are constructed from all partons with $|\eta|<5$ using the infrared-safe anti-$k_{T}$ jet algorithm \cite{Cacciari:2008gp} with the resolution parameter $R = 0.4$. All final states have to fulfill the following selection criteria that closely mimic the ATLAS and CMS detector response \cite{ATLAS:2023ajo,CMS:2023ftu}: 
\begin{equation}
\begin{array}{lll}
 p_{T,\,\ell}>25 ~{\rm GeV}\,,    
 &\quad \quad \quad \quad\quad|y_\ell|<2.5\,,&
\quad \quad \quad \quad \quad
\Delta R_{\ell
 \ell} > 0.4\,,\\[0.2cm]
p_{T,\,b}>25 ~{\rm GeV}\,,  
&\quad \quad\quad\quad\quad |y_b|<2.5 \,, 
 &\quad \quad\quad \quad \quad
\Delta R_{bb}>0.4\,,\\[0.2cm]
p_{T,\,j}>25 ~{\rm GeV}\,,  
&\quad \quad\quad\quad\quad |y_j|<2.5 \,, 
 &\quad \quad\quad \quad \quad
\Delta R_{bj}>0.4\,.
\end{array}
\end{equation}
We require exactly 3 charged leptons, at least two light jets and at least 4 $b$-jets in the final state. Furthermore, we require that at least one pair of light jets $(jj)$ with $\Delta R_{jj} > 0.4$ has an invariant mass $(M_{jj})$ in the following range
\begin{equation}\label{qcut25}
    |M_{jj} - m_W| < Q_{cut} \,,
\end{equation}
where $m_W$ is the nominal mass of the $W$ gauge boson. Our default setup utilizes $Q_{cut} = 25 \; \rm GeV$ but we also investigate the effect of varying the value of $Q_{cut}$ on the integrated and differential (fiducial) cross sections. The rationale behind the restriction given in Eq.~\eqref{qcut25} is that we want to suppress NLO configurations in which one of the light quarks from the $W^\pm$ decays undergoes recombination with another light or bottom quark.  In such cases, the additional jet, if resolved and passed all the cuts, can play the role of a second decay product of the $W$ gauge boson. It is already known that the absence of such a requirement in processes involving hadronic $W$ boson decays can lead to very large ${\cal K}= \sigma^{\rm NLO}/\sigma^{\rm LO}$ factors, see e.g. Refs.  \cite{Melnikov:2011ta,Denner:2017kzu,Stremmer:2023kcd}. This is because the topologies described above do not mimic corrections to the born-level amplitude in which the $2$ light jets from the $W^\pm$ decays are always well-separated with an invariant mass equal to the $W$ mass. Finally, no restrictions are applied to the kinematics of the additional light- or $b$-jet (if resolved), as well as the total missing transverse momentum from the three neutrinos.

%
\section{Integrated fiducial cross sections}
\label{sec:integrated}
%
%

We start with the LO and NLO integrated fiducial cross-section results for the $pp \to t\bar{t} t \bar{t}+X$ process in the 3$\ell$  decay channel at the LHC Run III with $\sqrt{s} = 13.6 \; \rm TeV$.  Our default setup utilizes the MSHT20 PDF set as well as the expanded NWA version already described in detail in Ref. \cite{Dimitrakopoulos:2024qib}. In this approach, the decay width of the top quark is treated as a perturbative parameter everywhere in the calculation. Thus, a proper expansion in the strong coupling is employed.  We will henceforth refer to this approach as $\sigma^{\rm NLO}_{\rm exp}$ or simply $\sigma^{\rm NLO}$.  In Table \ref{tab:lo_nlo} the LO and NLO results with  $|M_{jj}-m_W| <  25$ GeV are given. Our (N)LO cross sections are provided for (N)LO MSHT20, NNPDF3.1 and CT18 PDF sets and two different scale settings, along with their scale and PDF uncertainties. In the last column, the corresponding $\mathcal{K}$ factors are depicted. 
\begin{table}[t!]
\centering
\scalebox{0.9}{
\begin{tabular}{ccccccc}
\midrule\midrule
PDF & $\sigma^{\textrm{LO}}$ [ab] & $\delta_{scale}$ & $\sigma^{\textrm{NLO}}$ [ab] & $\delta_{scale}$ & $\delta_{PDF}$  &  $\mathcal{K} = \sigma^{\textrm{NLO}}/\sigma^{\textrm{LO}}$\\
\midrule\midrule
\multicolumn{7}{c}{\centering $\mu_R = \mu_F = \mu_0 = 2m_t$}\\
\midrule\midrule
MSHT20 & 35.575(2) & \begin{tabular}[c]{@{}c@{}}$+25.883$ (73\%)\\$-13.982$ (39\%)\end{tabular} & 42.25(2) & \begin{tabular}[c]{@{}c@{}}$+8.27$ (20\%)\\$-9.72$ (23\%)\end{tabular} & \begin{tabular}[c]{@{}c@{}}$+1.76$ (4\%)\\$-1.32$ (3\%) \end{tabular} & 1.19\\
\\
NNPDF3.1 & 30.581(2) & \begin{tabular}[c]{@{}c@{}}$+21.781$ (71\%)\\$-11.853$ (39\%)\end{tabular} & 41.89(2) & \begin{tabular}[c]{@{}c@{}}$+8.28$ (20\%)\\$-9.71$ (23\%)\end{tabular} & \begin{tabular}[c]{@{}c@{}}$+0.89$ (2\%)\\$-0.89$ (2\%) \end{tabular} & 1.37\\
\\
CT18 & 37.675(3) & \begin{tabular}[c]{@{}c@{}}$+26.858$ (71\%)\\$-14.655$ (39\%)\end{tabular} & 41.89(2) & \begin{tabular}[c]{@{}c@{}}$+8.20$ (20\%)\\$-9.62$ (23\%)\end{tabular} & \begin{tabular}[c]{@{}c@{}}$+2.42$ (6\%)\\$-2.00$ (5\%) \end{tabular} & 1.11\\
\midrule\midrule
\multicolumn{7}{c}{\centering $\mu_R = \mu_F = \mu_0 = E_T/4$}\\
\midrule\midrule
MSHT20 & 39.424(3) & \begin{tabular}[c]{@{}c@{}}$+29.111$ (74\%)\\$-15.632$ (40\%)\end{tabular} & 44.91(2) & \begin{tabular}[c]{@{}c@{}}$+7.91$ (18\%)\\$-10.16$ (23\%)\end{tabular} & \begin{tabular}[c]{@{}c@{}}$+1.84$ (4\%)\\$-1.38$ (3\%) \end{tabular} & 1.14\\
\\
NNPDF3.1 & 34.193(3) & \begin{tabular}[c]{@{}c@{}}$+24.868$ (73\%)\\$-13.413$ (39\%)\end{tabular} & 44.60(2) & \begin{tabular}[c]{@{}c@{}}$+7.87$ (18\%)\\$-10.16$ (23\%)\end{tabular} & \begin{tabular}[c]{@{}c@{}}$+0.94$ (2\%)\\$-0.94$ (2\%) \end{tabular} & 1.30\\
\\
CT18 & 41.306(3) & \begin{tabular}[c]{@{}c@{}}$+29.728$ (72\%)\\$-16.169$ (39\%)\end{tabular} & 44.50(2) & \begin{tabular}[c]{@{}c@{}}$+7.86$ (18\%)\\$-10.06$ (23\%)\end{tabular} & \begin{tabular}[c]{@{}c@{}}$+2.51$ (6\%)\\$-2.08$ (5\%) \end{tabular} & 1.08\\
\midrule\midrule
\end{tabular}}
\caption{\textit{Integrated fiducial cross-section results at LO and NLO in QCD for the $pp \to t\bar{t}t\bar{t}+X$ process in the $3\ell$ decay channel at the LHC with $\sqrt{s} = $ 13.6 TeV. At NLO QCD, at least one light jet pair that fulfills the criterion of Eq.~\eqref{qcut25} with $Q_{cut} = 25 \; GeV$ is also required. Results are presented for the (N)LO MSHT20, NNPDF3.1 and CT18 PDF sets. They are evaluated using $\mu_0=2m_t$ and $\mu_0=E_T/4$. The theoretical uncertainties due to $7$-point scale variation and internal PDF errors are also given. In the last column, the ${\cal K}$-factor is shown.}}
\label{tab:lo_nlo} 
\end{table}
We can clearly see that the LO results provide only a rough estimate of the cross section due to the large scale uncertainties that are of the order of $70\%$ for both scale choices. However, the situation improves considerably at the NLO level, where these uncertainties decrease to about $20\%$ for both scale settings. In addition, the NLO  cross sections obtained with $\mu_0=2m_t$ and $\mu_0=E_T/4$ differ by about $6\%$, which is well within their corresponding theoretical uncertainties. We can also observe that although at LO the central values for the cross section vary significantly for the three PDF sets, there is a high level of agreement among the results at NLO in QCD. Indeed, the differences among the three LO results are of the order of $20\%$, while at NLO in QCD they are reduced to $1\%$ only. A similar pattern has already been observed for this process in the $4\ell$ decay channel.
The reasons for this are the large differences in the LO and NLO gluon PDFs at small $x$ and the different experimental data used in the global PDF analyses for these two cases, see Ref.  \cite{Dimitrakopoulos:2024qib} for more details. The PDF uncertainties are independent of the scale setting and are of the order of  $4\%$ for MSHT20, $6\%$ for CT18 while for NNPDF3.1 they account for only $2\%$. Finally, the $\mathcal{K}$-factors are slightly smaller when the dynamical scale setting is employed. However, the size of the $\mathcal{K}$-factors highly depends on the PDF set under consideration. Consequently, higher-order QCD effects are in the range of $8\%-37\%$. This large range, discussed in detail in our previous work on the $4\ell$ decay channel, is primarily due to the substantial differences in LO cross sections.

\begin{figure}[t!]
        \centering        
        \includegraphics[width=0.48\linewidth]{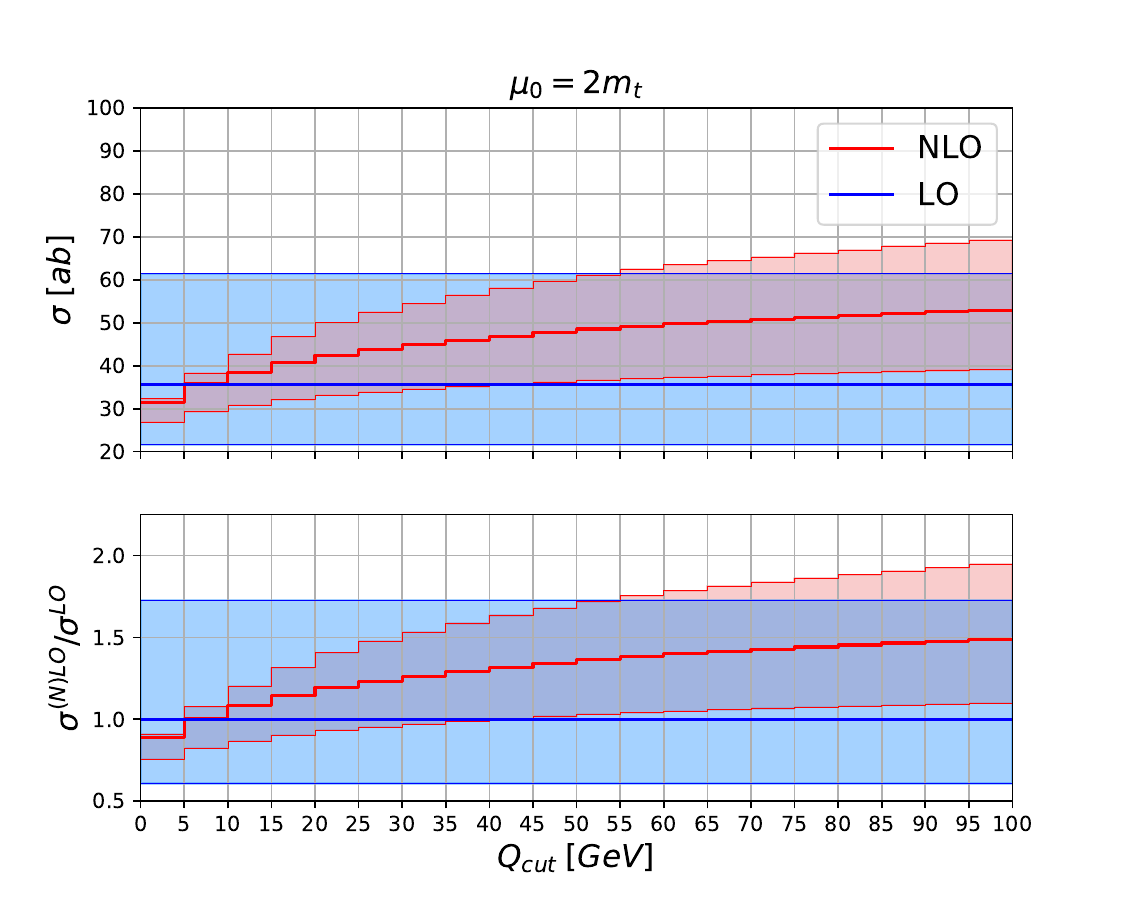}
        \includegraphics[width=0.48\linewidth]{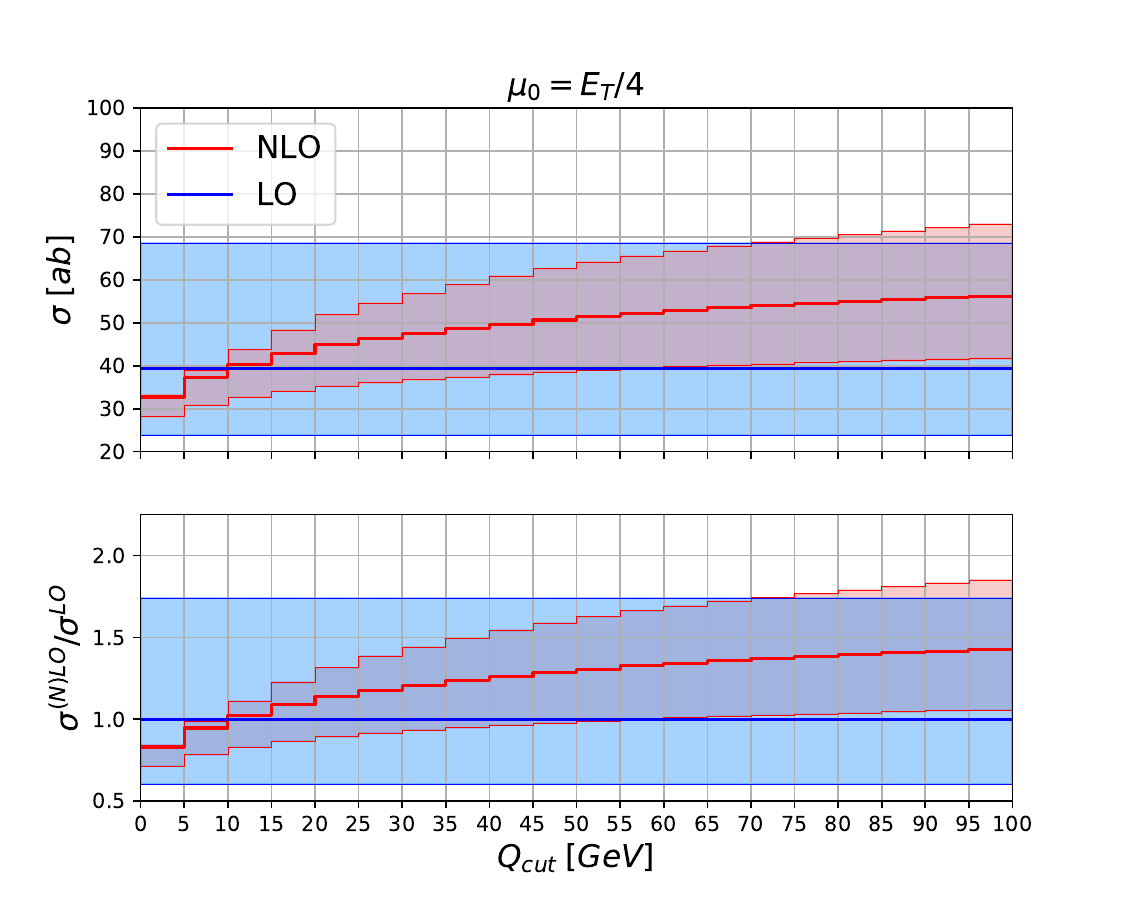}
        \caption{\textit{Integrated fiducial cross-section results at LO and NLO in QCD for the $pp \to t\bar{t}t\bar{t}+X$  process in the $3\ell$ decay channel at the LHC with $\sqrt{s} = $ 13.6 TeV, as a function of $Q_{cut}$ defined as $|M_{jj} - m_W| < Q_{cut}$. Results are shown for the (N)LO MSHT20 PDF set with $\mu_0 = 2m_t$  and $\mu_0 = E_T/4$. The theoretical uncertainties due to $7$-point scale variation are also given. The lower panel displays the $\mathcal{K}$-factor with its uncertainty band along with the relative scale uncertainties of the LO cross section.}}
         \label{fig:Qcut}
\end{figure}

In the next step, we examine the dependence of the results on the $Q_{cut}$ cut applied to the two light jets in the final state
\begin{equation}
    |M_{jj} - m_W| < Q_{cut} \,.
\end{equation}
As mentioned in Section \ref{sec:setup}, we expect a high sensitivity of the ${\cal K}$-factor to the choice of $Q_{cut}$, indicating the need for a closer examination of this issue. This dependence is illustrated in Figure \ref{fig:Qcut}. The top panels display the integrated (fiducial) cross section as a function of $Q_{cut}$ at LO  and NLO QCD for $\mu_R=\mu_F=\mu_0 = 2m_t$ and $\mu_R=\mu_F=\mu_0 = E_T/4$. The lower panels display the ${\cal K}$-factors together with their uncertainty bands and the relative scale uncertainties of the LO cross sections.
As expected, the LO cross section is independent of the $Q_{cut}$ parameter, since the two light jets always come from the decay of the $W$ boson. At NLO, the integrated fiducial cross section increases drastically when the value of $Q_{cut}$ rises, leading to large ${\cal K}$ factors close to $1.5$ when $Q_{cut} = 100$  GeV. This behavior can be explained by kinematic configurations in which one parton from the $W$ decays is recombined with another parton to form a jet, such that the total number of light-jets coming from the $W$ decays decreases. Although such configurations are not present at LO, they can manifest themselves at NLO. In this case, an additional light jet, originating from another decay or the production stage of the process, which successfully passes all selection criteria, can mimic the second decay product of the $W$ gauge boson. Despite the fact that such configurations can not be directly interpreted as real corrections to the born-level amplitude for the $pp \to t\bar{t}t\bar{t}+X$ process in the $3\ell$ decay channel, they have to be taken into account because they contain two light jets in the final state. Exactly these configurations, which can be described as a background process, cause a significant increase in the ${\cal K}$ factors. Furthermore, for large values of $Q_{cut}$, NLO scale uncertainties become similar in size to LO scale uncertainties, emphasizing the fact that such topologies are mainly governed by LO dynamics.
\begin{figure}[t!]
        \centering        
        \includegraphics[width=0.48\linewidth]{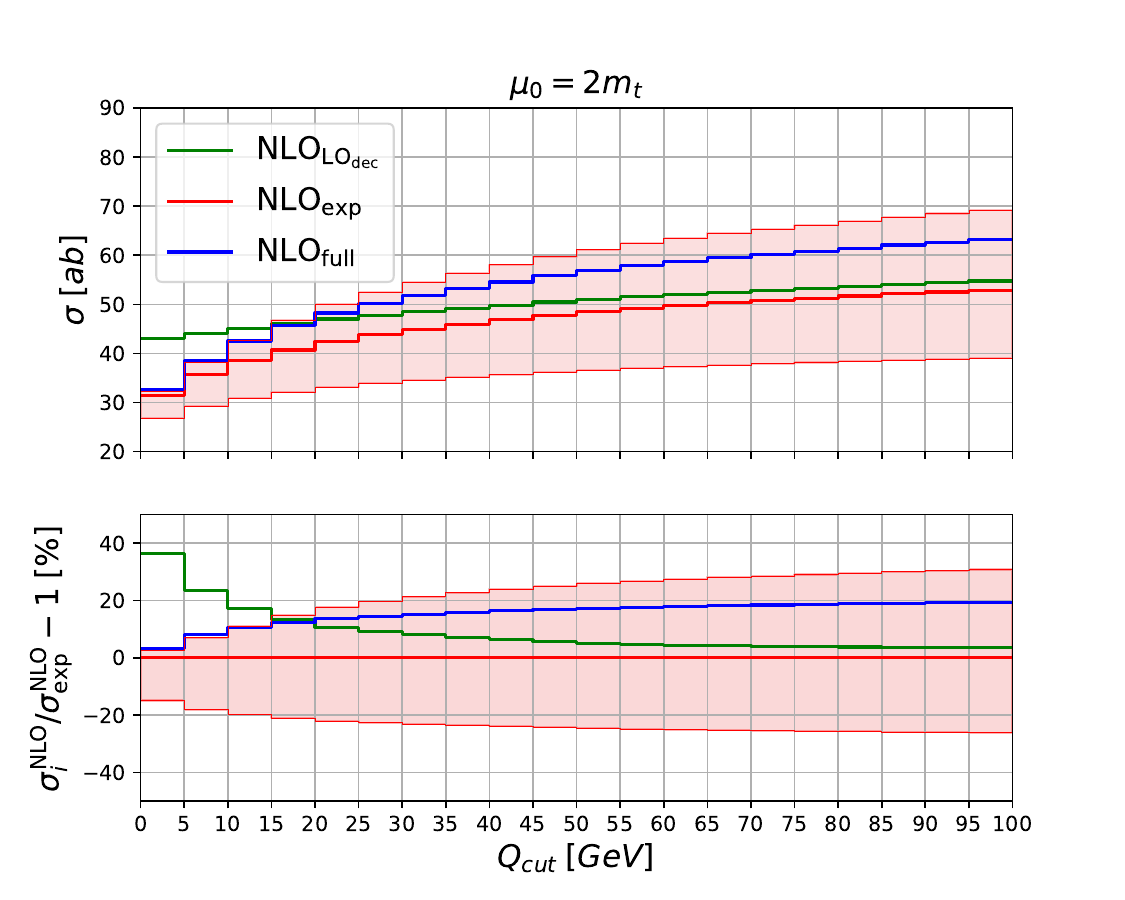}
        \includegraphics[width=0.48\linewidth]{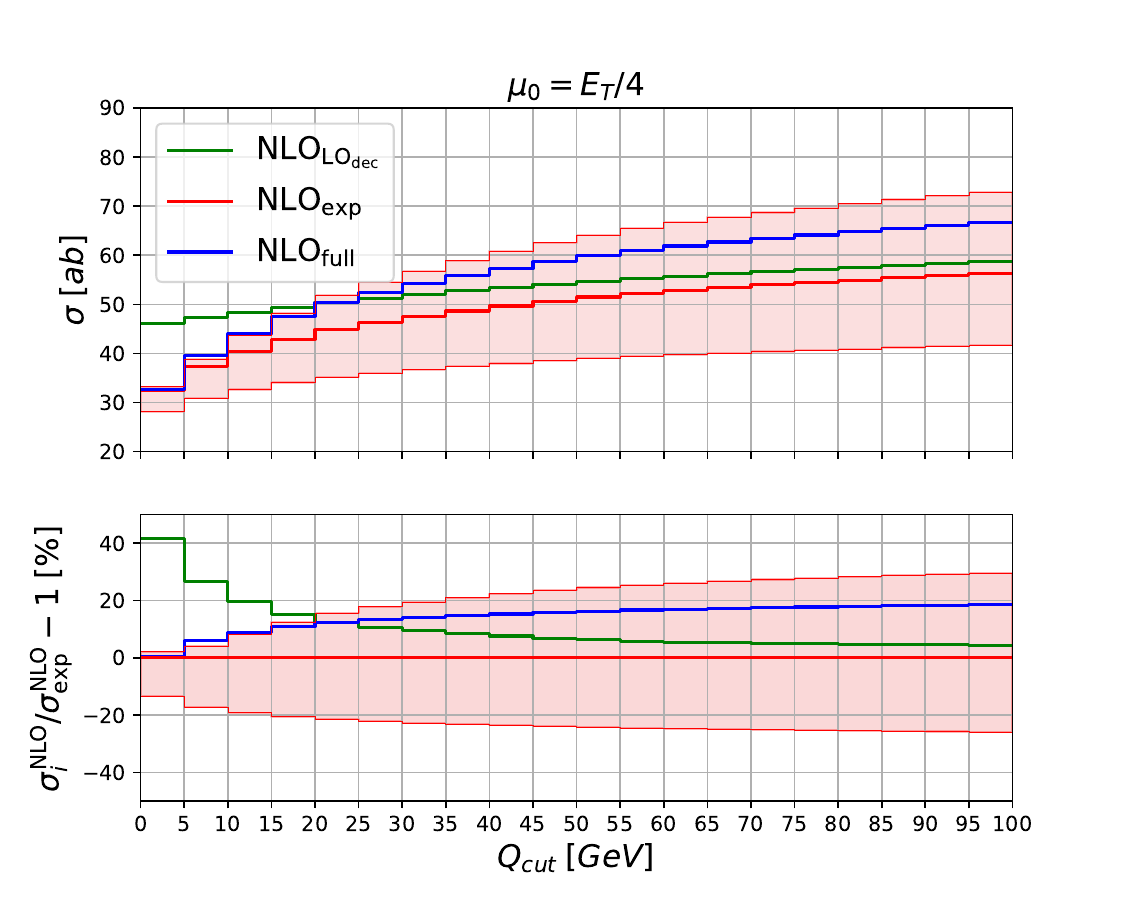}
        \caption{\textit{Integrated fiducial cross section for three different scenarios, $\sigma^{\rm NLO}_{\rm exp}$, $\sigma^{\rm NLO}_{\rm LO_{dec}}$  and $\sigma^{\rm NLO}_{\rm full}$ at NLO in QCD for the $pp \to t\bar{t}t\bar{t}+X$  process in the $3\ell$ decay channel at the LHC with $\sqrt{s} = $ 13.6 TeV, as a function of $Q_{cut}$ defined as $|M_{jj} - m_W| < Q_{cut}$. Results are shown for the NLO MSHT20 PDF set with  $\mu_0 = 2m_t$ and  $\mu_0 = E_T/4$. The upper plots show absolute predictions while the lower ones illustrate the percentage differences compared to the $\sigma^{\rm NLO}_{\rm exp}$ case. In all panels, uncertainty bands for the $\sigma^{\rm NLO}_{\rm exp}$ case are also provided.}}
         \label{fig:Qcut_comp}
\end{figure}

Another interesting feature worth investigating is how our NLO results change if we consider QCD corrections only at the production stage, while treating top quark decays with LO accuracy.  We will denote this approach as $\sigma^{\rm NLO}_{\rm LO_{dec}}$. In addition, we would like to investigate the case where the width of the top quark is treated as a fixed parameter in the NLO calculation. We will denote such an approach as $\sigma^{\rm NLO}_{\rm full}$. Both approaches were described in detail in Ref. \cite{Dimitrakopoulos:2024qib}. Since our results are very sensitive to the $Q_{cut}$ cut, we first plot the $Q_{cut}$ dependence of the $\sigma^{\rm NLO}_{\rm exp}$, $\sigma^{\rm NLO}_{\rm LO_{dec}}$ and $\sigma^{\rm NLO}_{\rm full}$ cross sections for  $\mu_0 = 2m_t$ and  $\mu_0 = E_T/4$. These results are presented in  Figure \ref{fig:Qcut_comp}, where the upper panels show the absolute predictions, while the lower panels depict the percentage differences of the various scenarios compared to our default $\sigma^{\rm NLO}_{\rm exp}$ case. In all panels, the scale uncertainties for the expanded NWA case are also displayed. The increase in the cross sections for high values of $Q_{cut}$  is mainly due to the extra radiation emitted at the production stage. This is evident, as the result for $\sigma^{\rm NLO}_{\rm LO_{dec}}$ goes closer to the corresponding result for $\sigma^{\rm NLO}_{\rm exp}$ in this region, indicating that QCD corrections at the production stage are dominant there.  Indeed, for $Q_{cut}$ close to 100 GeV, the differences between the two approaches are less than $5\%$. Conversely, with more stringent cuts applied to the invariant mass of the two light jets, the two approaches can exhibit large differences. For example, when $Q_{cut} = 5$  GeV these differences are of the order of $40\%$, independently of the scale setting. In addition, for small values of the $Q_{cut}$ cut, the $\sigma^{\rm NLO}_{\rm LO_{dec}}$ predictions lie completely outside of the uncertainty bands for the expanded case. This is because the invariant mass of the light jets is significantly affected when QCD corrections are applied at the decay stage, particularly when additional radiation is emitted from one of the quarks originating from the $W^\pm$ decays. This leads to a broad distribution in the $M_{jj}$ observable for the $\sigma^{\rm NLO}_{\rm exp}$ case. In contrast, when QCD corrections are only applied at the production stage, the $M_{jj}$ observable exhibits a sharper peak around the $W$ boson mass. As a consequence, choosing a smaller value for the parameter $Q_{cut}$ removes substantial contributions in the $\sigma^{\rm NLO}_{\rm exp}$ case, while the $\sigma^{\rm NLO}_{\rm LO_{dec}}$ case is less affected by this choice. In addition, for values above $Q_{cut} = (20-25)$  GeV the $\sigma^{\rm NLO}_{\rm full}$ result is larger by $15\%-20\%$ compared to the $\sigma^{\rm NLO}_{\rm exp}$ one for both scale choices. The corresponding differences below $(20-25)$ GeV are smaller and almost vanish when the two light jets have an invariant mass very close to the nominal value of the mass of the $W$ gauge boson. We can also see that choosing $Q_{cut} = 25$ GeV as the default value to perform our calculations seems to be a very reasonable choice, as it helps mitigate both the substantial increase in NLO QCD cross-section values, which would lead to significant ${\cal K}$-factors, and the large differences between the various NWA approaches.  
\begin{table}[t!]
\centering
\label{tab:lodec}
\resizebox{0.85\columnwidth}{!}{
\begin{tabular}{ccccc}
\midrule\midrule
Decay treatment & $\sigma_i^{\rm NLO}$ [ab] & $+\delta_{scale}$ [ab] & $-\delta_{scale}$ [ab] & $\sigma_i^{\rm NLO}/\sigma^{\rm NLO}_{\rm exp}-1$\\
\midrule\midrule 
\\
\multicolumn{5}{c}{\centering  $Q_{cut} = 25 \; \rm GeV$} \\ \\
\midrule\midrule
\multicolumn{5}{c}{\centering $\mu_R = \mu_F = \mu_0 = 2m_t$}\\
\midrule
${\rm full}$ & 47.92(2) & $+4.56$ (10\%) & $-9.22$ (19\%) & $+13.4\%$ \\
\\
${\rm LO_{dec}}$ & 46.76(2) & $+12.33$ (26\%) & $-11.51$ (25\%) & $+10.7\%$\\
\\
${\rm exp}$ & 42.25(2) & $+8.27$ (20\%)  & $-9.72$ (23\%) & $-$\\
\midrule
\multicolumn{5}{c}{\centering $\mu_R = \mu_F = \mu_0 = E_T/4$}\\
\midrule
${\rm full}$ & 50.45(3) & $+3.47\;(7\%)$ & $-9.32\;(19\%) $ & $+12.3\%$\\
\\
${\rm LO_{dec}}$ & 50.17(3) & $+12.95\;(26\%)$ & $-12.34\;(25\%)$ & $+11.7\%$\\
\\
${\rm exp}$ & 44.91(2) & $+7.91\;(18\%)$ & $-10.16\;(23\%)$ & $-$ \\
\midrule
\midrule
\\
\multicolumn{5}{c}{\centering  $Q_{cut} \to \infty$} \\ \\
\midrule\midrule
\multicolumn{5}{c}{\centering $\mu_R = \mu_F = \mu_0 = 2m_t$}\\
\midrule
${\rm full}$ & 82.45(7) & $+33.16$ (40\%) & $-24.17$ (29\%) & $+24.2\%$ \\
\\
${\rm LO_{dec}}$ & 67.80(3) & $+30.49$ (45\%) & $-20.64$ (30\%) & $+2.2\%$\\
\\
${\rm exp}$ & 66.36(5) & $+29.01$ (44\%)  & $-20.17$ (30\%) & $-$\\
\midrule
\multicolumn{5}{c}{\centering $\mu_R = \mu_F = \mu_0 = E_T/4$}\\
\midrule
${\rm full}$ & 86.7(1) & $+33.1\;(38\%)$ & $-25.0\;(29\%) $ & $+23.5\%$\\
\\
${\rm LO_{dec}}$ & 72.05(3) & $+31.83\;(44\%)$ & $-21.83\;(30\%)$ & $+2.6\%$\\
\\
${\rm exp}$ & 70.19(7) & $+29.69\;(42\%)$ & $-21.12\;(30\%)$ & $-$ \\
\midrule
\midrule
\end{tabular}
}
\caption{\textit{Integrated fiducial cross sections at NLO in QCD for  the $pp \to t\bar{t}t\bar{t}+X$  process in the $3\ell$ decay channel at the LHC with $\sqrt{s} =  13.6$ TeV for three different scenarios: $\sigma^{\rm NLO}_{\rm full}$, $\sigma^{\rm NLO}_{\rm LO_{dec}}$ and  $\sigma^{\rm NLO}_{\rm exp}$. The results are given with and without the criterion of 
Eq.~\eqref{qcut25}. In the former case, $Q_{cut} = 25$ GeV is required, while in the  $Q_{cut} \to \infty$ scenario there is no restriction on $M_{jj}$. All results are provided with the NLO MSHT20 PDF set and for $\mu_0=2m_t$ and $\mu_0=E_T/4$. Also given are theoretical uncertainties coming from the scale variation. In the last column the percentage difference to the $\sigma^{\rm NLO}_{\rm exp}$ result is also shown.}}
\label{tab:scenarios} 
\end{table}

To conclude this section, in Table \ref{tab:scenarios} we present the integrated fiducial cross-section results along with their scale uncertainties for the various NWA treatments both for our default setup that utilizes $|M_{jj}-m_W| < Q_{cut} = 25$ GeV and for the case where no restriction on the invariant mass of the two light jets is imposed. This extreme scenario is equivalent to putting $Q_{cut} \to \infty$ in Eq.~\eqref{qcut25}. By comparing the two cases mentioned above, we will be able to directly assess the impact of the $Q_{cut}$ cut in our results. In both cases, results are provided for  $\mu_0 = 2m_t$ and $\mu_0 = E_T/4$. Starting our discussion with $Q_{cut} = 25$ GeV, we notice that neglecting higher-order corrections in the top-quark decays overestimates the cross section by about  $11\%-12\%$ depending on the scale choice. The difference between $\sigma^{\rm NLO}_{\rm full}$ and $\sigma^{\rm NLO}_{\rm exp}$  is of similar magnitude and is at the level of $12\%-13\%$.  These effects are within the corresponding theoretical uncertainties, that are of the order of $23\%$, $26\%$ and $19\%$ for $\sigma^{\rm NLO}_{\rm exp}$, $\sigma^{\rm NLO}_{\rm LO_{dec}}$ and $\sigma^{\rm NLO}_{\rm full}$, respectively. Shifting our focus to the $Q_{cut}\to \infty$ scenario, we observe larger deviations between $\sigma^{\rm NLO}_{\rm full}$ and $\sigma^{\rm NLO}_{\rm exp}$ as compared to the $Q_{cut} = 25$ GeV case. These differences 
may now be as high as $24\%$, but are still within the corresponding scale uncertainties. Neglecting higher-order effects in the top-quark decays has less impact when $Q_{cut}\to \infty$. Indeed, the differences between $\sigma^{\rm NLO}_{\rm LO_{dec}}$ and $\sigma^{\rm NLO}_{\rm exp}$ are at the  $2\%-3\%$ level only. Most importantly, without any restriction on the invariant mass of the two light jets, the magnitude of the scale uncertainties is substantial for all NWA treatments, reaching values up to $38\%-40\%$ even for the $\sigma^{\rm NLO}_{\rm full}$ case. The corresponding scale uncertainties for $\sigma^{\rm NLO}_{\rm LO_{dec}}$ and $\sigma^{\rm NLO}_{\rm exp}$ are in the range of $42\%-45\%$, depending on the scale choice. Such sizes of theoretical errors are usually present for LO predictions. Finally, we would like to underline the fact, that in the $Q_{cut}\to \infty$ case the  ${\cal K}$-factors increase significantly.  Indeed, for the  ${\cal K}$-factors defined according to ${\cal K}=\sigma^{\rm NLO}_{\rm exp}/\sigma^{\rm LO}$ we obtain ${\cal K}=(1.8-1.9)$, whereas for  ${\cal K}=\sigma^{\rm NLO}_{\rm full}/\sigma^{\rm LO}$, the corresponding values are even up to  ${\cal K} =  (2.2-2.3)$. Thus, for the $Q_{cut}\to \infty$ case not only theoretical uncertainties are substantial, up to even $45\%$, but also higher-order effects are of the order of  $80\%-130\%$. This emphasizes the importance of employing the $Q_{cut}$ cut 
to restore perturbative convergence in the calculation of higher-order corrections for the $pp \to t\bar{t}t\bar{t}+X$ process in the $3\ell$ decay channel. 

%
\section{Differential fiducial cross-section distributions}
\label{sec:differential}
%
%

\begin{figure}[!t]
        \includegraphics[width=0.5\linewidth]{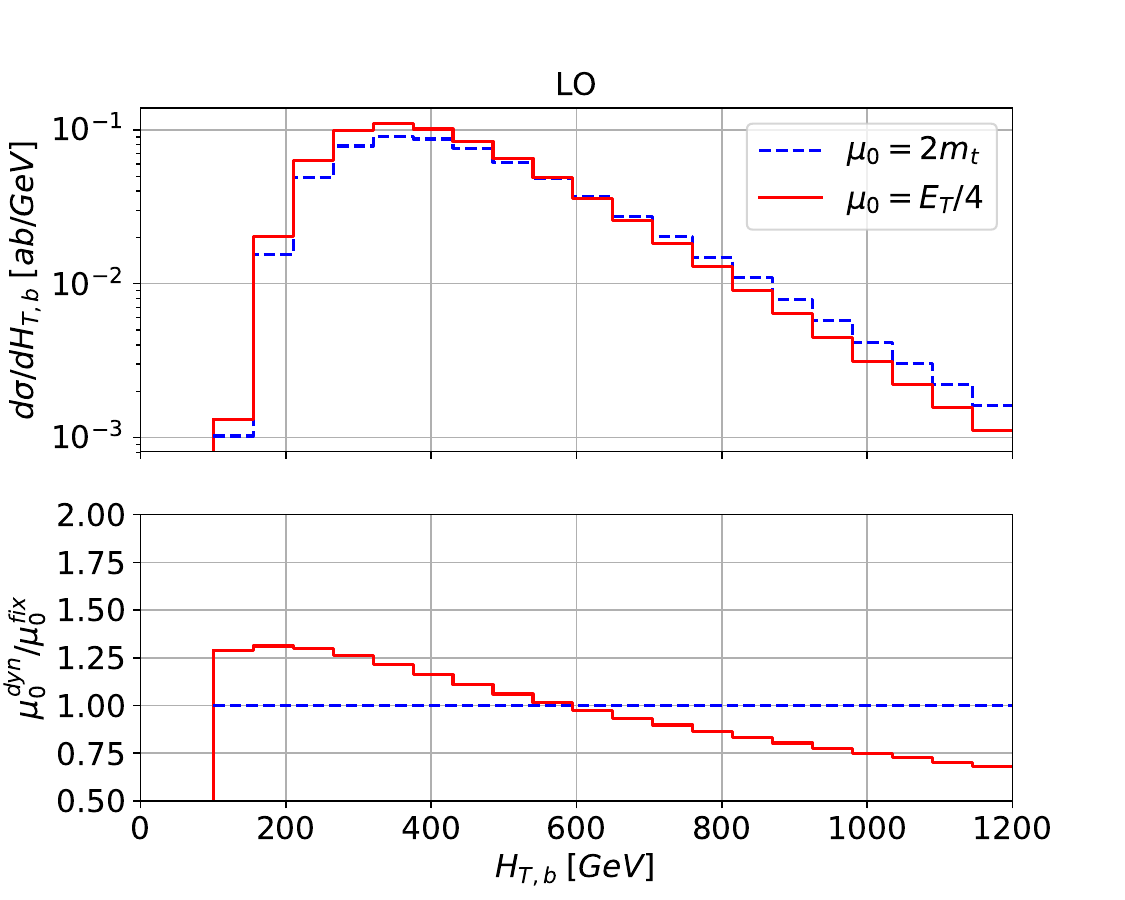}
        \includegraphics[width=0.5\linewidth]{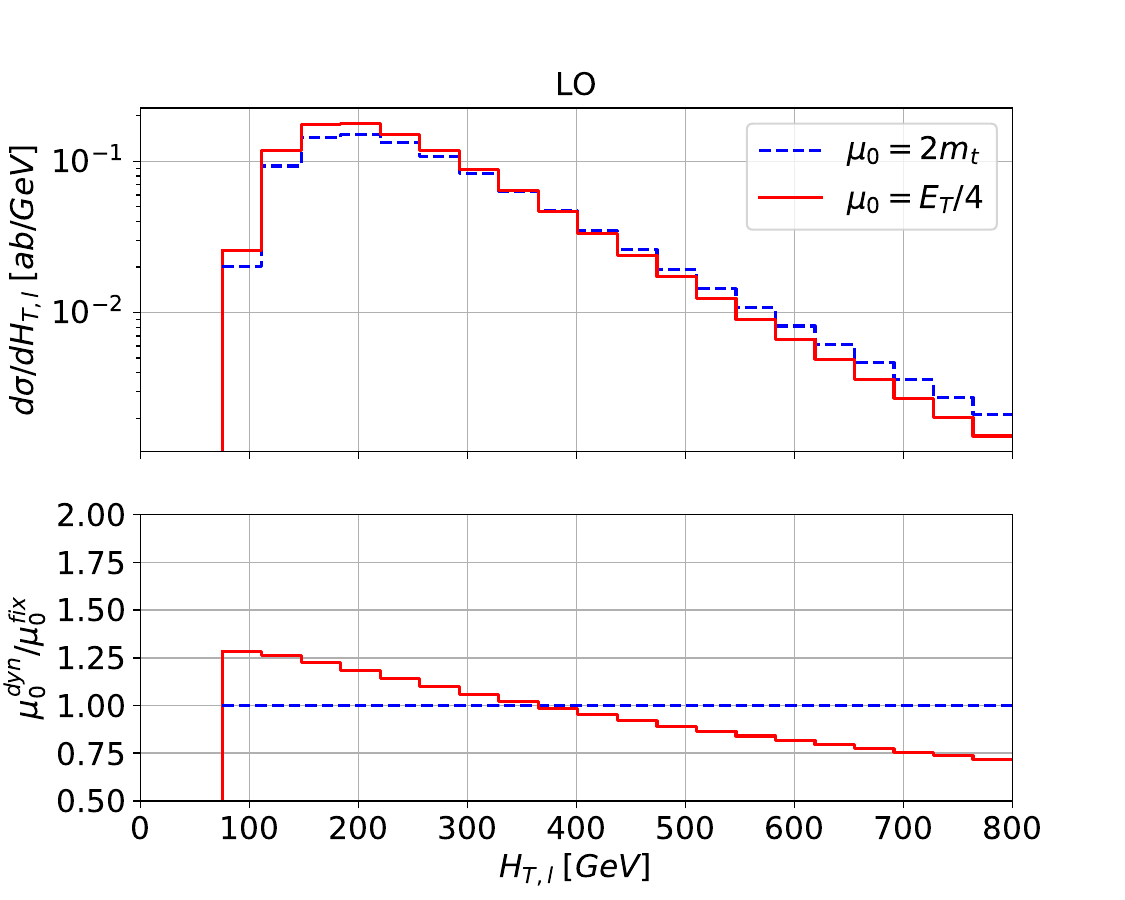}
        \includegraphics[width=0.5\textwidth]{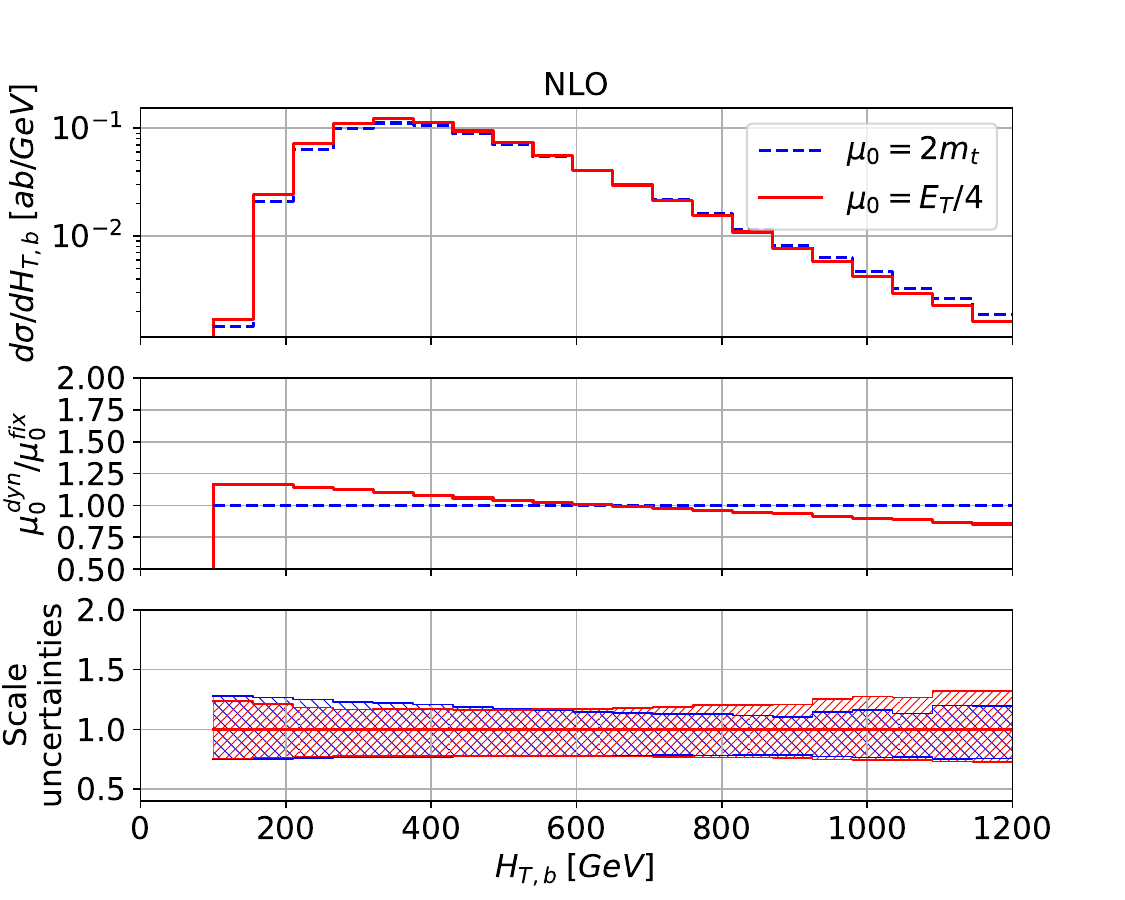} 
        \includegraphics[width=0.5\textwidth]{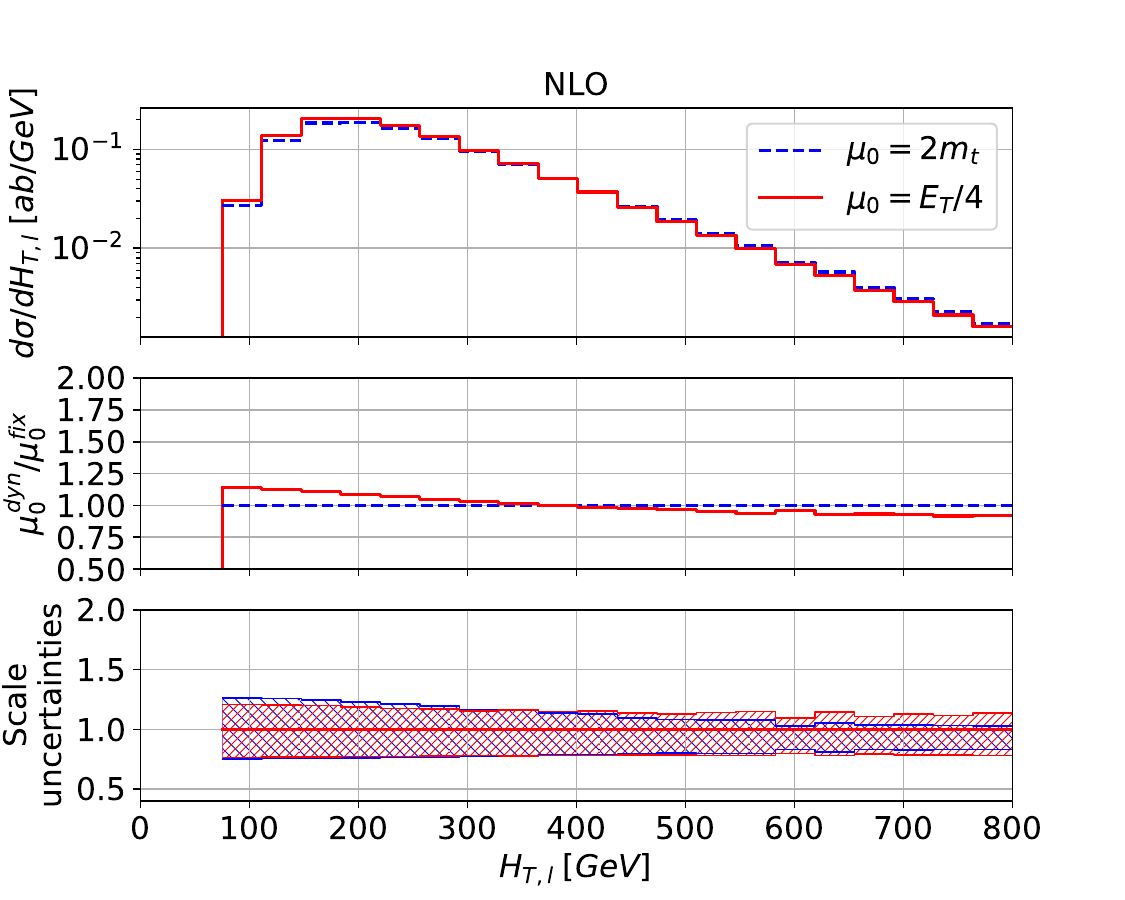}
\caption{\textit{Differential cross-section distributions for $H_{T,\,b}$  and  $H_{T,\,l}$  for the $pp \to t\bar{t}t\bar{t}+X$ process in the $3\ell$ decay channel at the LHC with $\sqrt{s} = 13.6$ TeV.  The upper panels display the absolute (N)LO predictions for $\mu_0 = 2m_t$ and $\mu_0 = E_T/4$. The lower (middle) panels illustrate ratios to the result with $\mu_0 = 2m_t$. For the NLO case, the absolute magnitude of the scale uncertainty for both scale settings is also  shown in the bottom panels. All results are presented with $|M_{jj}-m_W|< Q_{cut} = 25$ GeV  and for the (N)LO MSHT20 PDF set.}}
    \label{fig:compare_Scales_lo_nlo}
\end{figure}

We turn our attention to the differential cross-section distributions. In order to get a full picture of the impact of NLO QCD corrections on different phase-space regions, not just those sensitive to the threshold for $t\bar{t}t\bar{t}$  production, it is crucial to analyze various dimensionful and dimensionless observables. All results presented in this section are obtained with the default (N)LO MSHT20 PDF set for the (N)LO calculations. Nevertheless, as in the case of the $pp\to t\bar{t}t\bar{t}+X$ process in the $4\ell$ decay channel, also for the $3\ell$ decay channel we have examined the stability of our predictions with respect to changing PDF sets. Using the NLO MSHT20, CT18 and NNPDF3.1 PDF sets for the numerous observables, we could conclude that any observed differences are within the estimated theoretical uncertainties due to scale variation. We could also note that the intrinsic PDF uncertainties are more significant in the tails of some dimensionful observables, reaching half the size of the theoretical error due to scale variation. Therefore, also at the differential cross-section level the PDF uncertainties should be carefully considered in the final estimate of the total theoretical error. Returning to the default setup, we begin our discussion by presenting the results at both the LO and NLO levels for $\mu_0=2m_t$ and $\mu_0=E_T/4$ using the  $|M_{jj}-m_W|< Q_{cut} = 25$ GeV cut. Our goal is to assess the level of agreement between the two scale settings at different perturbative orders as well as to determine the size of NLO theoretical uncertainties associated with these scale settings.

In Figure \ref{fig:compare_Scales_lo_nlo} we  show the sum of the transverse momenta of the four hardest $b$-jets, $H_{T,\, b}$,  and the sum of the transverse momenta of all charged leptons, $H_{T, \, \ell}$, defined according to 
\begin{equation}
    H_{T, \, b} = \sum_{i=1}^{4}p_{T,\, b_i} \,,
    \quad  \quad \quad \quad \quad \quad
    H_{T, \, \ell} = \sum_{i=1}^{3}p_{T, \, \ell_i} \,.
    \label{HTbl}
\end{equation}
At NLO in QCD, the potential 5-th (hardest) $b$-jet, even if resolved and passed all the cuts, is not taken into account
in the definition of $H_{T,\,b}$ to ensure that the same observable is considered at LO and NLO. Otherwise, one would expect large shape differences for this observable at the NLO QCD level. The upper panels display the absolute LO and NLO predictions for $\mu_0 = 2m_t$ and $\mu_0 =  E_T/4$, while the lower panels illustrate the corresponding ratios to the results obtained with $\mu_0 = 2m_t$. In addition, for the NLO case only, additional panels emphasise the absolute magnitude of scale uncertainties for both scale choices. At LO, significant shape distortions up to even $60\%$ are evident for both observables. This confirms the well-known fact that LO predictions are very sensitive to the scale choice.  However, as illustrated in the middle panels of the NLO plots, these large differences are mitigated when higher-order effects are included. Focusing now on the bottom panels of the NLO predictions, we can also observe that the scale uncertainties are of the order of $20\%-25\%$.  More interestingly, both scale settings behave very similarly within the plotted range. A comparable pattern has been observed for most of the observables we examined. Contrary to the expectations, the fixed scale setting is also sufficient to describe differential cross-sectional distributions. This is due to the very large center-of-mass energy that is required to produce four tops, indicating that even higher-$p_T$ tails beyond the TeV range are needed to find potential differences. Such phase-space regions, however, would not be accessible at the LHC. Since both scale settings produce similar results, we use the dynamic scale setting for the remaining distributions, keeping in mind that the differential cross-section results for the two scale choices are consistent with each other within their corresponding theoretical errors.
\begin{figure}[!t]
        \includegraphics[width=0.5\linewidth]{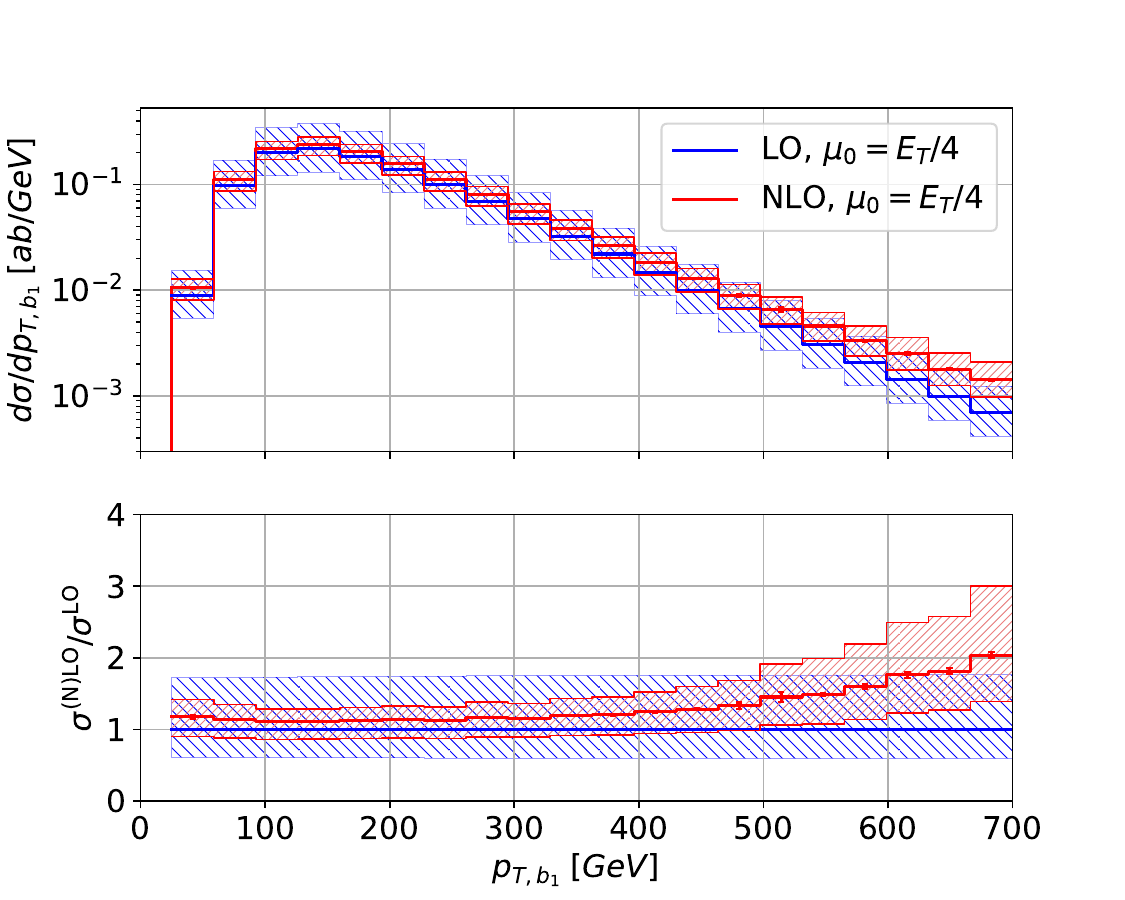}
        \includegraphics[width=0.5\linewidth]{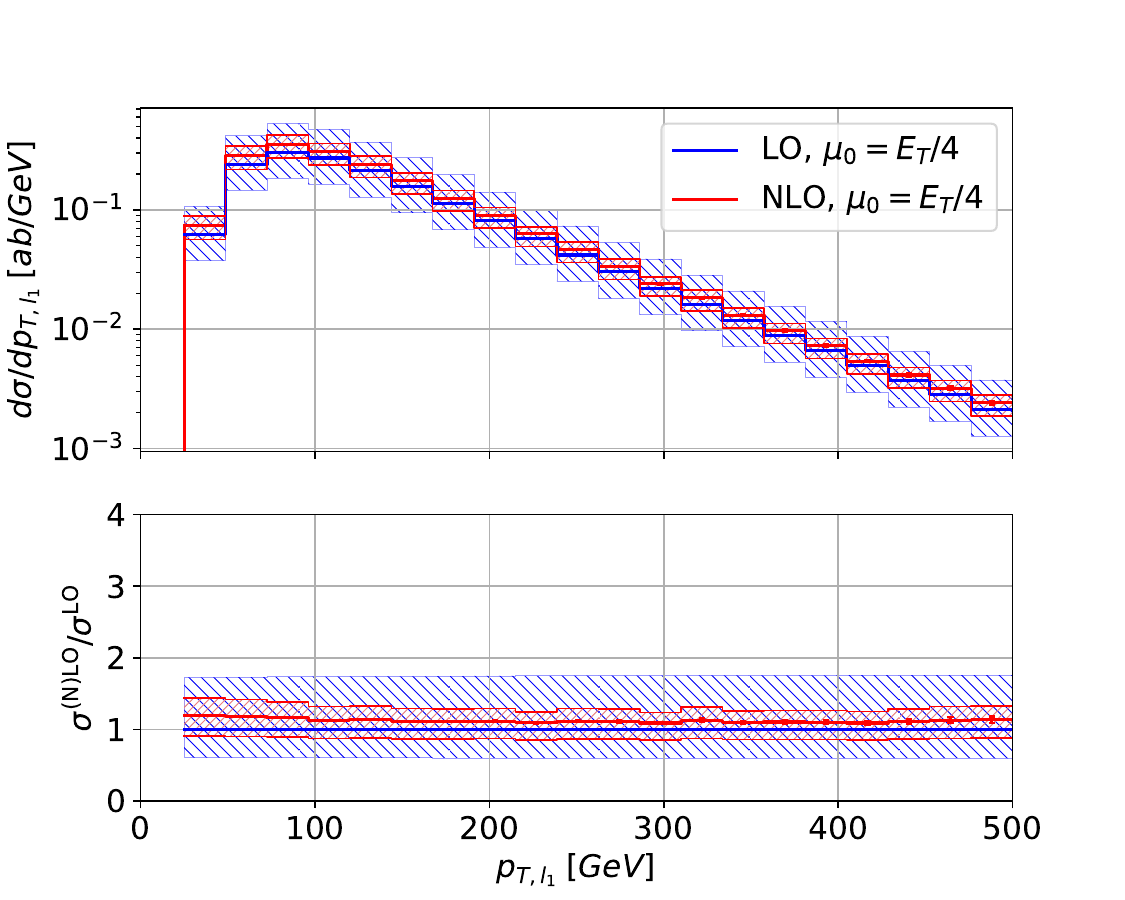}
\caption{\textit{Differential cross-section distributions for  $p_{T, \, b_1}$ and $p_{T, \, \ell_1}$ for the $pp \to t\bar{t}t\bar{t}+X$ process in the $3\ell$ decay channel at the LHC with $\sqrt{s} = 13.6$ TeV. Results are shown for $\mu_0=E_T/4$ with $|M_{jj}-m_W|<Q_{cut} = 25$ GeV and for the (N)LO MSHT20 PDF set. The upper panels show the absolute  (N)LO results. The lower panels present the differential ${\cal K}$-factors together with their uncertainty bands and the relative scale uncertainties of the LO cross sections. Monte Carlo errors are also displayed.}}
    \label{fig:normal_lo_nlo}
\end{figure}

In Figure \ref{fig:normal_lo_nlo} we show the transverse momentum of the hardest $b$-jet, $p_{T, \,b_1}$, and the transverse momentum of the hardest charged lepton, $p_{T,\, \ell_1}$. The upper panels display absolute LO and NLO predictions, including theoretical uncertainties. The lower panels show the differential ${\cal K}$-factors together with their uncertainty bands and the relative scale uncertainties of the LO cross sections. Even with $Q_{cut}=25$  GeV, the higher-order QCD effects for $p_{T, \,b_1}$  can be notably large in the tail of the distribution, reaching values up to even $100\%$. Moreover, in this phase-space region, the NLO result lies outside the LO uncertainty bands. In addition, the size of the NLO scale uncertainties increases significantly and becomes comparable to the size of the LO scale uncertainties. The QCD corrections for $p_{T,\, \ell_1}$, on the other hand, are less pronounced, with the differential $\mathcal{K}$ factor in the range of  $1.10-1.20$. 
\begin{figure}[!t]
        \includegraphics[width=0.5\linewidth]{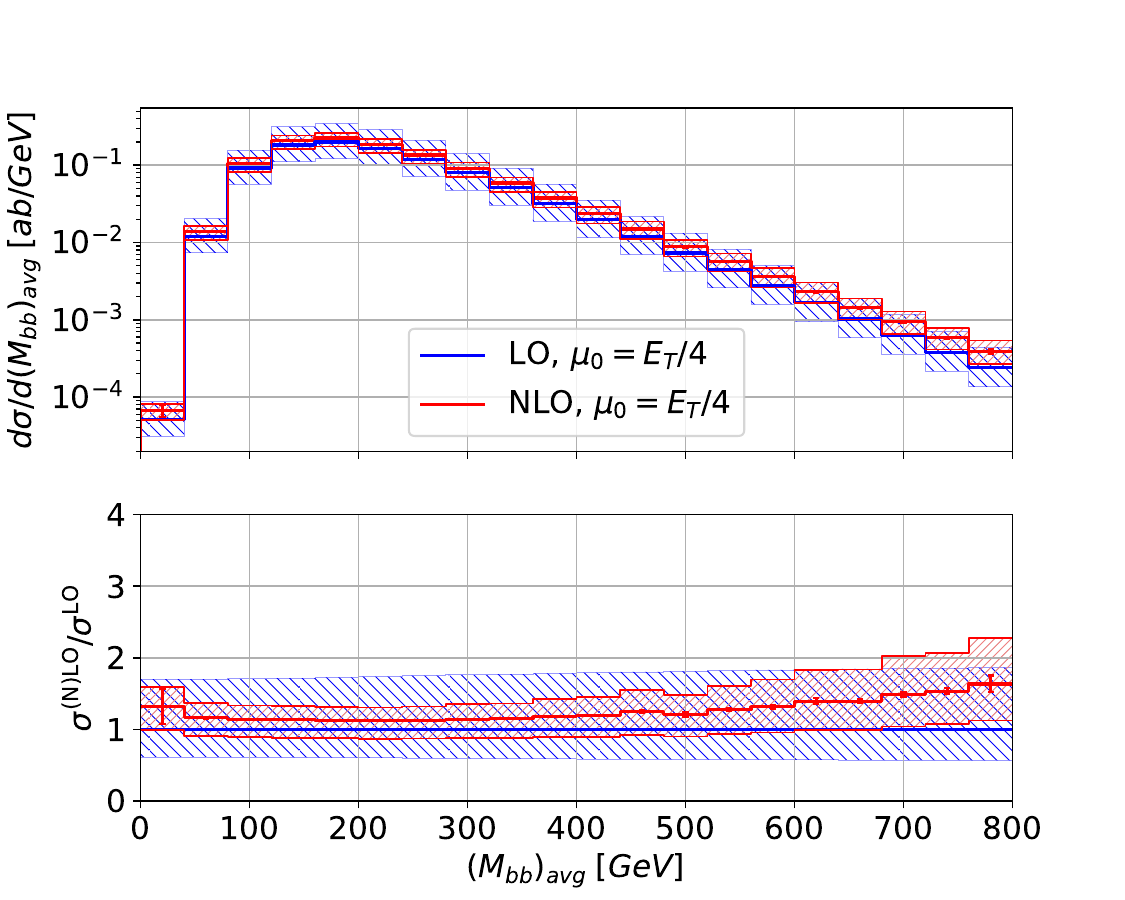}
        \includegraphics[width=0.5\linewidth]{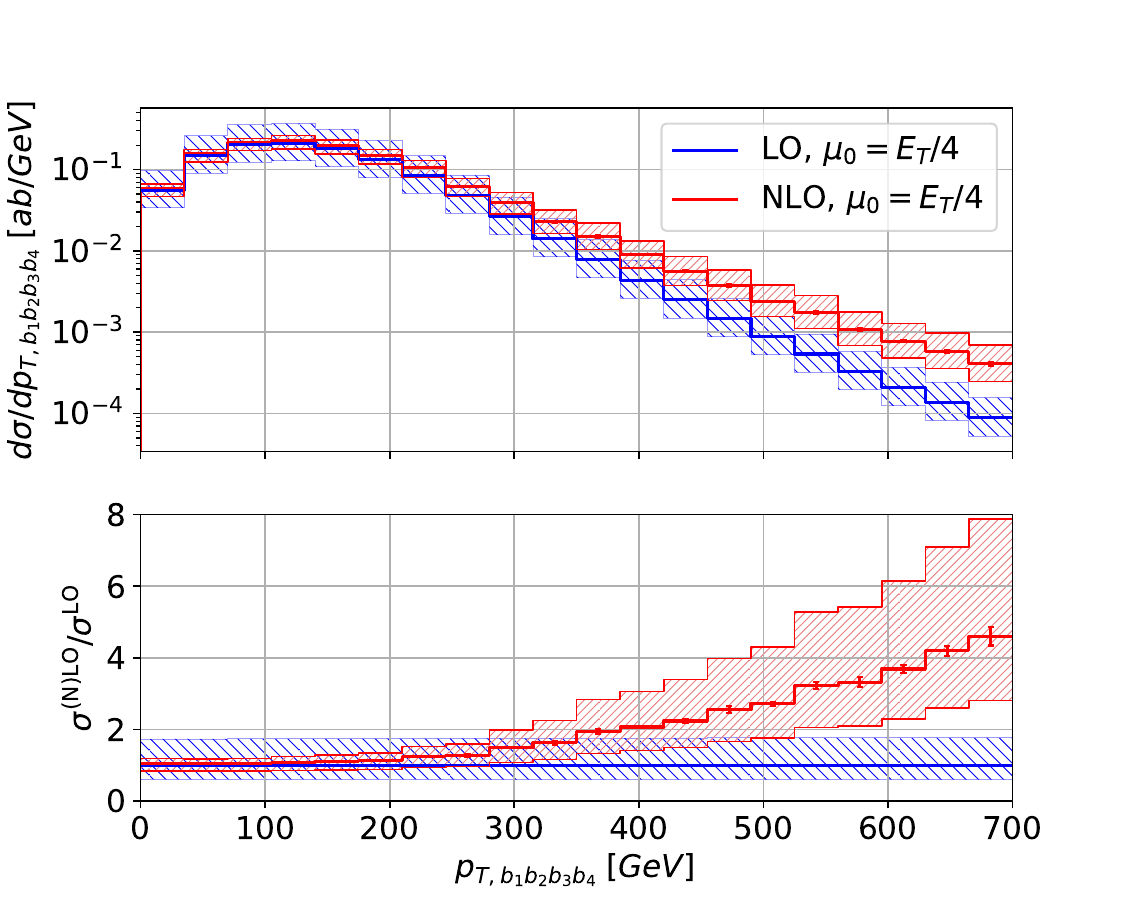}
        \begin{center}
        \includegraphics[width=0.5\textwidth]{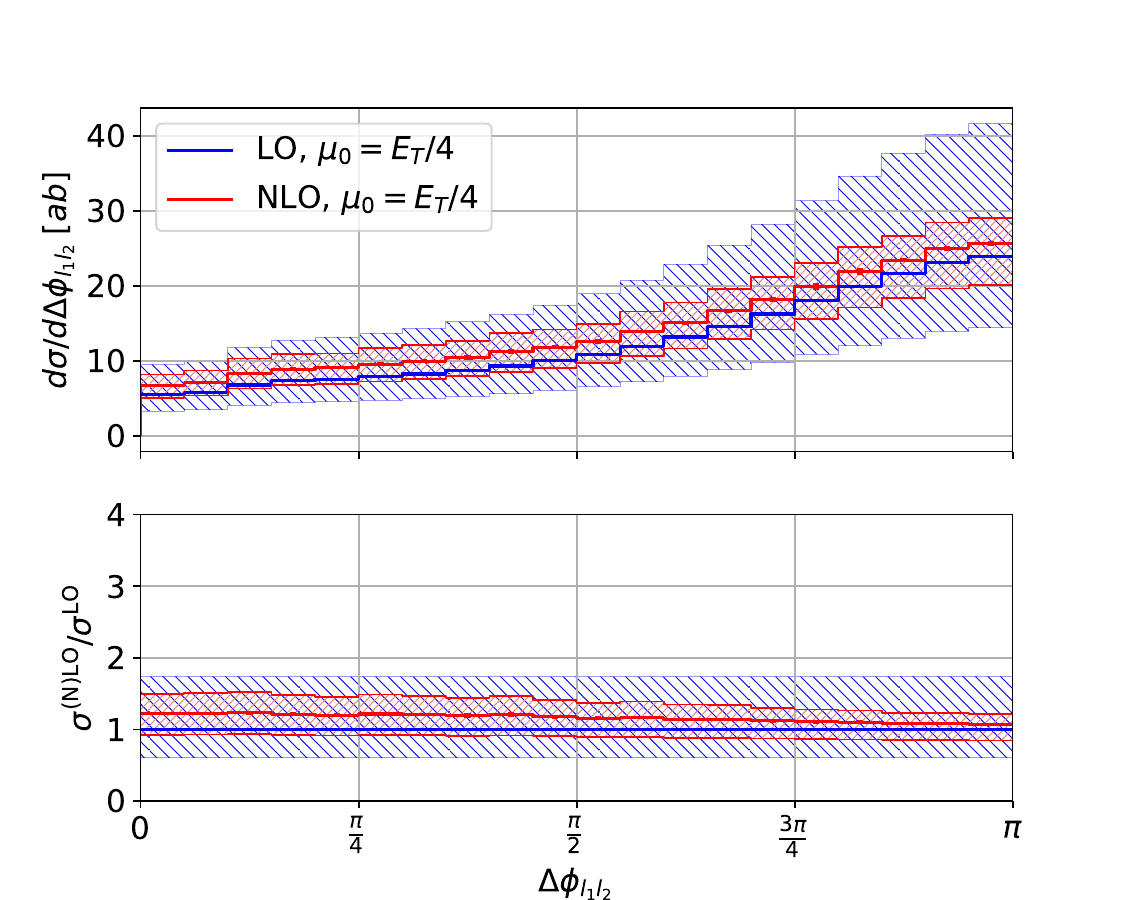} 
        \end{center}
\caption{\textit{Same as Figure \ref{fig:normal_lo_nlo} but for the observables $(M_{bb})_{avg}$, $p_{T,\, b_1b_2b_3b_4}$ and $\Delta \phi_{l_1l_2}$.}}
    \label{fig:more_lo_nlo}
\end{figure}

In Figure \ref{fig:more_lo_nlo} we present additional observables related to the kinematics of the $b$-jets and charged leptons. We plot the average invariant mass of the two $b$-jets, $(M_{bb})_{avg}$, the transverse momentum of the system of the four hardest $b$-jets, $p_{T,b_1b_2b_3b_4}$, and the azimuthal angle between the two hardest charged leptons, $\Delta \phi_{l_1l_2}$. In general, we could observe that all observables associated with $b$-jets are more sensitive to higher-order QCD corrections, while those constructed of charged leptons are less susceptible.  Specifically, for both $(M_{bb})_{avg}$ and $p_{T,b_1b_2b_3b_4}$, NLO QCD corrections become significant in the tails of the distributions, reaching values of up to $65\%$ and $360\%$, respectively. For $(M_{bb})_{avg}$, the NLO scale uncertainties are of the order of  $40\%$ in the distribution's tail, where additionally the NLO predictions are within the LO uncertainty bands. For the $p_{T,b_1b_2b_3b_4}$ observable, NLO predictions lie almost entirely outside of the LO uncertainty bands. Moreover, the NLO uncertainties grow rapidly in the tails of the distribution. The gigantic ${\cal K}$-factor,  which is also present in the $4\ell$ decay channel, can be attributed to the presence of a highly energetic light jet emitted during the production stage, which recoils against the system of four top quarks (the system of four $b$-jets). For the dimensionless observable $\Delta \phi_{l_1l_2}$, the size of the NLO QCD corrections ranges between $1\%-20\%$. To summarize this part, we can note that, on the one hand, the kinematics of the three charged leptons are very similar to the corresponding kinematics in the $4\ell$ decay channel \cite{Dimitrakopoulos:2024qib}. Indeed, for leptonic observables, the NLO QCD corrections are relatively small and the NLO predictions fall within the corresponding LO uncertainty bands. Nevertheless, there are some exceptions, e.g. the transverse momentum of the $l_1l_2l_3$ system or the total missing transverse momentum from the three neutrinos, where large QCD corrections are obtained in the tails. Similar effects have already been observed for these observables in the $4\ell$ decay channel. On the other hand, the observables involving $b$-jets are uniformly affected by much larger higher-order QCD effects compared to the $4\ell$ decay channel. 
\begin{figure}[!t]
        \includegraphics[width=0.5\linewidth]{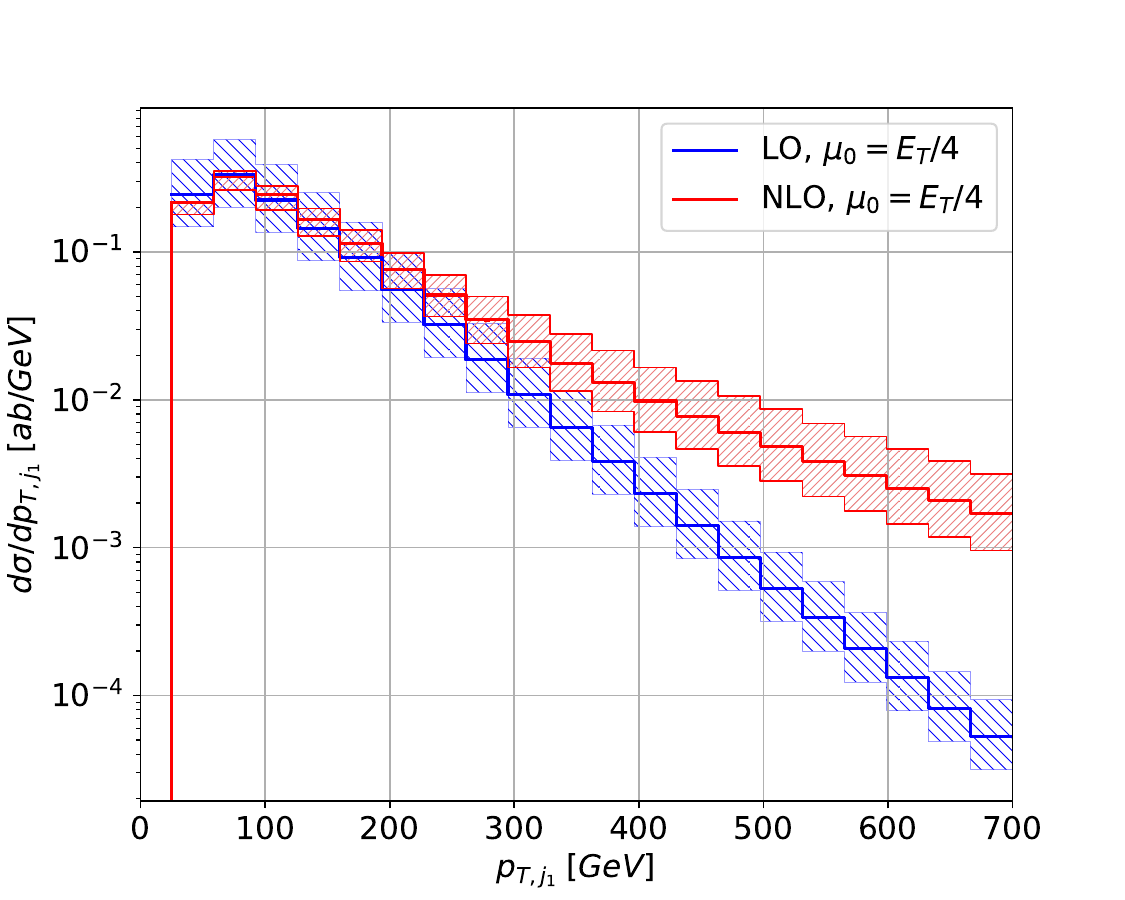}
        \includegraphics[width=0.5\linewidth]{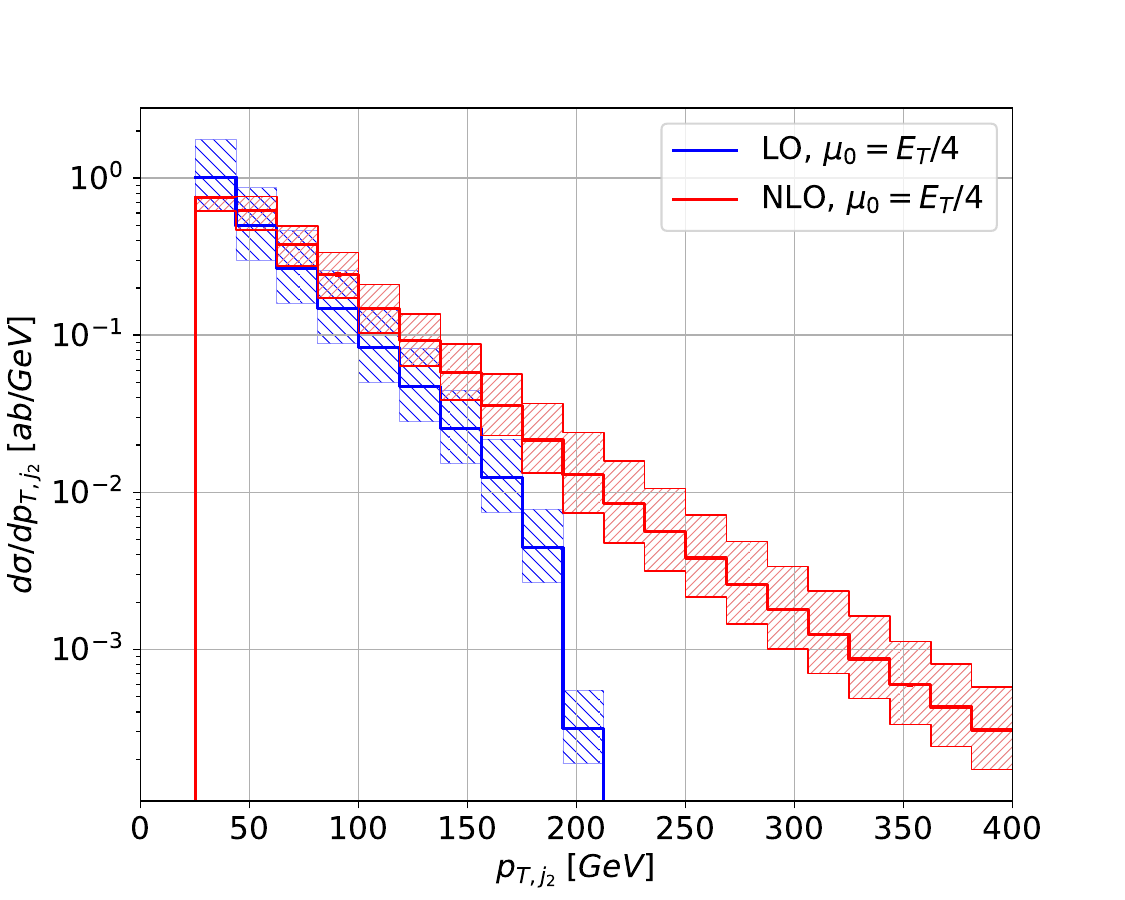}
        \begin{center}
            \includegraphics[width=0.5\textwidth]{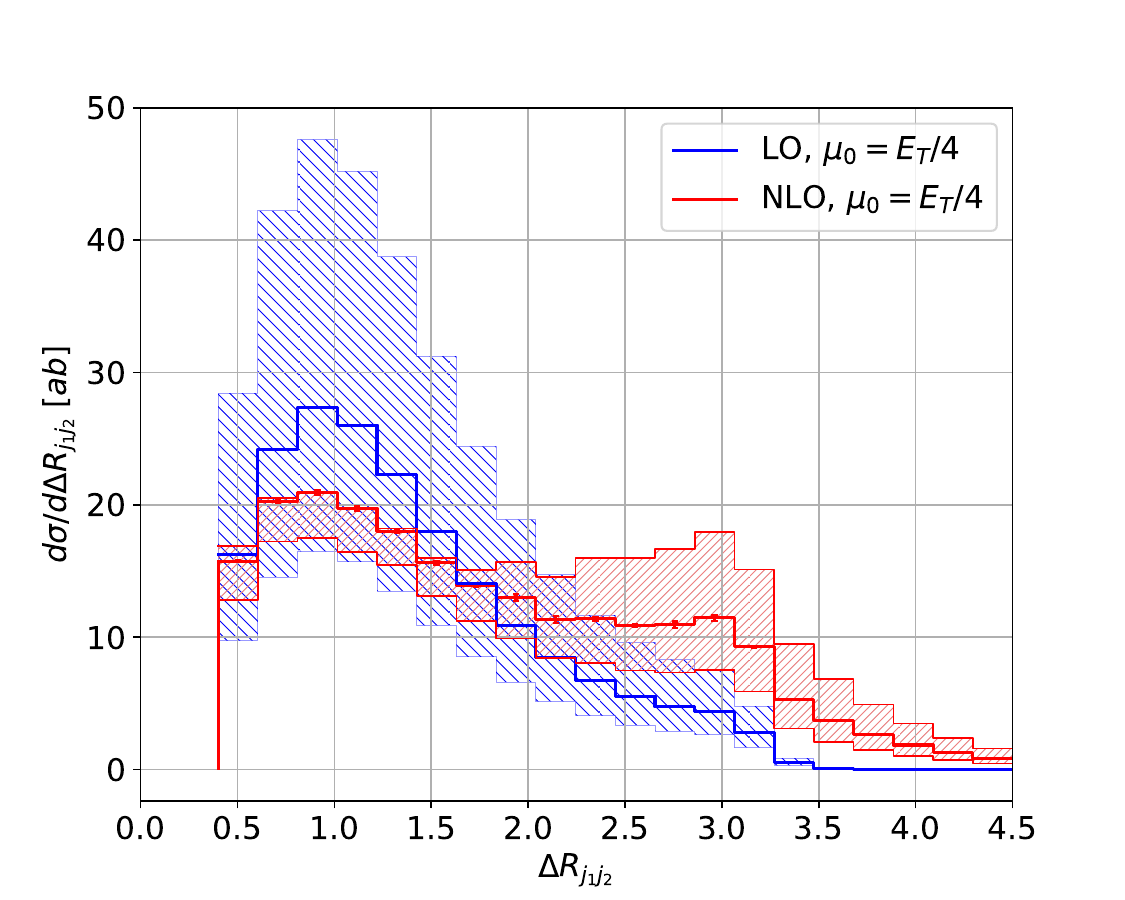}
        \end{center}
\caption{\textit{Differential cross-section distributions for $p_{T,j_1}$, $p_{T,j_2}$ and $\Delta R_{j_1j_2}$ for the $pp \to t\bar{t}t\bar{t}+X$  process in the $3\ell$ decay channel at the LHC with $\sqrt{s} = 13.6$ TeV. The (N)LO results, along with their uncertainties due to scale variation, are shown for $\mu_0=E_T/4$ with  $|M_{jj}-m_W|< Q_{cut} = 25$ GeV and for the (N)LO MSHT20 PDF set. Monte Carlo errors are also displayed.}}
    \label{fig:jet_kinematics_NLO}
\end{figure}
\begin{figure}[!t]
        \includegraphics[width=0.5\textwidth]{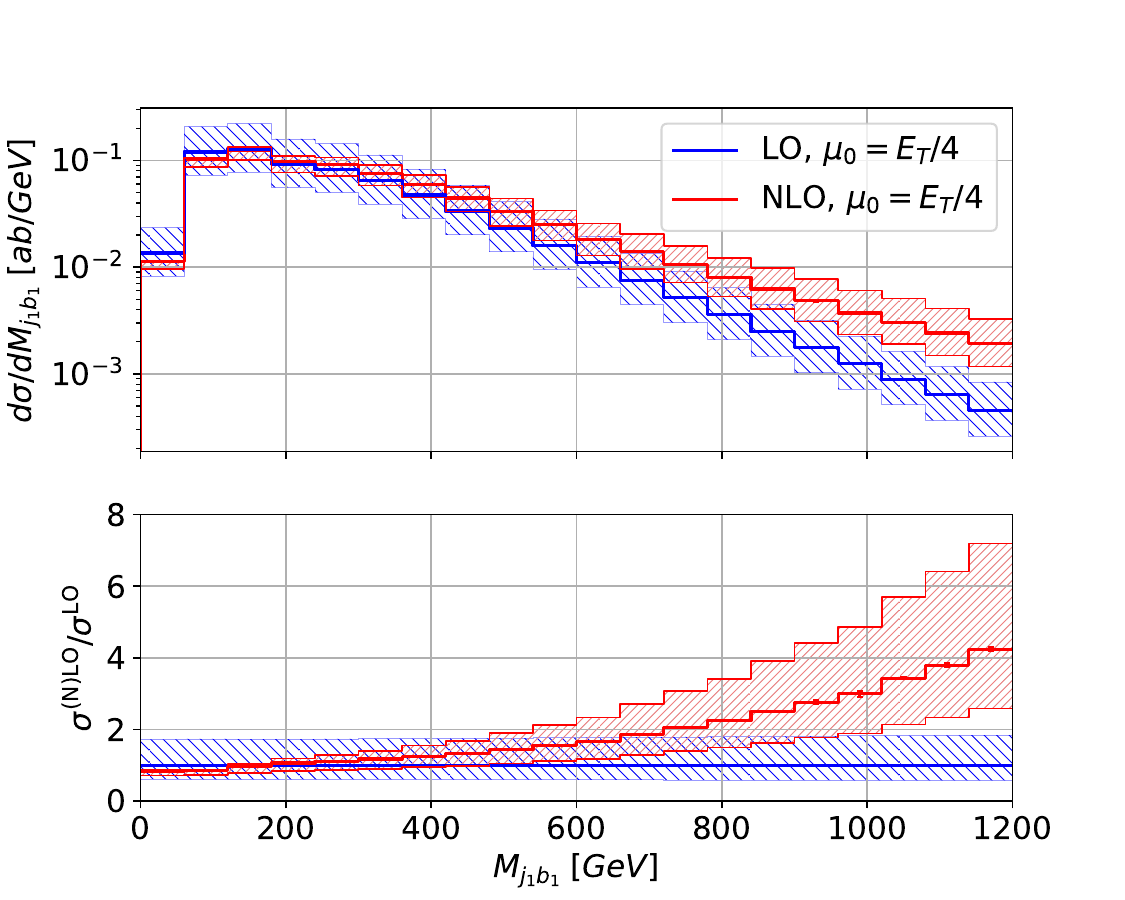} 
        \includegraphics[width=0.5\textwidth]{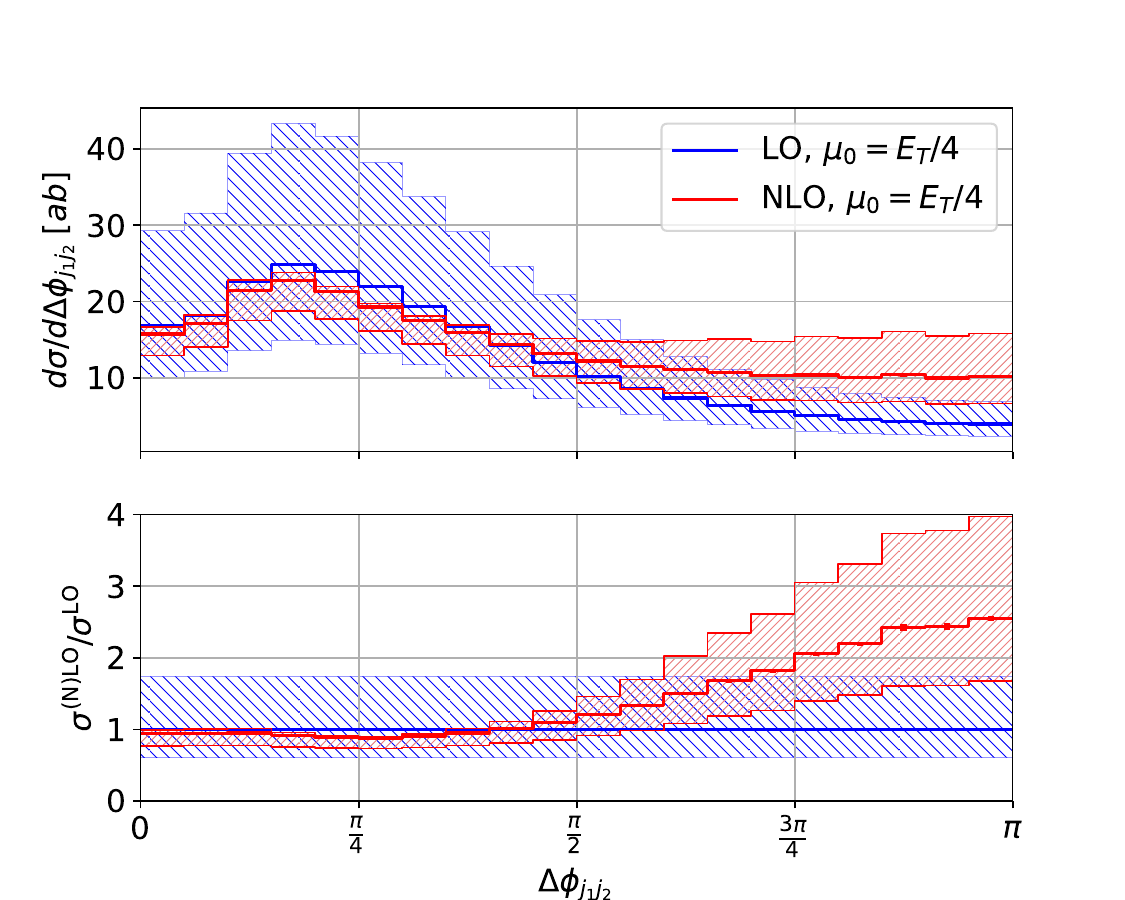} 
        \begin{center}
            \includegraphics[width=0.5\textwidth]{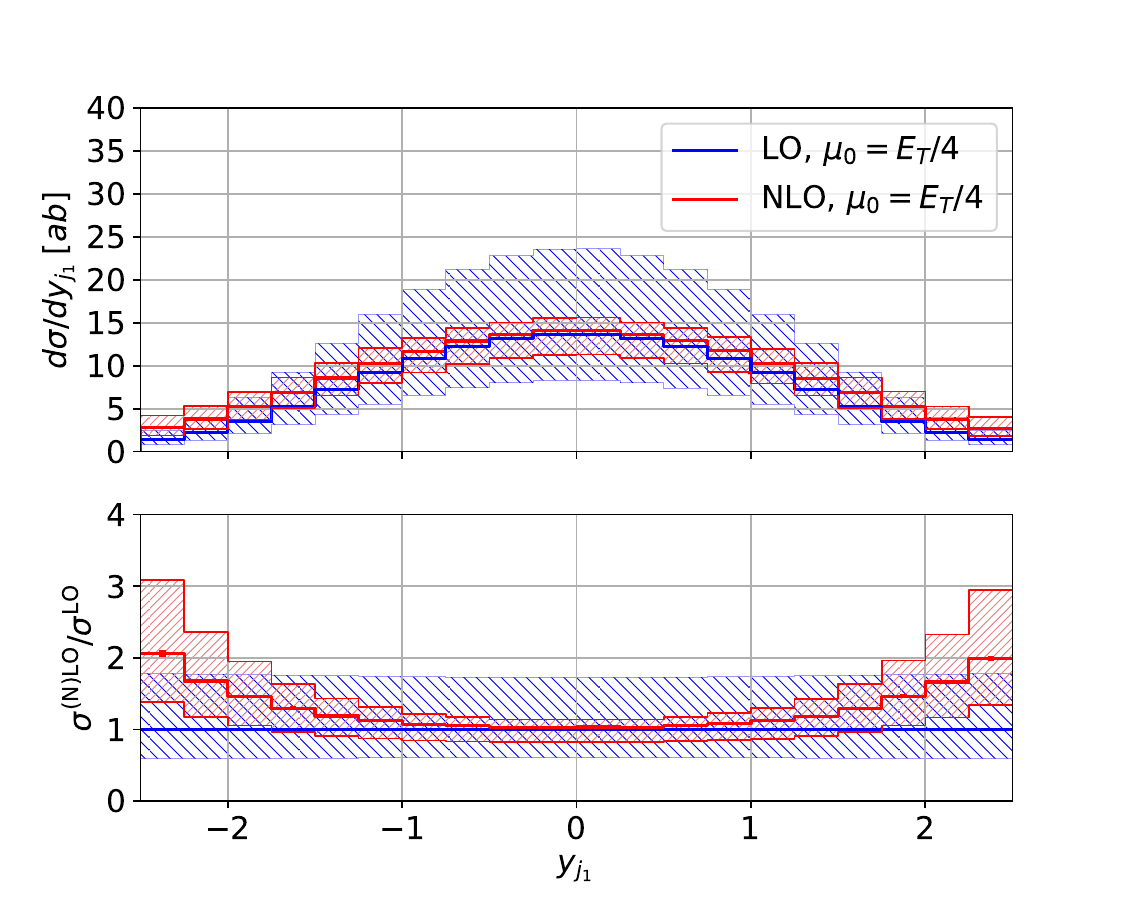}
        \end{center}
\caption{\textit{Same as Figure \ref{fig:jet_kinematics_NLO} but for the following observables: $M_{j_1b_1}$, $\Delta \phi_{j_1j_2}$ and $y_{j_1}$.  The lower panels present the differential ${\cal K}$-factors together with their uncertainty bands and the relative scale uncertainties of the LO cross sections. }}
    \label{fig:jet_kinematics_NLO_2}
\end{figure}
\begin{figure}[!t]
        \includegraphics[width=0.5\linewidth]{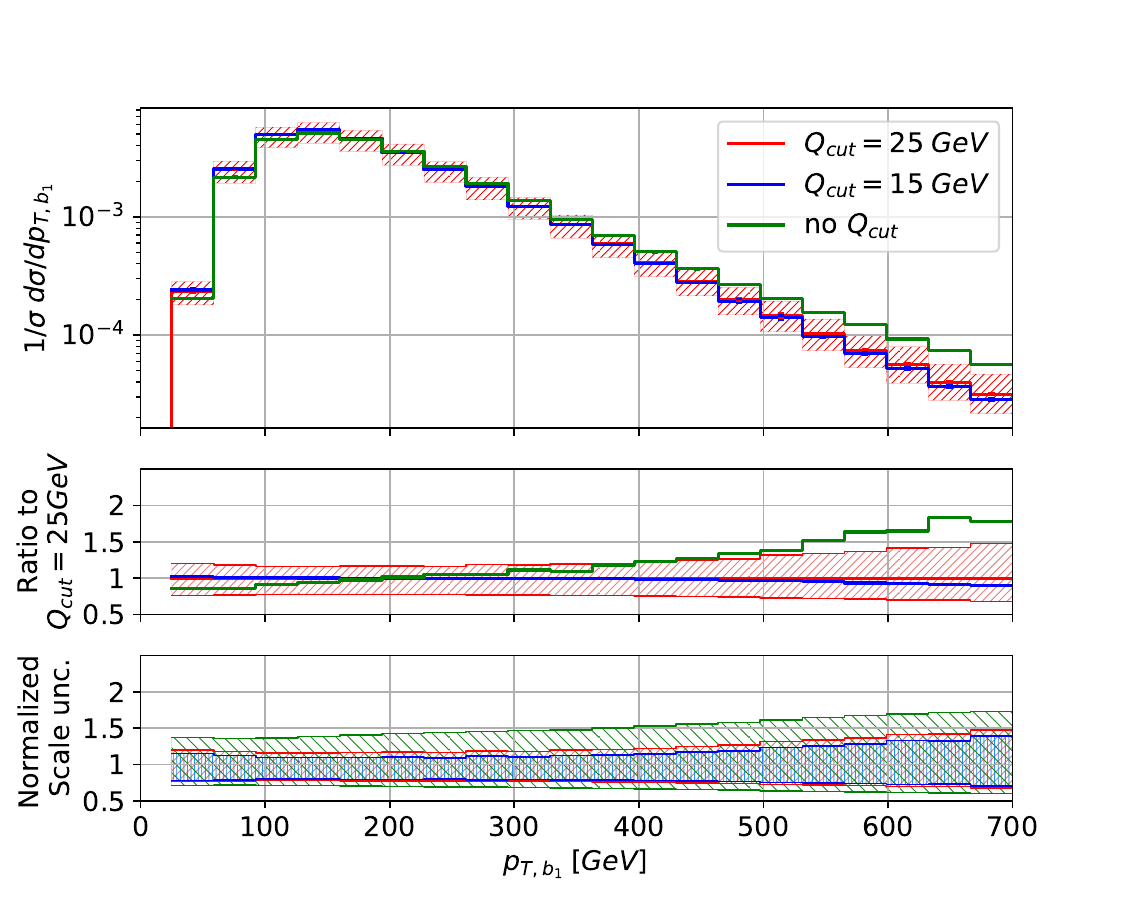}
        \includegraphics[width=0.5\linewidth]{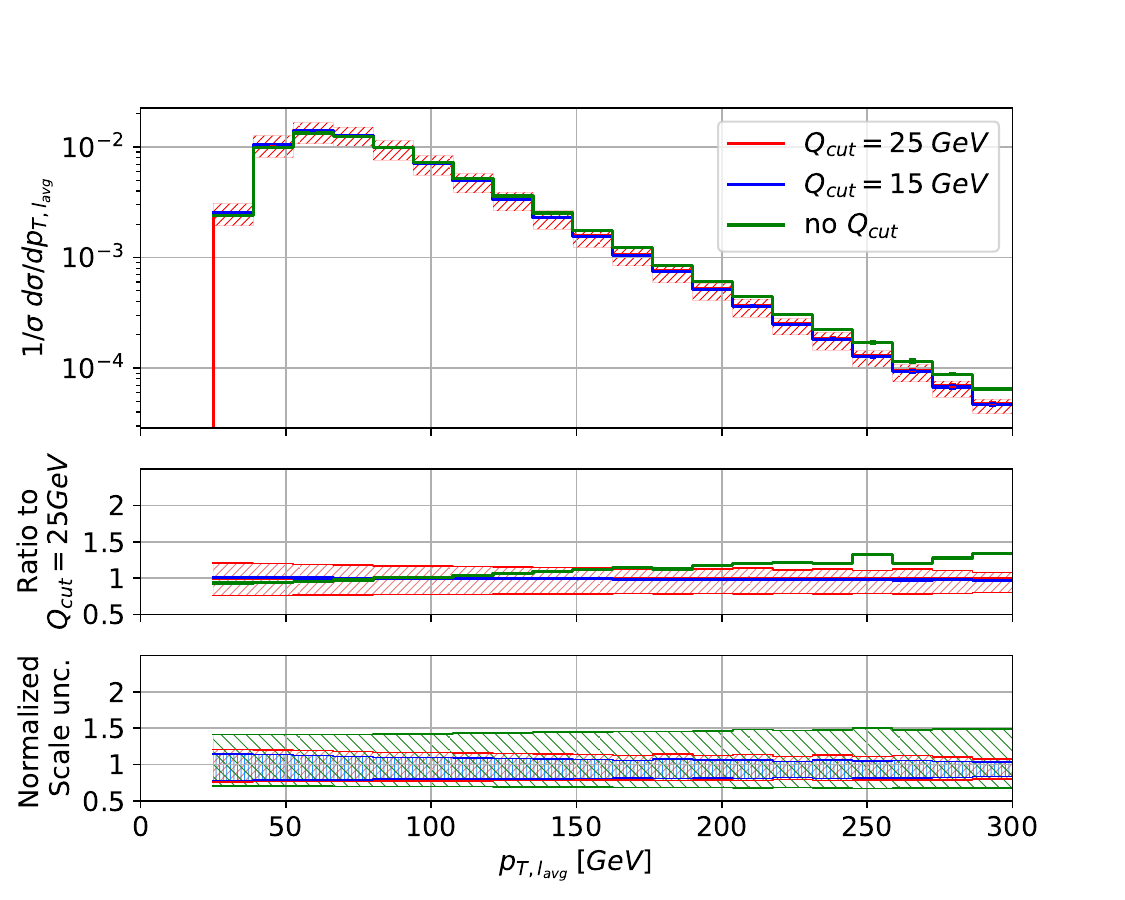}
        \includegraphics[width=0.5\textwidth]{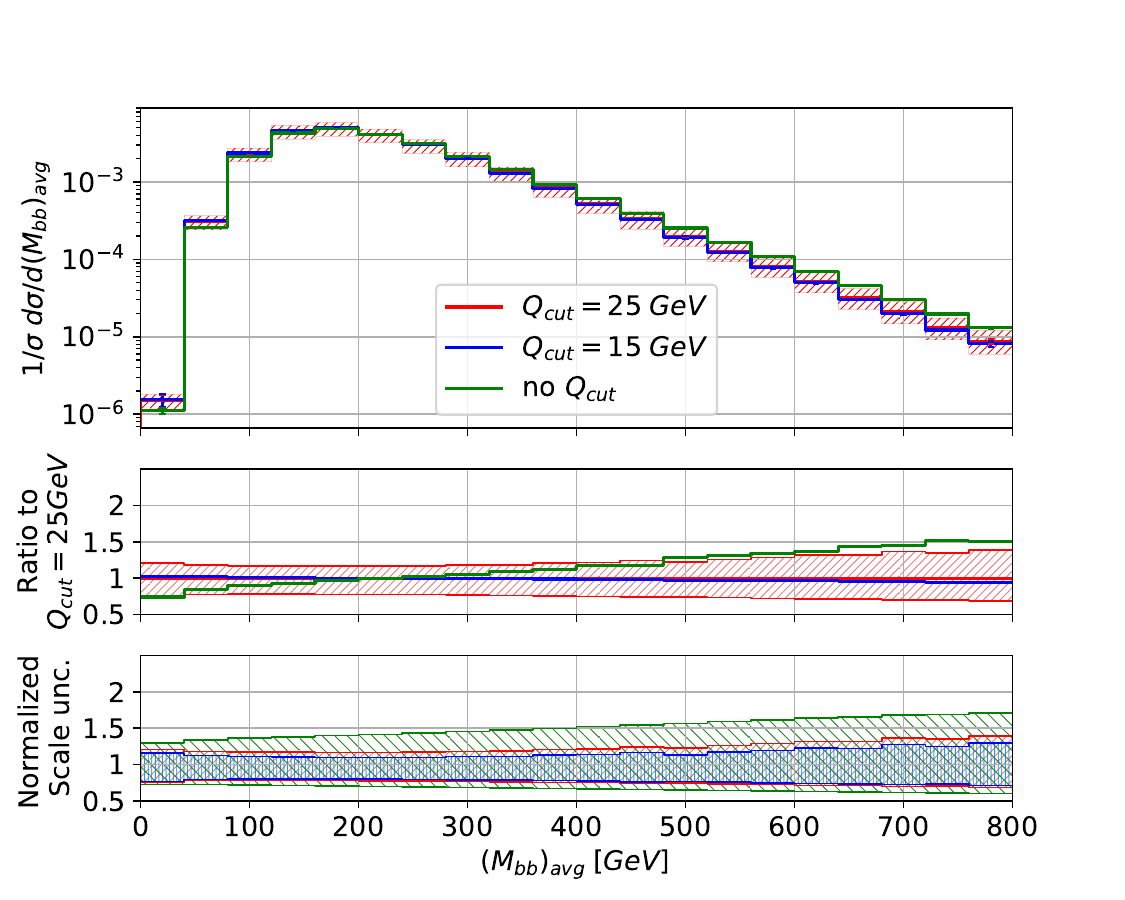} 
        \includegraphics[width=0.5\textwidth]{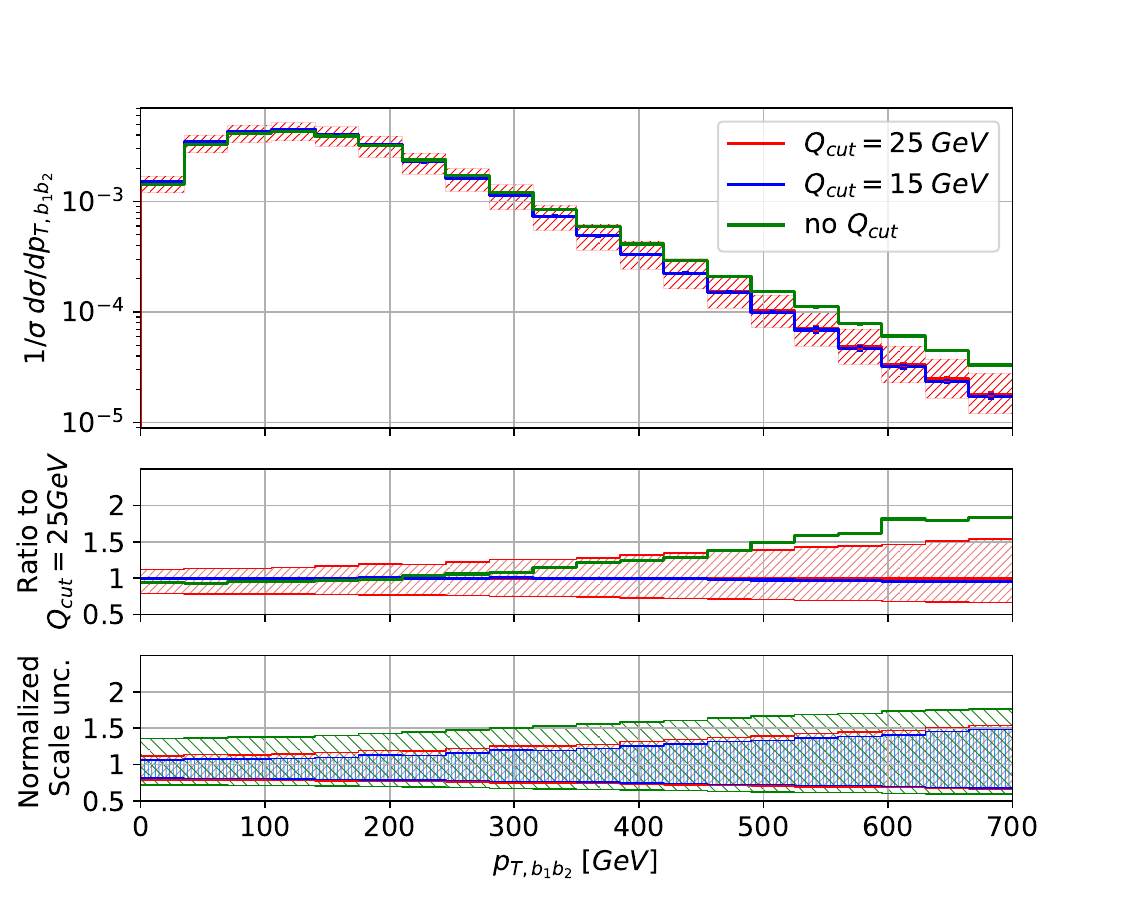}
\caption{\textit{Normalized differential cross-section distributions for $p_{T,\, b_1}$, $p_{T,\, \ell_{avg}}$, $(M_{bb})_{avg}$ and $p_{T,\,b_1b_2}$, for the $pp \to t\bar{t}t\bar{t}+X$ process in the $3\ell$ decay channel at the LHC with $\sqrt{s} = 13.6$ TeV for three different scenarios of the $|M_{jj}-m_W|<Q_{cut}$ cut:  $Q_{cut} = 15, 25$ GeV and $Q_{cut} \to \infty$, the latter case is labeled as no $Q_{cut}$. Results are given for  $\mu_0 = E_T/4$ and the NLO MSHT20 PDF set.  The upper panels show the normalised NLO predictions for the three cases including the scale uncertainties for the $Q_{cut} = 25$ GeV case.  The middle panels present the ratios to the default case along with its uncertainties. The bottom panels display the size of the normalized  NLO scale uncertainties for the three results.}}
    \label{fig:normalized}
\end{figure}
\begin{figure}[!t]
        \includegraphics[width=0.5\linewidth]{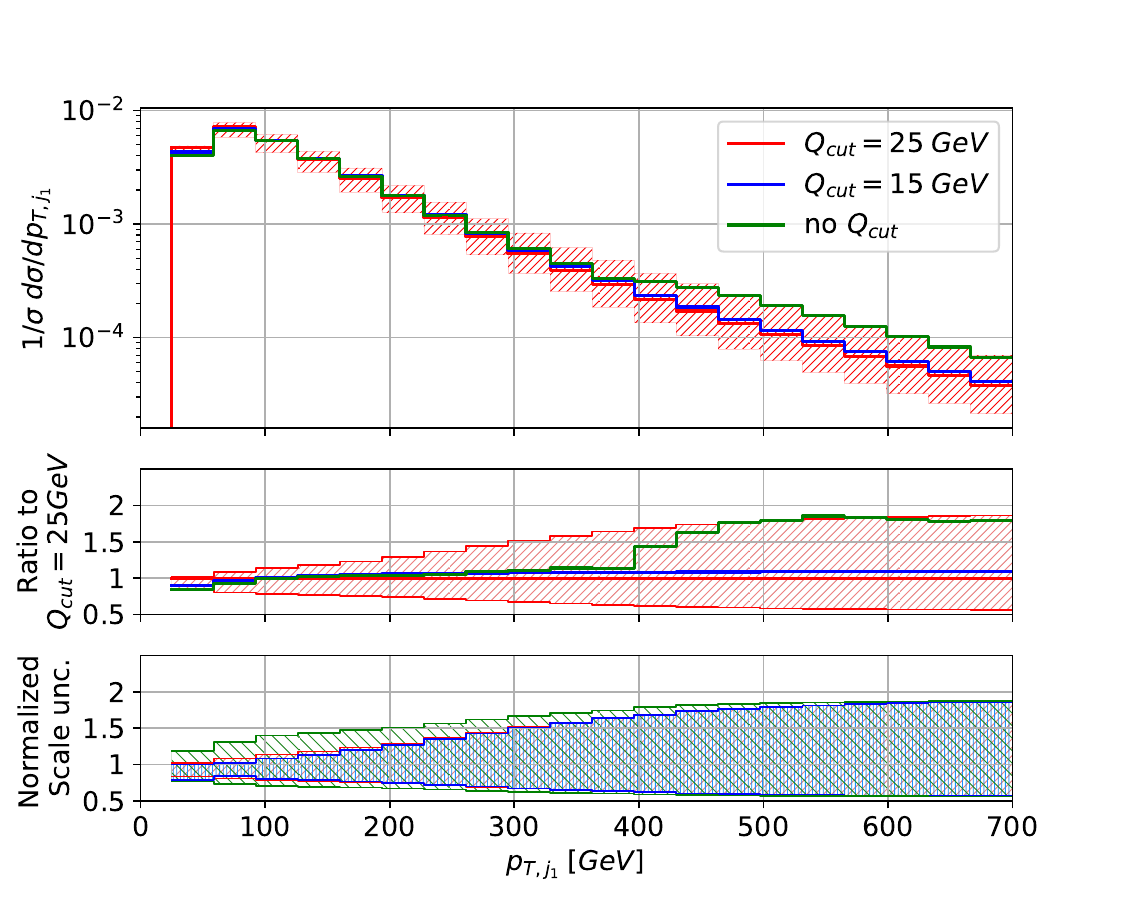}
        \includegraphics[width=0.5\linewidth]{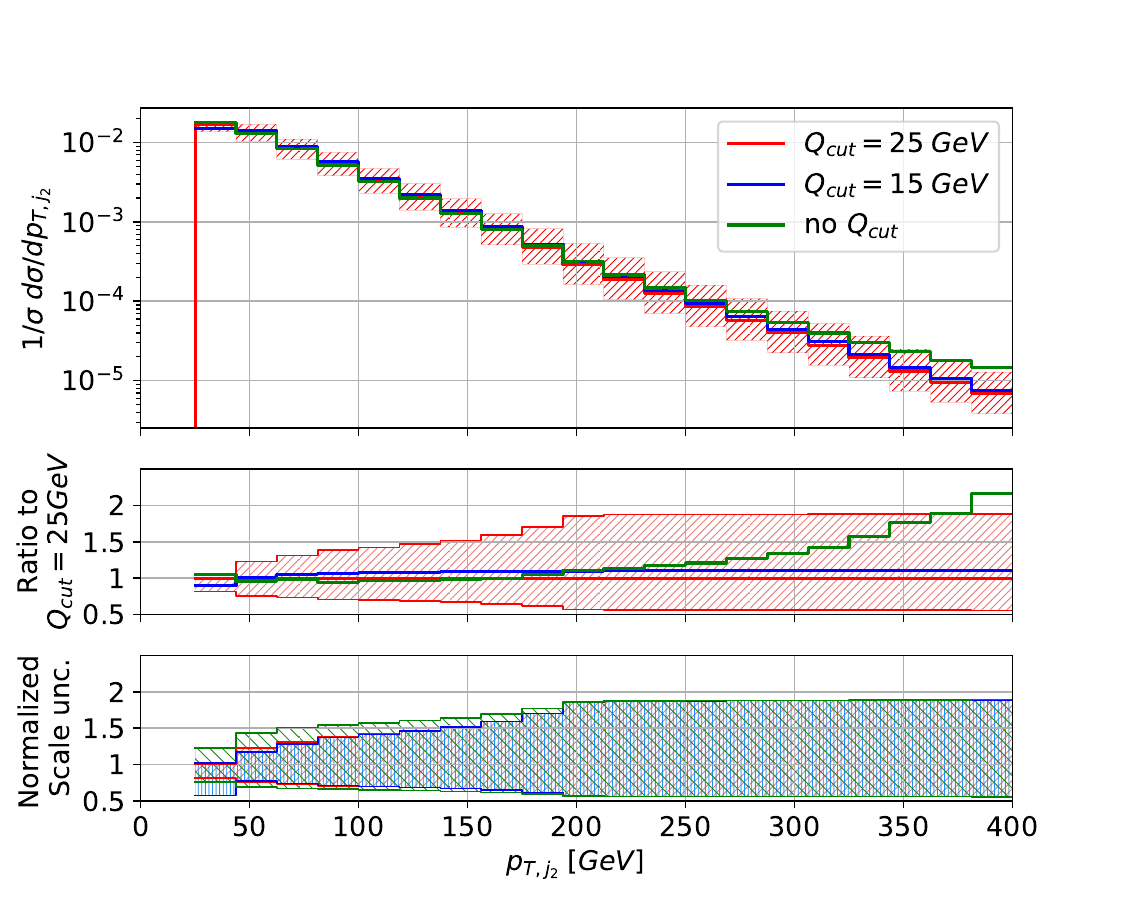}
        \includegraphics[width=0.5\textwidth]{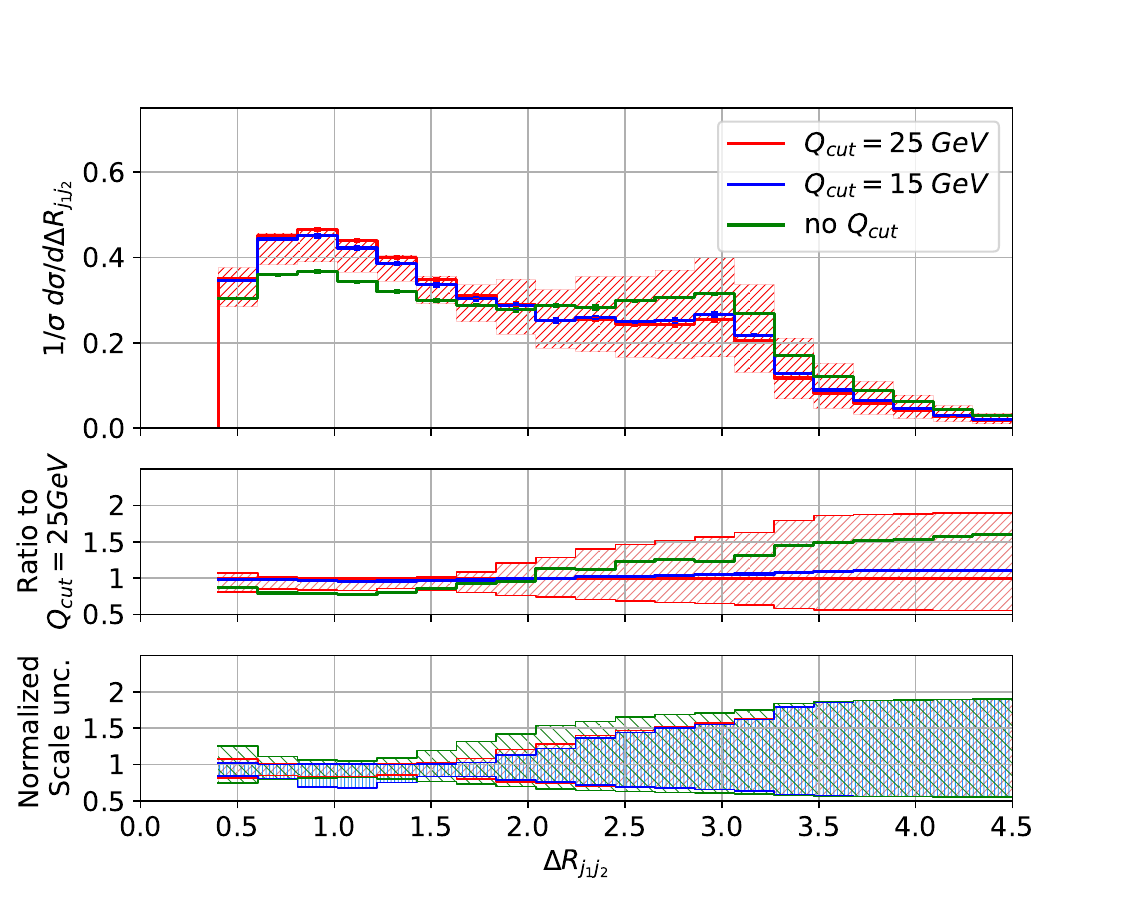} 
        \includegraphics[width=0.5\textwidth]{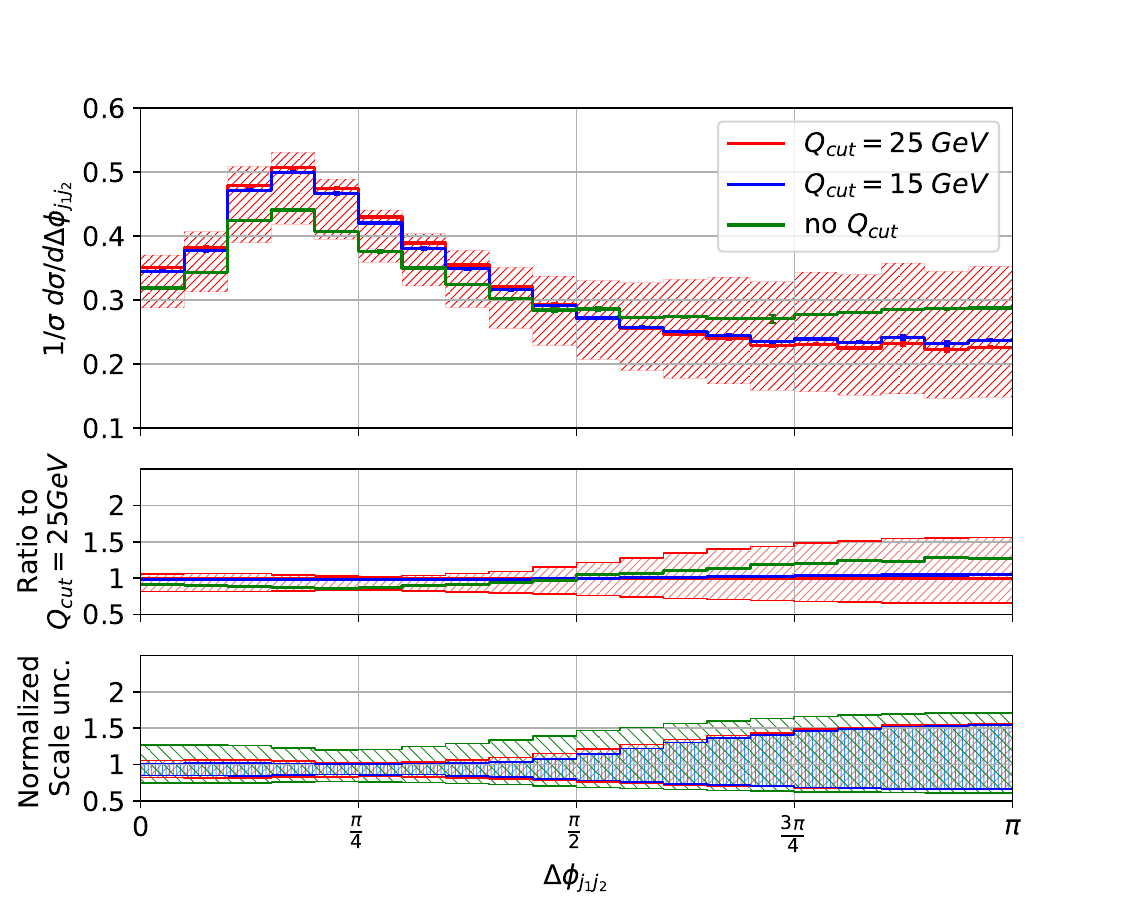}
\caption{\textit{Same as Figure \ref{fig:normalized} but for the  observables $p_{T, \, j_1}$, $p_{T,\, j_2}$, $\Delta R_{j_1j_2}$ and $\Delta \phi_{j_1j_2}$.}}
    \label{fig:normalized_ljets}
\end{figure}

In the $3\ell$ decay channel, we observe significant shape distortions between the LO and NLO differential cross-section distributions for certain types of observables that involve the light jets. To illustrate this, in  Figure \ref{fig:jet_kinematics_NLO} and Figure \ref{fig:jet_kinematics_NLO_2} we present the transverse momentum of the first and the second hardest light jet, $p_{T, \, j_1}$ and $p_{T, \, j_2}$ respectively, the angular distance (or angular separation) in the rapidity-azimuthal angle $(y-\phi)$ phase space between the first and second hardest light jets, $\Delta R_{j_1j_2}$, the invariant mass of the hardest light and $b$-jet, $M_{j_1b_1}$,  the azimuthal angle separation between the first and the second hardest light jet, $\Delta \phi_{j_1j_2}$ and the rapidity of the hardest light jet, $y_{j_1}$. We emphasize that the differential ${\cal K}$-factors are not shown in Figure \ref{fig:jet_kinematics_NLO}. This is due to the fact that for $p_{T, \, j_1}$, $p_{T,\, j_2}$ and $\Delta R_{j_1j_2}$ these factors become too large,  which makes the results difficult to read. Therefore, we decided not to include them in these cases. For $p_{T, \, j_1}$, notable shape distortions are observed at NLO in QCD, especially towards the tail of the distribution. Again, they are caused by the presence of an additional high-energetic jet originating from the production stage of the $pp\to t\bar{t}t\bar{t}+X$ process. This extra light jet can attain large $p_{T}$ values and emerge as either the hardest or second-hardest light jet. On the contrary, large values of $p_{T,j_1}$ are strongly suppressed at LO by the kinematics of the $W$ gauge boson. These constraints are particularly evident for $p_{T,\,j_2}$, where at LO there is a cut around $200$ GeV which can be calculated from
\begin{equation}
\left(p_{T,\, j_2}\right)_{max} \approx \frac{m_W}{\left(\Delta R_{j_1j_2}\right)_{min}} \approx 200 \; \textrm{GeV}\,. 
\end{equation}
A similar cut can also be derived for the hardest light jet 
\begin{equation}
\left(p_{T,\, j_1}\right)_{max} \approx \dfrac{m_W^2}{\left(p_{T, \,j_2}\right)_{min} \left(\Delta R_{j_1j_2}\right)_{min}^2} \approx 1615 \; \textrm{GeV} \,.
\end{equation}
However, this particular value is well outside the plotted range shown for $p_{T,\, j_1}$. Moreover, NLO scale uncertainties for both observables increase significantly at high-$p_T$ tails. They are similar in magnitude or even larger than the corresponding LO uncertainties, which highlights the fact that in these phase-space regions, these observations only have LO accuracy. At NLO, significant changes in the shape of the $\Delta R_{j_1j_2}$ distribution are also visible. In particular, while at LO there is a single peak around ${\Delta R_{j_1j_2}} \approx 1$, at NLO we notice the appearance of a second peak near $\Delta R_{j_1j_2} \approx 3$. As a result, substantial higher-order QCD corrections are evident across the entire plotted range.  Again, this is due to the additional light jet emitted at the production stage that breaks the LO kinematical restriction which favors $\Delta R_{j_1j_2} \approx 1$. Therefore, as expected, in the phase-space regions where two light jets are produced in a back-to-back configuration, which are described by LO-like dynamics, the magnitude of the NLO scale uncertainties exceeds the magnitude of the LO scale uncertainties, reaching values of up to $60\%$.

For the invariant mass of the hardest light and $b$-jet, we can observe that higher-order  QCD corrections become very large in the tail of the distribution, where the ${\cal K}$-factor can exceed the value of $4$. In addition, the magnitude of the NLO theoretical uncertainties in these phase-space regions is substantial, of the order of  $70\%$. To explain the size of the large ${\cal K}$-factors, we note that at LO, in the rest frame of the (anti-)top  quark, the maximum transverse momentum of the resulting $b$-jet is given by 
\begin{equation}
\left(p_{T, \, b}\right)_{max} 
= \frac{(m_t^2 - m_W^2)}{2m_t} \approx 67.5 \; \rm GeV\,. 
\end{equation}
Given that the majority of (anti-)top quarks are produced near the $4m_t$ threshold, pure kinematics imply that events above the value of
\begin{equation}
   \left( M_{j_1b_1}\right)_{max} \approx 2\, \Big( \left(p_{T,\, b}\right)_{max} \left(p_{T, \, j_1} \right)_{max} \Big)^{1/2} \approx 660 \; \rm GeV\,,
\end{equation}
are kinematically strongly disfavored at LO. The absence of such a restriction at NLO explains the sharp rise in the ${\cal K}$-factor roughly above this value. Finally, the azimuthal angle between the first and second hardest light jet receives large NLO QCD corrections up to $160\%$ in the phase-space region where $\Delta \phi_{j_1 j_2} \approx \pi$, while for the last observable $y_{j_1}$ we notice NLO QCD corrections up to $100\%$ near $|y_{j_1}| \approx 2.5$. 
\begin{figure}[!t]
        \includegraphics[width=0.5\linewidth]{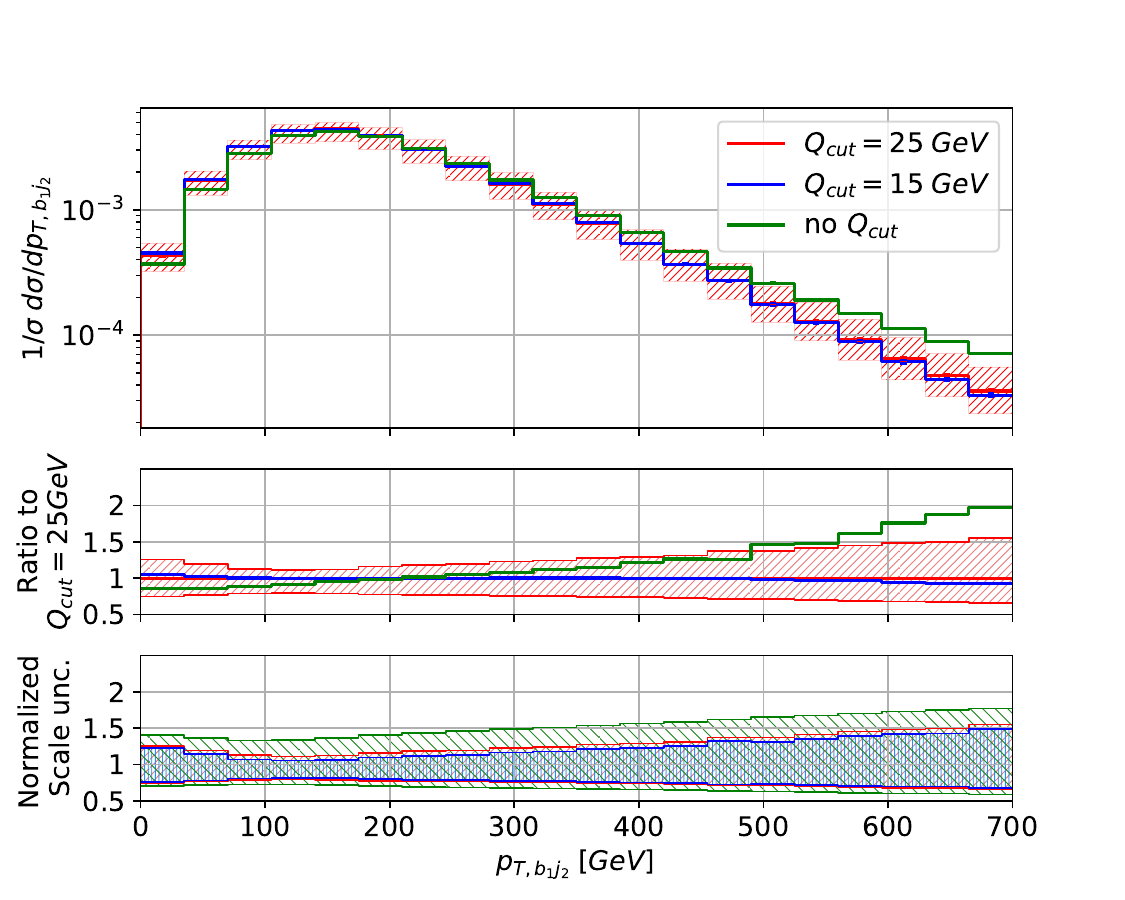}
        \includegraphics[width=0.5\linewidth]{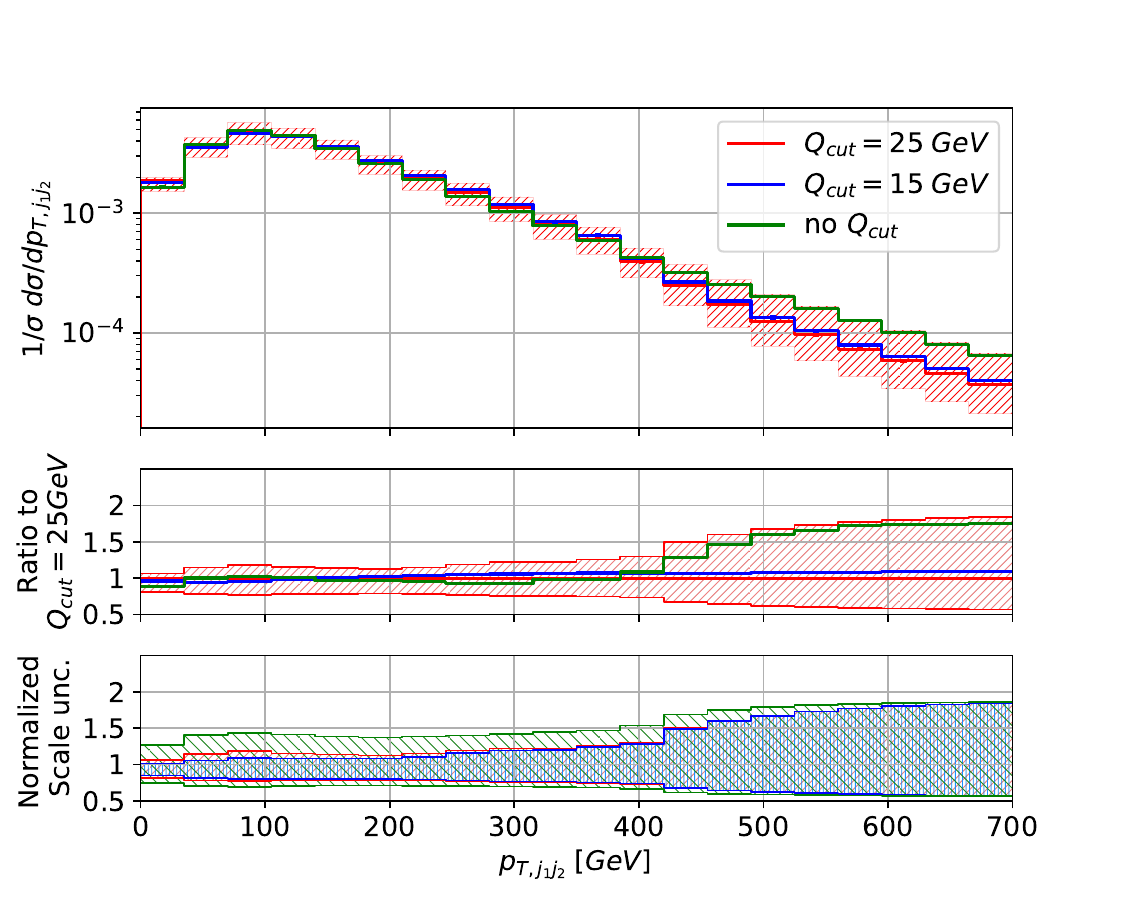}
\caption{\textit{Same as Figure \ref{fig:normalized} but for the  observables $p_{T, \, b_1j_2}$ and $p_{T, \,j_1j_2}$.}}
    \label{fig:normalized_ljets_2}
\end{figure}
\begin{figure}[!t]
        \includegraphics[width=0.5\linewidth]{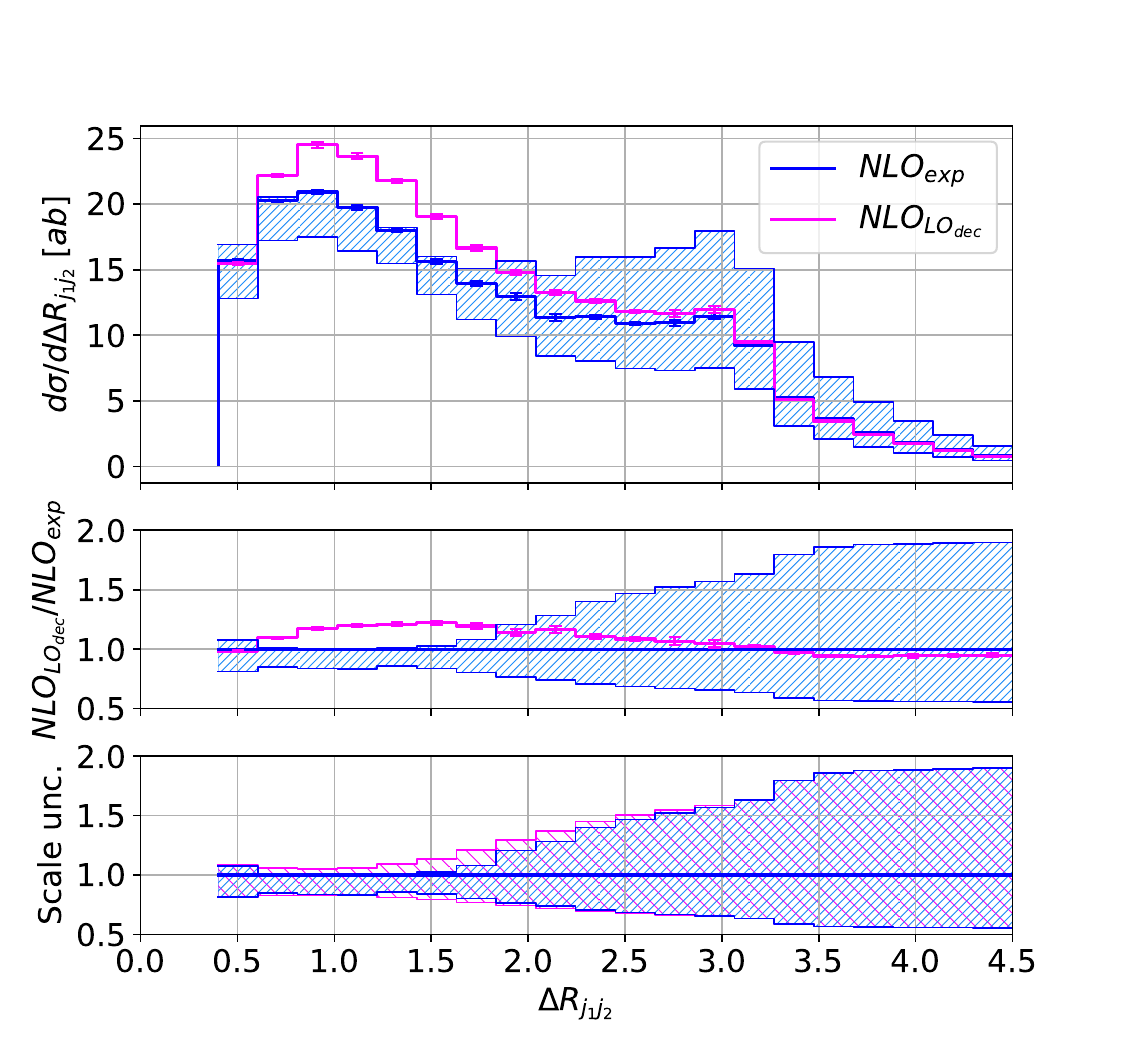}
        \includegraphics[width=0.5\linewidth]{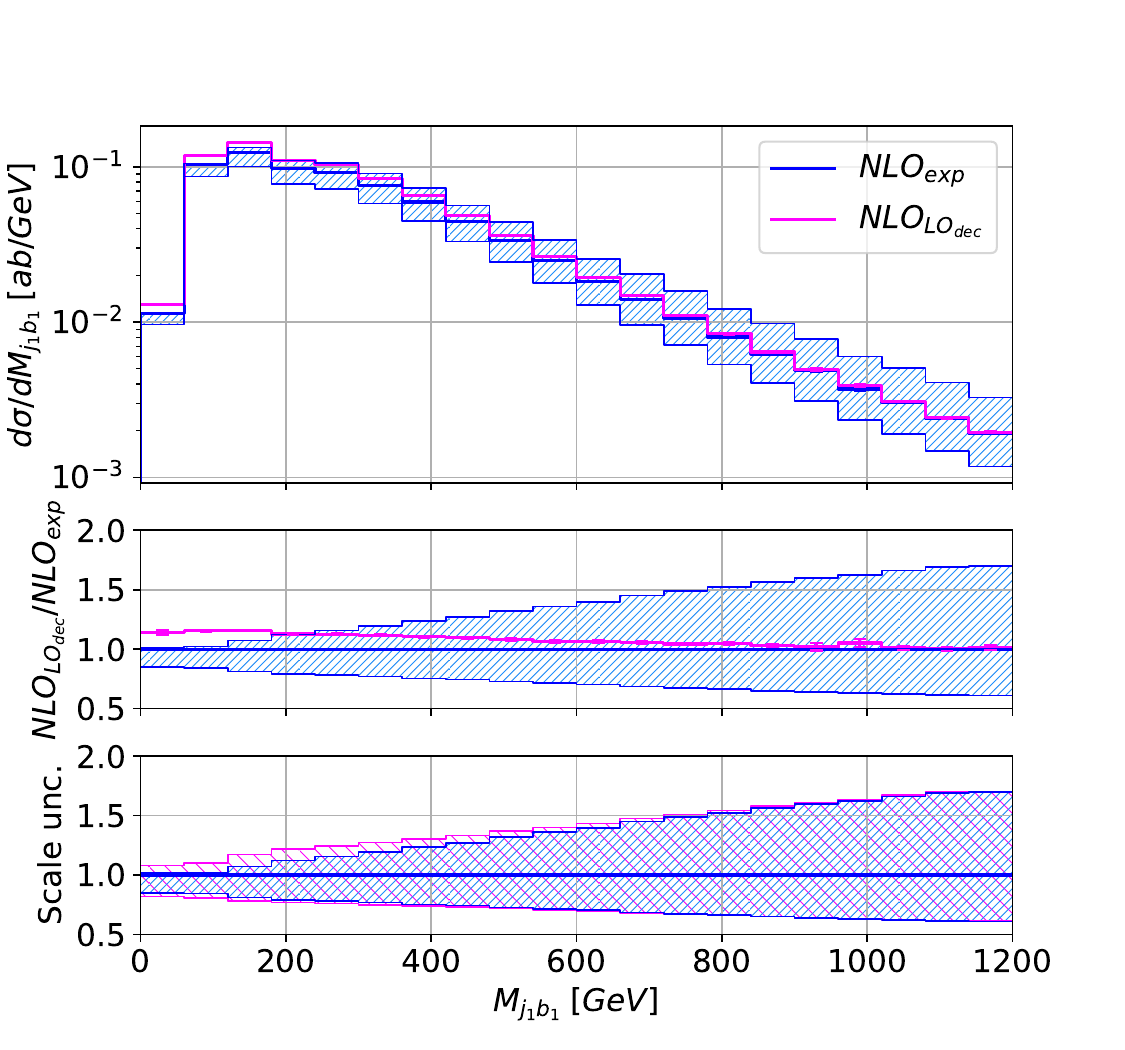}
        \includegraphics[width=0.5\textwidth]{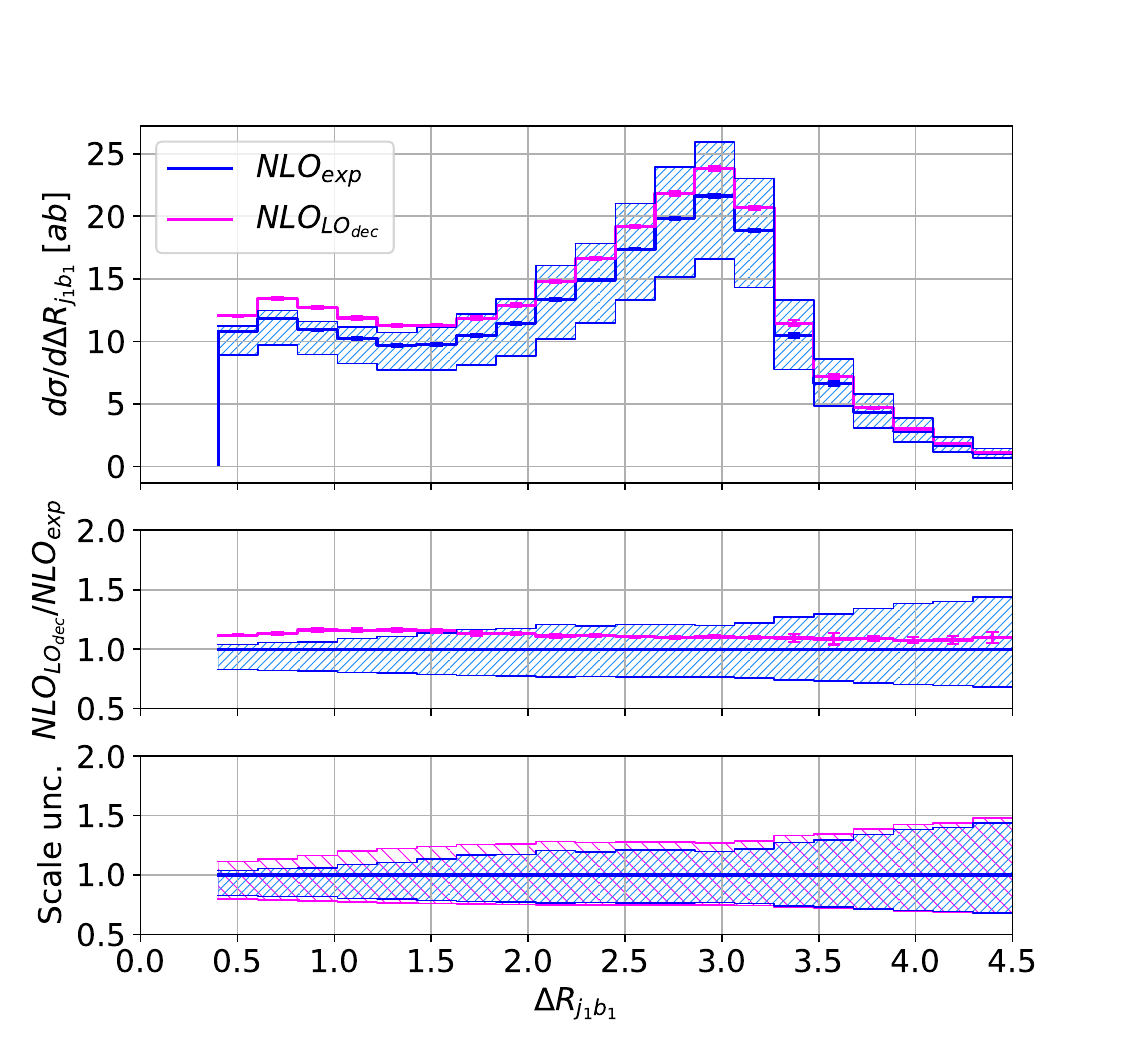} 
        \includegraphics[width=0.5\textwidth]{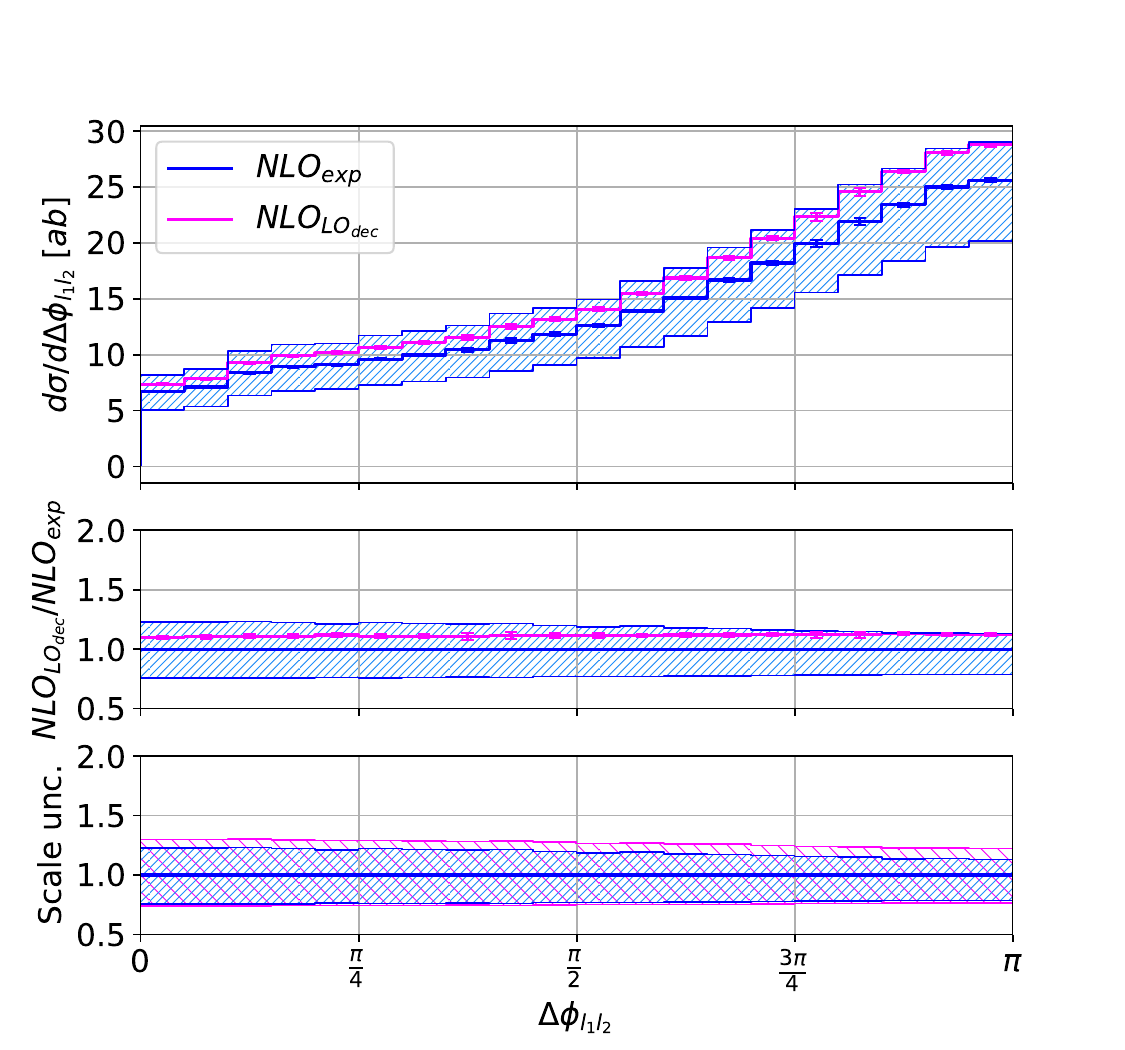} 
\caption{\textit{NLO differential cross-section distributions for $d\sigma^{\rm NLO}_{\rm exp}/dX$ and  $d\sigma^{\rm NLO}_{\rm LO_{dec}}/dX$, where $X= \Delta R_{j_1j_2}, \,M_{j_1b_1}, \, \Delta R_{j_1b_1}, \,\Delta \phi_{\ell_1\ell_2}$ for the $pp \to t\bar{t}t\bar{t}+X$ process in the $3\ell$ decay channel at the LHC with $\sqrt{s} = 13.6$ TeV. Results are shown for $\mu_0 = E_T/4$ with $|M_{jj} -m_W|< Q_{cut} =25$ GeV and for the NLO MSHT20 PDF set.  The upper panels  display the absolute NLO predictions for $d\sigma^{\rm NLO}_{\rm exp}/dX$ and  $d\sigma^{\rm NLO}_{\rm LO_{dec}}/dX$ together with the scale uncertainties of the $d\sigma^{\rm NLO}_{\rm exp}/dX$  case. The middle panels illustrate the corresponding ratio to $d\sigma^{\rm NLO}_{\rm exp}/dX$. The bottom panels provide the size of the scale uncertainties for both cases normalized to their corresponding NLO results. Monte Carlo errors are also displayed.}}
    \label{fig:diff_lodec}
\end{figure}

All the results presented so far in this Section have been obtained with $|M_{jj} -m_W|<Q_{cut}=25$ GeV. However, we would like to study the effect of changing $Q_{cut}$ also at the differential cross-section level. As a reminder, at the integrated cross-section level, our results strongly depended on the value of the parameter $Q_{cut}$, with higher values of $Q_{cut}$ resulting in much larger cross sections. In the following, we study three different scenarios: $Q_{cut} = 15, \,25$ GeV and the case with no restriction at all on $M_{jj}$, labeled as no $Q_{cut}$ or the $Q_{cut} \to \infty$ case. Furthermore, due to large differences in normalization, we only analyze normalized differential cross-section distributions. In this way, we shall identify potential shape distortions,  which can be very important for many experimental analyses targeting the $pp\to t\bar{t}t\bar{t}$ process in the $3\ell$ decay channel. 

In Figure \ref{fig:normalized} we present the transverse momentum of the hardest $b$-jet, $p_{T, \, b_1}$, the averaged transverse momentum of the lepton, $p_{T,\, \ell_{avg}}$,  the averaged invariant mass of the two $b$-jet system,  $(M_{bb})_{avg}$, and the transverse momentum of the system comprising the first and second hardest $b$-jets, $p_{T,\, b_1b_2}$. In each case,  the upper panels show the normalized NLO predictions for $Q_{cut}=15, \, 25$ GeV and $Q_{cut} \to \infty$, including the scale uncertainties for the  $Q_{cut} = 25$ GeV case. The middle panels present the ratios to the default case along with their uncertainties. The bottom panels display the size of the normalized NLO scale uncertainties for the three results. In all four cases, large shape distortions can be observed in the tails of the distributions when $Q_{cut}\to \infty$. In these phase-space regions the central predictions for the no $Q_{cut}$ case are not even within the uncertainty bands of the default results, resulting in differences up to $80\%$ for $p_{T, \,b_1}$ and $p_{T, \,b_1b_2}$, $50\%$ for  $(M_{bb})_{avg}$ and $30\%$ for $p_{T,\, \ell_{avg}}$. When it comes to comparing the results for $Q_{cut} = 15 $ GeV and $Q_{cut} = 25 $ GeV, we note that their normalized predictions agree very well with each other and no significant distortion can be observed. 
Another important observation is the significant reduction in the size of the scale uncertainties for $Q_{cut} = 15, \,25$ GeV. This reduction is evident 
throughout the plotted range of all distributions.  Specifically, at the beginning of the spectrum, the scale uncertainties for the scenario without $Q_{cut}$ reach $35\%-45\%$, while when $Q_{cut} = 15, \, 25$ GeV is applied, the corresponding uncertainties vary between $15\%-25\%$ only. On the other hand, in the tails of these distributions, we can observe uncertainties of the order of $50\%-80\%$ for the $Q_{cut} \to \infty$ case. 

In Figure \ref{fig:normalized_ljets}, we show additional cross-section distributions focusing this time on the kinematics of the light jets. We display anew $p_{T, \,j_1}, \,  p_{T, \,j_2}, \, \Delta R_{j_1j_2}$ and $\Delta \phi_{j_1j_2}$. Also here there are minimal differences between the two cases with $Q_{cut} = 15$ GeV and $Q_{cut} = 25$ GeV. On the other hand, the predictions for the scenario with no restriction on $M_{jj}$ tend to increase in the tails of the dimensionful observables. However, such predictions are still within the range of the large NLO uncertainty bands observed in these phase-space regions. The only exception is the last bin of the $p_{T, \,j_2}$ observable. For the angular separation between the two hardest light jets we observe differences of up to $20\%$ in the region near $\Delta R_{j_1j_2} \approx 1$, where most of the events are concentrated. In addition, the result without $Q_{cut}$ lies outside the uncertainty bands of the default case. Finally, for both observables $\Delta R_{j_1j_2}$ and $\Delta \phi_{j_1 j_2}$ the differences between the $Q_{cut} \to \infty$ case and the other two predictions are substantial for $\Delta R_{j_1j_2} \gtrsim 3 $  and $\Delta \phi_{j_1 j_2} > \pi/2$.  Still, the result without the $Q_{cut}$ cut is within the large NLO uncertainties that are LO-like for these phase-space regions. It is worth noting that the size of the scale uncertainties tends to be consistent for the three different scenarios in the phase-space regions where the NLO predictions are mainly driven by the LO dynamics, and is significantly reduced for low values of $p_T$, $\Delta R_{j_1j_2} < 3$ and $\Delta \phi_{j_1 j_2} < \pi/2$, when $Q_{cut}$ is applied. 

In Figure \ref{fig:normalized_ljets_2} we display  $p_{T, \, j_1j_2}$ and $p_{T, \, b_1j_2}$. These two observables are closely related because in the case of potential misidentification of $b$- and light-jets, $p_{T, \, b_1j_2}$ can be interpreted as $p_{T, \,j_1j_2}$.  For the $Q_{cut} \to \infty$ case shape distortions are evident in both observables, but they are more pronounced for $p_{T, \,b_1j_2}$, where the ratio to the default case of $Q_{cut} = 25$ GeV approaches $2$ in the tail of the distribution. 

So far, we have observed significant shape distortions in almost all observables we have studied when the $M_{jj}$ cut is not applied. We have also verified that these distortions are less evident for the dimensionless observables that do not involve light jets in their definition. Furthermore, the size of the scale uncertainties is significantly larger in the $Q_{cut}\to \infty$ scenario. Thus, restricting the value of $M_{jj}$ may prove beneficial for increasing the efficiency of future experimental analyses targeting the $3\ell$ channel.

In the last step, we examine the differences between $d\sigma^{\rm NLO}_{\rm exp}/dX$ and $d\sigma^{\rm NLO}_{\rm LO_{dec}}/dX$ theoretical predictions for $X= \Delta R_{j_1j_2}, \,M_{j_1b_1}, \, \Delta R_{j_1b_1}, \,\Delta \phi_{l_1l_2}$. In this way, we wish to assess the importance of higher-order effects in top-quark decays also at the differential cross-section level. In Figure \ref{fig:diff_lodec} we present these results for the default $|M_{jj} -m_W|< Q_{cut} =25$ GeV cut. The upper panels display the absolute predictions for $d\sigma^{\rm NLO}_{\rm exp}/dX$  and $d\sigma^{\rm NLO}_{\rm LO_{dec}}/dX$ along with the scale uncertainties of the $d\sigma^{\rm NLO}_{\rm exp}/dX$ result. The middle panels illustrate the ratios to $d\sigma^{\rm NLO}_{\rm exp}/dX$ including the scale uncertainties of the expanded predictions. The bottom panels give the size of the scale uncertainties for both cases normalized to the respective NLO values. Large differences up to $22\%$ are observed for the angular separation between the two hardest light jets for  $\Delta R_{j_1j_2} \approx 1$. In these phase-space regions the  $d\sigma^{\rm NLO}_{\rm LO_{dec}}$ theoretical predictions lie outside the uncertainty bands of the default case. The latter uncertainties are of the order of  $13\%$. The observables $M_{j_1b_1}$ and $\Delta R_{j_1b_1}$ show smaller differences between the two NWA cases. Still in some phase-space regions, i.e.  for $M_{j_1b_1}  \le 200$ GeV and for $\Delta R_{j_1b_1} \in (0.4,1.5)$, the $d\sigma^{\rm NLO}_{\rm LO_{dec}}/dX$ predictions are  not within  the scale uncertainty bands of $d\sigma^{\rm NLO}_{\rm exp}/dX$, which are of the order of $20\%$. The opposite is true for  $\Delta \phi_{l_1l_2}$ that show consistent differences of the order of $10\%$. Furthermore,  these effects are within the theoretical uncertainties of the expanded result, which do not exceed $25\%$.  When analyzing the lower panels of  Figure \ref{fig:diff_lodec} we notice a universal behavior. Scale uncertainties decrease when QCD corrections are properly accounted for in both the production and decay stages of four top quarks.  The only exceptions are the phase-space regions where the NLO QCD uncertainties for $d\sigma^{\rm NLO}_{\rm exp}/dX$ are very large and where the theoretical errors of both NWA approaches are of the same size.

%
\section{Summary and Outlook}
\label{sec:summary}
%
%

In this paper we presented a comprehensive NLO QCD analysis of  the $pp \to t\bar{t}t\bar{t}+X$ process in the $3\ell$ decay channel  at the LHC with $\sqrt{s}=13.6$
TeV. We provided our NLO QCD predictions both at the integrated and differential (fiducial) cross-section level utilizing the NWA methods. Specifically, we studied
the so-called full NWA, the expanded version of the NWA and the case where the NLO QCD corrections are present only in the production stage of the $t\bar{t}t\bar{t}$ process. In
addition, we considered different scale choices and PDF sets in our studies. We showed the need to use the $|M_{jj}-m_W|<Q_{cut}$ cut to restore perturbative convergence in the calculations of the higher-order QCD corrections for the current process and argued that $Q_{cut}=25$ GeV is a very well-motivated choice.  With our input parameters and the set of selection cuts, as well as for  $|M_{jj}-m_W|< 25$ GeV we observed that the magnitude of the NLO QCD corrections strongly depends on the choice of the LO PDF set, and in particular on the $\alpha_s(m_Z)$ value used there. Accordingly, the ${\cal K}$-factors defined as  ${\cal K} = \sigma^{\rm NLO}_{\rm exp}/\sigma^{\rm LO}$ are within the range of ${\cal K} \in (1.08-1.37)$. However, at the NLO level in QCD, theoretical predictions for various PDF sets agree with each other at the $1\sigma$ level, separately for  $\mu_0 = 2m_t$ and $\mu_0 = E_T/4$. In addition, the PDF and scale uncertainties are independent of the scale setting and are of the order of $2\%-6\%$ and $23\%$, respectively. The various NWA models we included in our study provided us with additional information. In particular, we established that neglecting higher-order QCD corrections in top-quark decays resulted in an overestimation of the cross section by about $11\% - 12\%$, depending on the renormalization and factorization scale chosen, and led to an increase in scale uncertainties from $23\%$ to $26\%$. On the other hand, the difference between $\sigma^{\rm NLO}_{\rm full}$ and $\sigma^{\rm NLO}_{\rm exp}$ is at the $12\% - 13\%$ level, thus, well within the stated theoretical uncertainties. In addition, when  moving from $\sigma^{\rm NLO}_{\rm exp}$ to $\sigma^{\rm NLO}_{\rm full}$, theoretical uncertainties are reduced from $23\%$ to $19\%$. Furthermore, we investigated the dependence of NLO QCD results on the $Q_{cut}$ used. We observed a high sensitivity of the ${\cal K}$ factor on the choice of $Q_{cut}$. In particular, in the most extreme case, i.e., for $Q_{cut} \to \infty$, we obtained not only significant theoretical uncertainties, up to even $45\%$, but also higher-order QCD effects of the order of  $80\%-130\%$. 

At the differential (fiducial) cross-section level even with the $|M_{jj}-m_W|< 25$ GeV cut, we observed giant differential ${\cal K}$-factors and huge shape distortions between the LO and NLO results for certain types of observables that are associated with light jets. Such effects are caused by the presence of an additional jet coming from the production stage of the $pp \to t\bar{t}t\bar{t}+X$ process, which can reach large $p_T$ values and appear as the hardest or second-hardest light jet. Moreover, it can mimic one of the decay products of the $W$ boson. In addition, we studied the effect of changing $Q_{cut}$ on the differential cross-section distributions. We examined the following three cases: $Q_{cut}=15$  GeV,  $25$ GeV and the $Q_{cut} \to \infty$ case. For all observables considered, the normalized predictions for $Q_{cut} = 15$ GeV and $Q_{cut} = 25$ GeV agreed very well with each other and no significant distortion could be observed. On the other hand, large shape distortions could be detected in the tails of the dimensionful distributions when $Q_{cut} \to \infty$. This was true even for some angular distributions.  In a final step, to estimate the size of higher-order QCD effects in top-quark decays, we compared  $d\sigma^{\rm NLO}_{\rm exp} /dX$ and $d\sigma^{\rm NLO}_{\rm LO_{dec}} /dX$. We observed significant differences for both dimensionless and dimensionful observables. Moreover, for some observables, these shape distortions were not covered by the theoretical uncertainties.   

In summary, higher-order corrections are undoubtedly required for the $pp \to t\bar{t}t\bar{t}+X$ process in the $3\ell$ decay channel at the LHC. They should be consistently included in both the production and decay stages of the $4t$ process as they impact not only the overall normalization, but also the size of the theoretical uncertainties due to scale dependence. To control the size of NLO QCD corrections and theoretical uncertainties, the $Q_{cut}$ cut is mandatory. But even after applying such a cut, giant higher-order QCD effects are observed for some dimensionless and dimensionful observables  in particular (fiducial) phase-space regions. Since NNLO QCD predictions for this process are currently out of reach,  a  (N)LO  merged result, as obtained with  $pp \to t\bar{t}t\bar{t}$, $pp \to t\bar{t}t\bar{t}+j$ and $pp \to t\bar{t}t\bar{t}+jj$ predictions in the $3\ell$ channel, could be used instead to improve precision of the results in these particular phase-space regions. To this end, the schemes like for example  MLM \cite{Mangano:2006rw,Alwall:2007fs}, CKKW \cite{Catani:2001cc,Krauss:2002up} and/or MINLO \cite{Hamilton:2012np} can be employed. Such studies are left for the future. 
  
We might add at this point that, for the  High Luminosity Large Hadron Collider (HL-LHC) we expect $N \approx (150-200)$ events per experiment assuming $\ell^\pm=e^\pm, \mu^\pm$ only. Thus, the predicted uncertainties for this process, estimated solely on the basis of the integrated luminosity planned for the HL-LHC, which is $(3-4) \, {\rm ab}^{-1}$, and when combining the  ATLAS and  CMS results, are in the range of $\delta = 1/\sqrt{N} \in (6\%-5\%)$. This further highlights the importance of precise calculations for this process.

It is worth repeating that although at the  total cross-section level the subleading NLO corrections for the  $pp\to t\bar{t}t\bar{t}+X$ process are negligible, they can be enhanced for various differential cross-section distributions in specific regions of the phase space.  Similar behavior has already been observed for simpler processes, such as for example $pp\to t\bar{t}\gamma$, $pp\to t\bar{t}Z$  and $pp\to t\bar{t}\gamma\gamma$ in the di-lepton top-quark decay channel. In these cases the substantial subleading contributions have been found in the tails of dimensionful observables due to one-loop EW Sudakov logarithms \cite{Denner:2023eti,Stremmer:2024ecl}.  Therefore, it would be recommended  to perform similar complete NLO calculations also for this complex final state.

Finally, it would be beneficial to compare  NLO QCD  results in the NWA to results obtained from the on-shell $pp\to t\bar{t}t\bar{t}$ calculation, with approximate spin-correlations in top-quark and $W$ decays, matched to parton showers. Such a comparison could assess to what extent parton-shower effects can reproduce all the contributions required at the NLO level in QCD for the $pp \to t\bar{t}t\bar{t}+X$ process in the $3\ell$ decay channel. In short, we could check to what extent parton-shower programs can mimic higher-order effects in top-quark decays. It would also help to identify phase-space regions and various dimensionless and dimensionful observables that are sensitive to parton-shower effects and in particular to  top-quark (pair) spin correlations. To this end, the various angular distributions such as for example $\Delta \phi_{\ell_1 \ell_2}$, $\Delta \phi_{b_1b_2}$  and $\Delta \phi_{j_1j_2}$ need to be carefully analyzed. One obvious observable that would require a parton shower  is the invariant mass of the two hardest light jets, since $M_{j_1j_2}$ is not correctly modeled with fixed-order predictions only. Similar effects have already been seen for the simpler  $pp \to t\bar{t}$ process in the $\ell+j$ decay channel, see e.g. Ref. \cite{Denner:2017kzu}. In addition, by comparing our predictions with the NLO theoretical prediction matched to parton showers, we would be able to inspect and thus verify the magnitude of the theoretical uncertainties arising from scale dependence. Indeed, in the  case of the theoretical prediction matched to parton showers these uncertainties are based solely on the production part of the $pp\to t\bar{t}t\bar{t}+X$ process. We can also check whether the cut $Q_{cut}$ used in our NLO QCD predictions is really needed or whether it is just an artifact of fixed-order computations and its use can only be phenomenologically motivated. Such analyses and comparisons are already at an advanced stage.

\acknowledgments{
This work was supported by the Deutsche Forschungsgemeinschaft (DFG) under grant 396021762 - TRR 257: \textit{Particle Physics Phenomenology after the Higgs Discovery}, and grant 400140256 - GRK 2497: \textit{The Physics of the Heaviest Particles at the LHC}.

Support by a grant of the Bundesministerium f\"ur Bildung und Forschung (BMBF) is additionally acknowledged.

The authors gratefully acknowledge the computing time provided to them at the NHR Center NHR4CES at RWTH Aachen University (project number \texttt{p0020216}). This is funded by the Federal Ministry of Education and Research, and the state governments participating on the basis of the resolutions of the GWK for national high performance computing at universities. }


\providecommand{\href}[2]{#2}\begingroup\raggedright\begin{thebibliography}{10}

\bibitem{Bevilacqua:2012em}
G.~Bevilacqua and M.~Worek, \emph{{Constraining BSM Physics at the LHC: Four
  top final states with NLO accuracy in perturbative QCD}},
  \href{https://doi.org/10.1007/JHEP07(2012)111}{\emph{JHEP} {\bfseries 07}
  (2012) 111} [\href{https://arxiv.org/abs/1206.3064}{{\ttfamily 1206.3064}}].

\bibitem{Cao:2016wib}
Q.-H.~Cao, S.-L.~Chen and Y.~Liu, \emph{{Probing Higgs Width and Top Quark
  Yukawa Coupling from $t\bar{t}H$ and $t\bar{t}t\bar{t}$ Productions}},
  \href{https://doi.org/10.1103/PhysRevD.95.053004}{\emph{Phys. Rev. D}
  {\bfseries 95} (2017) 053004}
  [\href{https://arxiv.org/abs/1602.01934}{{\ttfamily 1602.01934}}].

\bibitem{Cao:2019ygh}
Q.-H.~Cao, S.-L.~Chen, Y.~Liu, R.~Zhang and Y.~Zhang, \emph{{Limiting top
  quark-Higgs boson interaction and Higgs-boson width from multitop
  productions}}, \href{https://doi.org/10.1103/PhysRevD.99.113003}{\emph{Phys.
  Rev. D} {\bfseries 99} (2019) 113003}
  [\href{https://arxiv.org/abs/1901.04567}{{\ttfamily 1901.04567}}].

\bibitem{Guchait:2007jd}
M.~Guchait, F.~Mahmoudi and K.~Sridhar, \emph{{Associated production of a
  Kaluza-Klein excitation of a gluon with a t anti-t pair at the LHC}},
  \href{https://doi.org/10.1016/j.physletb.2008.07.085}{\emph{Phys. Lett. B}
  {\bfseries 666} (2008) 347}
  [\href{https://arxiv.org/abs/0710.2234}{{\ttfamily 0710.2234}}].

\bibitem{Pomarol:2008bh}
A.~Pomarol and J.~Serra, \emph{{Top Quark Compositeness: Feasibility and
  Implications}}, \href{https://doi.org/10.1103/PhysRevD.78.074026}{\emph{Phys.
  Rev. D} {\bfseries 78} (2008) 074026}
  [\href{https://arxiv.org/abs/0806.3247}{{\ttfamily 0806.3247}}].

\bibitem{Plehn:2008ae}
T.~Plehn and T.M.P.~Tait, \emph{{Seeking Sgluons}},
  \href{https://doi.org/10.1088/0954-3899/36/7/075001}{\emph{J. Phys. G}
  {\bfseries 36} (2009) 075001}
  [\href{https://arxiv.org/abs/0810.3919}{{\ttfamily 0810.3919}}].

\bibitem{Jung:2010ms}
S.~Jung and J.D.~Wells, \emph{{Low-scale warped extra dimension and its
  predilection for multiple top quarks}},
  \href{https://doi.org/10.1007/JHEP11(2010)001}{\emph{JHEP} {\bfseries 11}
  (2010) 001} [\href{https://arxiv.org/abs/1008.0870}{{\ttfamily 1008.0870}}].

\bibitem{Gregoire:2011ka}
T.~Gregoire, E.~Katz and V.~Sanz, \emph{{Four top quarks in extensions of the
  standard model}},
  \href{https://doi.org/10.1103/PhysRevD.85.055024}{\emph{Phys. Rev. D}
  {\bfseries 85} (2012) 055024}
  [\href{https://arxiv.org/abs/1101.1294}{{\ttfamily 1101.1294}}].

\bibitem{Calvet:2012rk}
S.~Calvet, B.~Fuks, P.~Gris and L.~Valery, \emph{{Searching for sgluons in
  multitop events at a center-of-mass energy of 8 TeV}},
  \href{https://doi.org/10.1007/JHEP04(2013)043}{\emph{JHEP} {\bfseries 04}
  (2013) 043} [\href{https://arxiv.org/abs/1212.3360}{{\ttfamily 1212.3360}}].

\bibitem{Arina:2016cqj}
C.~Arina et~al., \emph{{A comprehensive approach to dark matter studies:
  exploration of simplified top-philic models}},
  \href{https://doi.org/10.1007/JHEP11(2016)111}{\emph{JHEP} {\bfseries 11}
  (2016) 111} [\href{https://arxiv.org/abs/1605.09242}{{\ttfamily
  1605.09242}}].

\bibitem{Alvarez:2016nrz}
E.~Alvarez, D.A.~Faroughy, J.F.~Kamenik, R.~Morales and A.~Szynkman,
  \emph{{Four tops for LHC}},
  \href{https://doi.org/10.1016/j.nuclphysb.2016.11.024}{\emph{Nucl. Phys. B}
  {\bfseries 915} (2017) 19}
  [\href{https://arxiv.org/abs/1611.05032}{{\ttfamily 1611.05032}}].

\bibitem{Baer:2016wkz}
H.~Baer, V.~Barger, J.S.~Gainer, P.~Huang, M.~Savoy, D.~Sengupta et~al.,
  \emph{{Gluino reach and mass extraction at the LHC in radiatively-driven
  natural SUSY}},
  \href{https://doi.org/10.1140/epjc/s10052-017-5067-3}{\emph{Eur. Phys. J. C}
  {\bfseries 77} (2017) 499}
  [\href{https://arxiv.org/abs/1612.00795}{{\ttfamily 1612.00795}}].

\bibitem{Baer:2017pba}
H.~Baer, V.~Barger, J.S.~Gainer, H.~Serce and X.~Tata, \emph{{Reach of the
  high-energy LHC for gluinos and top squarks in SUSY models with light
  Higgsinos}}, \href{https://doi.org/10.1103/PhysRevD.96.115008}{\emph{Phys.
  Rev. D} {\bfseries 96} (2017) 115008}
  [\href{https://arxiv.org/abs/1708.09054}{{\ttfamily 1708.09054}}].

\bibitem{Alvarez:2019uxp}
E.~Alvarez, A.~Juste and R.M.S.~Seoane, \emph{{Four-top as probe of light
  top-philic New Physics}},
  \href{https://doi.org/10.1007/JHEP12(2019)080}{\emph{JHEP} {\bfseries 12}
  (2019) 080} [\href{https://arxiv.org/abs/1910.09581}{{\ttfamily
  1910.09581}}].

\bibitem{Anisha:2023xmh}
Anisha, O.~Atkinson, A.~Bhardwaj, C.~Englert, W.~Naskar and P.~Stylianou,
  \emph{{BSM reach of four-top production at the LHC}},
  \href{https://doi.org/10.1103/PhysRevD.108.035001}{\emph{Phys. Rev. D}
  {\bfseries 108} (2023) 035001}
  [\href{https://arxiv.org/abs/2302.08281}{{\ttfamily 2302.08281}}].

\bibitem{Degrande:2010kt}
C.~Degrande, J.-M.~Gerard, C.~Grojean, F.~Maltoni and G.~Servant,
  \emph{{Non-resonant New Physics in Top Pair Production at Hadron Colliders}},
  \href{https://doi.org/10.1007/JHEP03(2011)125}{\emph{JHEP} {\bfseries 03}
  (2011) 125} [\href{https://arxiv.org/abs/1010.6304}{{\ttfamily 1010.6304}}].

\bibitem{Zhang:2017mls}
C.~Zhang, \emph{{Constraining $qqtt$ operators from four-top production: a case
  for enhanced EFT sensitivity}},
  \href{https://doi.org/10.1088/1674-1137/42/2/023104}{\emph{Chin. Phys. C}
  {\bfseries 42} (2018) 023104}
  [\href{https://arxiv.org/abs/1708.05928}{{\ttfamily 1708.05928}}].

\bibitem{Banelli:2020iau}
G.~Banelli, E.~Salvioni, J.~Serra, T.~Theil and A.~Weiler, \emph{{The Present
  and Future of Four Top Operators}},
  \href{https://doi.org/10.1007/JHEP02(2021)043}{\emph{JHEP} {\bfseries 02}
  (2021) 043} [\href{https://arxiv.org/abs/2010.05915}{{\ttfamily
  2010.05915}}].

\bibitem{Aoude:2022deh}
R.~Aoude, H.~El~Faham, F.~Maltoni and E.~Vryonidou, \emph{{Complete SMEFT
  predictions for four top quark production at hadron colliders}},
  \href{https://doi.org/10.1007/JHEP10(2022)163}{\emph{JHEP} {\bfseries 10}
  (2022) 163} [\href{https://arxiv.org/abs/2208.04962}{{\ttfamily
  2208.04962}}].

\bibitem{ATLAS:2020hpj}
{\scshape ATLAS} collaboration, \emph{{Evidence for $t\bar{t}t\bar{t}$
  production in the multilepton final state in proton\textendash{}proton
  collisions at $\sqrt{s}=13$ $\text {TeV}$ with the ATLAS detector}},
  \href{https://doi.org/10.1140/epjc/s10052-020-08509-3}{\emph{Eur. Phys. J. C}
  {\bfseries 80} (2020) 1085}
  [\href{https://arxiv.org/abs/2007.14858}{{\ttfamily 2007.14858}}].

\bibitem{CMS:2023zdh}
{\scshape CMS} collaboration, \emph{{Evidence for Four-Top Quark Production in
  Proton-Proton Collisions at $\sqrt{s}=13 \; TeV$}},
  \href{https://doi.org/10.1016/j.physletb.2023.138076}{\emph{Phys. Lett. B}
  {\bfseries 844} (2023) 138076}
  [\href{https://arxiv.org/abs/2303.03864}{{\ttfamily 2303.03864}}].

\bibitem{ATLAS:2023ajo}
{\scshape ATLAS} collaboration, \emph{{Observation of four-top-quark production
  in the multilepton final state with the ATLAS detector}},
  \href{https://doi.org/10.1140/epjc/s10052-023-11573-0}{\emph{Eur. Phys. J. C}
  {\bfseries 83} (2023) 496}
  [\href{https://arxiv.org/abs/2303.15061}{{\ttfamily 2303.15061}}].

\bibitem{CMS:2023ftu}
{\scshape CMS} collaboration, \emph{{Observation of four top quark production
  in proton-proton collisions at $\sqrt{s}=13$ TeV}},
  \href{https://doi.org/10.1016/j.physletb.2023.138290}{\emph{Phys. Lett. B}
  {\bfseries 847} (2023) 138290}
  [\href{https://arxiv.org/abs/2305.13439}{{\ttfamily 2305.13439}}].

\bibitem{Alwall:2014hca}
J.~Alwall, R.~Frederix, S.~Frixione, V.~Hirschi, F.~Maltoni, O.~Mattelaer
  et~al., \emph{{The automated computation of tree-level and next-to-leading
  order differential cross sections, and their matching to parton shower
  simulations}}, \href{https://doi.org/10.1007/JHEP07(2014)079}{\emph{JHEP}
  {\bfseries 07} (2014) 079} [\href{https://arxiv.org/abs/1405.0301}{{\ttfamily
  1405.0301}}].

\bibitem{Maltoni:2015ena}
F.~Maltoni, D.~Pagani and I.~Tsinikos, \emph{{Associated production of a
  top-quark pair with vector bosons at NLO in QCD: impact on $t\bar{t}H$
  searches at the LHC}},
  \href{https://doi.org/10.1007/JHEP02(2016)113}{\emph{JHEP} {\bfseries 02}
  (2016) 113} [\href{https://arxiv.org/abs/1507.05640}{{\ttfamily
  1507.05640}}].

\bibitem{Frederix:2017wme}
R.~Frederix, D.~Pagani and M.~Zaro, \emph{{Large NLO corrections in
  $t\bar{t}W^{\pm}$ and $t\bar{t}t\bar{t}$ hadroproduction from supposedly
  subleading EW contributions}},
  \href{https://doi.org/10.1007/JHEP02(2018)031}{\emph{JHEP} {\bfseries 02}
  (2018) 031} [\href{https://arxiv.org/abs/1711.02116}{{\ttfamily
  1711.02116}}].

\bibitem{vanBeekveld:2022hty}
M.~van Beekveld, A.~Kulesza and L.M.~Valero, \emph{{Threshold Resummation for
  the Production of Four Top Quarks at the LHC}},
  \href{https://doi.org/10.1103/PhysRevLett.131.211901}{\emph{Phys. Rev. Lett.}
  {\bfseries 131} (2023) 211901}
  [\href{https://arxiv.org/abs/2212.03259}{{\ttfamily 2212.03259}}].

\bibitem{Jezo:2021smh}
T.~Je\v{z}o and M.~Kraus, \emph{{Hadroproduction of four top quarks in the
  powheg box}}, \href{https://doi.org/10.1103/PhysRevD.105.114024}{\emph{Phys.
  Rev. D} {\bfseries 105} (2022) 114024}
  [\href{https://arxiv.org/abs/2110.15159}{{\ttfamily 2110.15159}}].

\bibitem{Dimitrakopoulos:2024qib}
N.~Dimitrakopoulos and M.~Worek, \emph{{Four top final states with NLO accuracy
  in perturbative QCD: 4 lepton channel}},
  \href{https://arxiv.org/abs/2401.10678}{{\ttfamily 2401.10678}}.

\bibitem{Fadin:1993kt}
V.S.~Fadin, V.A.~Khoze and A.D.~Martin, \emph{{How suppressed are the radiative
  interference effects in heavy instable particle production?}},
  \href{https://doi.org/10.1016/0370-2693(94)90837-0}{\emph{Phys. Lett. B}
  {\bfseries 320} (1994) 141}
  [\href{https://arxiv.org/abs/hep-ph/9309234}{{\ttfamily hep-ph/9309234}}].

\bibitem{Bevilacqua:2019quz}
G.~Bevilacqua, H.B.~Hartanto, M.~Kraus, T.~Weber and M.~Worek, \emph{{Off-shell
  vs on-shell modelling of top quarks in photon associated production}},
  \href{https://doi.org/10.1007/JHEP03(2020)154}{\emph{JHEP} {\bfseries 03}
  (2020) 154} [\href{https://arxiv.org/abs/1912.09999}{{\ttfamily
  1912.09999}}].

\bibitem{Bevilacqua:2020pzy}
G.~Bevilacqua, H.-Y.~Bi, H.B.~Hartanto, M.~Kraus and M.~Worek, \emph{{The
  simplest of them all: $t\bar{t} W^\pm$ at NLO accuracy in QCD}},
  \href{https://doi.org/10.1007/JHEP08(2020)043}{\emph{JHEP} {\bfseries 08}
  (2020) 043} [\href{https://arxiv.org/abs/2005.09427}{{\ttfamily
  2005.09427}}].

\bibitem{Stremmer:2021bnk}
D.~Stremmer and M.~Worek, \emph{{Production and decay of the Higgs boson in
  association with top quarks}},
  \href{https://doi.org/10.1007/JHEP02(2022)196}{\emph{JHEP} {\bfseries 02}
  (2022) 196} [\href{https://arxiv.org/abs/2111.01427}{{\ttfamily
  2111.01427}}].

\bibitem{Bevilacqua:2022nrm}
G.~Bevilacqua, H.B.~Hartanto, M.~Kraus, J.~Nasufi and M.~Worek, \emph{{NLO QCD
  corrections to full off-shell production of $ t\overline{t}Z $ including
  leptonic decays}}, \href{https://doi.org/10.1007/JHEP08(2022)060}{\emph{JHEP}
  {\bfseries 08} (2022) 060}
  [\href{https://arxiv.org/abs/2203.15688}{{\ttfamily 2203.15688}}].

\bibitem{Hermann:2022vit}
J.~Hermann, D.~Stremmer and M.~Worek, \emph{{$ \mathcal{CP} $ structure of the
  top-quark Yukawa interaction: NLO QCD corrections and off-shell effects}},
  \href{https://doi.org/10.1007/JHEP09(2022)138}{\emph{JHEP} {\bfseries 09}
  (2022) 138} [\href{https://arxiv.org/abs/2205.09983}{{\ttfamily
  2205.09983}}].

\bibitem{Denner:2017kzu}
A.~Denner and M.~Pellen, \emph{{Off-shell production of top-antitop pairs in
  the lepton+jets channel at NLO QCD}},
  \href{https://doi.org/10.1007/JHEP02(2018)013}{\emph{JHEP} {\bfseries 02}
  (2018) 013} [\href{https://arxiv.org/abs/1711.10359}{{\ttfamily
  1711.10359}}].

\bibitem{Bevilacqua:2011xh}
G.~Bevilacqua, M.~Czakon, M.V.~Garzelli, A.~van Hameren, A.~Kardos,
  C.G.~Papadopoulos et~al., \emph{{HELAC-NLO}},
  \href{https://doi.org/10.1016/j.cpc.2012.10.033}{\emph{Comput. Phys. Commun.}
  {\bfseries 184} (2013) 986}
  [\href{https://arxiv.org/abs/1110.1499}{{\ttfamily 1110.1499}}].

\bibitem{vanHameren:2009dr}
A.~van Hameren, C.G.~Papadopoulos and R.~Pittau, \emph{{Automated one-loop
  calculations: A Proof of concept}},
  \href{https://doi.org/10.1088/1126-6708/2009/09/106}{\emph{JHEP} {\bfseries
  09} (2009) 106} [\href{https://arxiv.org/abs/0903.4665}{{\ttfamily
  0903.4665}}].

\bibitem{Czakon:2009ss}
M.~Czakon, C.G.~Papadopoulos and M.~Worek, \emph{{Polarizing the Dipoles}},
  \href{https://doi.org/10.1088/1126-6708/2009/08/085}{\emph{JHEP} {\bfseries
  08} (2009) 085} [\href{https://arxiv.org/abs/0905.0883}{{\ttfamily
  0905.0883}}].

\bibitem{Ossola:2007ax}
G.~Ossola, C.G.~Papadopoulos and R.~Pittau, \emph{{CutTools: A Program
  implementing the OPP reduction method to compute one-loop amplitudes}},
  \href{https://doi.org/10.1088/1126-6708/2008/03/042}{\emph{JHEP} {\bfseries
  03} (2008) 042} [\href{https://arxiv.org/abs/0711.3596}{{\ttfamily
  0711.3596}}].

\bibitem{Ossola:2008zzb}
G.~Ossola, C.G.~Papadopoulos and R.~Pittau, \emph{{Reduction of one-loop
  amplitudes at the integrand level: NLO QCD calculations}}, {\emph{Acta Phys.
  Polon. B} {\bfseries 39} (2008) 1685}.

\bibitem{Ossola:2008xq}
G.~Ossola, C.G.~Papadopoulos and R.~Pittau, \emph{{On the Rational Terms of the
  one-loop amplitudes}},
  \href{https://doi.org/10.1088/1126-6708/2008/05/004}{\emph{JHEP} {\bfseries
  05} (2008) 004} [\href{https://arxiv.org/abs/0802.1876}{{\ttfamily
  0802.1876}}].

\bibitem{Draggiotis:2009yb}
P.~Draggiotis, M.V.~Garzelli, C.G.~Papadopoulos and R.~Pittau, \emph{{Feynman
  Rules for the Rational Part of the QCD 1-loop amplitudes}},
  \href{https://doi.org/10.1088/1126-6708/2009/04/072}{\emph{JHEP} {\bfseries
  04} (2009) 072} [\href{https://arxiv.org/abs/0903.0356}{{\ttfamily
  0903.0356}}].

\bibitem{vanHameren:2010cp}
A.~van Hameren, \emph{{OneLOop: For the evaluation of one-loop scalar
  functions}}, \href{https://doi.org/10.1016/j.cpc.2011.06.011}{\emph{Comput.
  Phys. Commun.} {\bfseries 182} (2011) 2427}
  [\href{https://arxiv.org/abs/1007.4716}{{\ttfamily 1007.4716}}].

\bibitem{Catani:1996vz}
S.~Catani and M.H.~Seymour, \emph{{A General algorithm for calculating jet
  cross-sections in NLO QCD}},
  \href{https://doi.org/10.1016/S0550-3213(96)00589-5}{\emph{Nucl. Phys. B}
  {\bfseries 485} (1997) 291}
  [\href{https://arxiv.org/abs/hep-ph/9605323}{{\ttfamily hep-ph/9605323}}].

\bibitem{Catani:2002hc}
S.~Catani, S.~Dittmaier, M.H.~Seymour and Z.~Trocsanyi, \emph{{The Dipole
  formalism for next-to-leading order QCD calculations with massive partons}},
  \href{https://doi.org/10.1016/S0550-3213(02)00098-6}{\emph{Nucl. Phys. B}
  {\bfseries 627} (2002) 189}
  [\href{https://arxiv.org/abs/hep-ph/0201036}{{\ttfamily hep-ph/0201036}}].

\bibitem{Bevilacqua:2013iha}
G.~Bevilacqua, M.~Czakon, M.~Kubocz and M.~Worek, \emph{{Complete Nagy-Soper
  subtraction for next-to-leading order calculations in QCD}},
  \href{https://doi.org/10.1007/JHEP10(2013)204}{\emph{JHEP} {\bfseries 10}
  (2013) 204} [\href{https://arxiv.org/abs/1308.5605}{{\ttfamily 1308.5605}}].

\bibitem{vanHameren:2007pt}
A.~van Hameren, \emph{{PARNI for importance sampling and density estimation}},
  {\emph{Acta Phys. Polon. B} {\bfseries 40} (2009) 259}
  [\href{https://arxiv.org/abs/0710.2448}{{\ttfamily 0710.2448}}].

\bibitem{vanHameren:2010gg}
A.~van Hameren, \emph{{Kaleu: A General-Purpose Parton-Level Phase Space
  Generator}},  \href{https://arxiv.org/abs/1003.4953}{{\ttfamily 1003.4953}}.

\bibitem{Nagy:1998bb}
Z.~Nagy and Z.~Trocsanyi, \emph{{Next-to-leading order calculation of four jet
  observables in electron positron annihilation}},
  \href{https://doi.org/10.1103/PhysRevD.62.099902}{\emph{Phys. Rev. D}
  {\bfseries 59} (1999) 014020}
  [\href{https://arxiv.org/abs/hep-ph/9806317}{{\ttfamily hep-ph/9806317}}].

\bibitem{Nagy:2003tz}
Z.~Nagy, \emph{{Next-to-leading order calculation of three jet observables in
  hadron hadron collision}},
  \href{https://doi.org/10.1103/PhysRevD.68.094002}{\emph{Phys. Rev. D}
  {\bfseries 68} (2003) 094002}
  [\href{https://arxiv.org/abs/hep-ph/0307268}{{\ttfamily hep-ph/0307268}}].

\bibitem{Bevilacqua:2009zn}
G.~Bevilacqua, M.~Czakon, C.G.~Papadopoulos, R.~Pittau and M.~Worek,
  \emph{{Assault on the NLO Wishlist: $pp \to t\bar{t}b\bar{b}$}},
  \href{https://doi.org/10.1088/1126-6708/2009/09/109}{\emph{JHEP} {\bfseries
  09} (2009) 109} [\href{https://arxiv.org/abs/0907.4723}{{\ttfamily
  0907.4723}}].

\bibitem{Czakon:2015cla}
M.~Czakon, H.B.~Hartanto, M.~Kraus and M.~Worek, \emph{{Matching the Nagy-Soper
  parton shower at next-to-leading order}},
  \href{https://doi.org/10.1007/JHEP06(2015)033}{\emph{JHEP} {\bfseries 06}
  (2015) 033} [\href{https://arxiv.org/abs/1502.00925}{{\ttfamily
  1502.00925}}].

\bibitem{Badger:2010nx}
S.~Badger, B.~Biedermann and P.~Uwer, \emph{{NGluon: A Package to Calculate
  One-loop Multi-gluon Amplitudes}},
  \href{https://doi.org/10.1016/j.cpc.2011.04.008}{\emph{Comput. Phys. Commun.}
  {\bfseries 182} (2011) 1674}
  [\href{https://arxiv.org/abs/1011.2900}{{\ttfamily 1011.2900}}].

\bibitem{Actis:2016mpe}
S.~Actis, A.~Denner, L.~Hofer, J.-N.~Lang, A.~Scharf and S.~Uccirati,
  \emph{{RECOLA: REcursive Computation of One-Loop Amplitudes}},
  \href{https://doi.org/10.1016/j.cpc.2017.01.004}{\emph{Comput. Phys. Commun.}
  {\bfseries 214} (2017) 140}
  [\href{https://arxiv.org/abs/1605.01090}{{\ttfamily 1605.01090}}].

\bibitem{Alwall:2006yp}
J.~Alwall et~al., \emph{{A Standard format for Les Houches event files}},
  \href{https://doi.org/10.1016/j.cpc.2006.11.010}{\emph{Comput. Phys. Commun.}
  {\bfseries 176} (2007) 300}
  [\href{https://arxiv.org/abs/hep-ph/0609017}{{\ttfamily hep-ph/0609017}}].

\bibitem{Antcheva:2009zz}
I.~Antcheva et~al., \emph{{ROOT: A C++ framework for petabyte data storage,
  statistical analysis and visualization}},
  \href{https://doi.org/10.1016/j.cpc.2009.08.005}{\emph{Comput. Phys. Commun.}
  {\bfseries 180} (2009) 2499}
  [\href{https://arxiv.org/abs/1508.07749}{{\ttfamily 1508.07749}}].

\bibitem{Bern:2013zja}
Z.~Bern, L.J.~Dixon, F.~Febres~Cordero, S.~H\"oche, H.~Ita, D.A.~Kosower
  et~al., \emph{{Ntuples for NLO Events at Hadron Colliders}},
  \href{https://doi.org/10.1016/j.cpc.2014.01.011}{\emph{Comput. Phys. Commun.}
  {\bfseries 185} (2014) 1443}
  [\href{https://arxiv.org/abs/1310.7439}{{\ttfamily 1310.7439}}].

\bibitem{Bevilacqua:HEPlot}
G.~Bevilacqua, ``unpublished.'' 2019.

\bibitem{Denner:2012yc}
A.~Denner, S.~Dittmaier, S.~Kallweit and S.~Pozzorini, \emph{{NLO QCD
  corrections to off-shell top-antitop production with leptonic decays at
  hadron colliders}},
  \href{https://doi.org/10.1007/JHEP10(2012)110}{\emph{JHEP} {\bfseries 10}
  (2012) 110} [\href{https://arxiv.org/abs/1207.5018}{{\ttfamily 1207.5018}}].

\bibitem{Butterworth:2015oua}
J.~Butterworth et~al., \emph{{PDF4LHC recommendations for LHC Run II}},
  \href{https://doi.org/10.1088/0954-3899/43/2/023001}{\emph{J. Phys. G}
  {\bfseries 43} (2016) 023001}
  [\href{https://arxiv.org/abs/1510.03865}{{\ttfamily 1510.03865}}].

\bibitem{Bailey:2020ooq}
S.~Bailey, T.~Cridge, L.A.~Harland-Lang, A.D.~Martin and R.S.~Thorne,
  \emph{{Parton distributions from LHC, HERA, Tevatron and fixed target data:
  MSHT20 PDFs}},
  \href{https://doi.org/10.1140/epjc/s10052-021-09057-0}{\emph{Eur. Phys. J. C}
  {\bfseries 81} (2021) 341}
  [\href{https://arxiv.org/abs/2012.04684}{{\ttfamily 2012.04684}}].

\bibitem{NNPDF:2017mvq}
{\scshape NNPDF} collaboration, \emph{{Parton distributions from high-precision
  collider data}},
  \href{https://doi.org/10.1140/epjc/s10052-017-5199-5}{\emph{Eur. Phys. J. C}
  {\bfseries 77} (2017) 663}
  [\href{https://arxiv.org/abs/1706.00428}{{\ttfamily 1706.00428}}].

\bibitem{Hou:2019efy}
T.-J.~Hou et~al., \emph{{New CTEQ global analysis of quantum chromodynamics
  with high-precision data from the LHC}},
  \href{https://doi.org/10.1103/PhysRevD.103.014013}{\emph{Phys. Rev. D}
  {\bfseries 103} (2021) 014013}
  [\href{https://arxiv.org/abs/1912.10053}{{\ttfamily 1912.10053}}].

\bibitem{Dulat:2015mca}
S.~Dulat, T.-J.~Hou, J.~Gao, M.~Guzzi, J.~Huston, P.~Nadolsky et~al.,
  \emph{{New parton distribution functions from a global analysis of quantum
  chromodynamics}},
  \href{https://doi.org/10.1103/PhysRevD.93.033006}{\emph{Phys. Rev. D}
  {\bfseries 93} (2016) 033006}
  [\href{https://arxiv.org/abs/1506.07443}{{\ttfamily 1506.07443}}].

\bibitem{Buckley:2014ana}
A.~Buckley, J.~Ferrando, S.~Lloyd, K.~Nordstr\"om, B.~Page, M.~R\"ufenacht
  et~al., \emph{{LHAPDF6: parton density access in the LHC precision era}},
  \href{https://doi.org/10.1140/epjc/s10052-015-3318-8}{\emph{Eur. Phys. J. C}
  {\bfseries 75} (2015) 132} [\href{https://arxiv.org/abs/1412.7420}{{\ttfamily
  1412.7420}}].

\bibitem{Cacciari:2008gp}
M.~Cacciari, G.P.~Salam and G.~Soyez, \emph{{The anti-$k_t$ jet clustering
  algorithm}}, \href{https://doi.org/10.1088/1126-6708/2008/04/063}{\emph{JHEP}
  {\bfseries 04} (2008) 063} [\href{https://arxiv.org/abs/0802.1189}{{\ttfamily
  0802.1189}}].

\bibitem{Melnikov:2011ta}
K.~Melnikov, M.~Schulze and A.~Scharf, \emph{{QCD corrections to top quark pair
  production in association with a photon at hadron colliders}},
  \href{https://doi.org/10.1103/PhysRevD.83.074013}{\emph{Phys. Rev. D}
  {\bfseries 83} (2011) 074013}
  [\href{https://arxiv.org/abs/1102.1967}{{\ttfamily 1102.1967}}].

\bibitem{Stremmer:2023kcd}
D.~Stremmer and M.~Worek, \emph{{Associated production of a top-quark pair with
  two isolated photons at the LHC through NLO in QCD}},
  \href{https://doi.org/10.1007/JHEP08(2023)179}{\emph{JHEP} {\bfseries 08}
  (2023) 179} [\href{https://arxiv.org/abs/2306.16968}{{\ttfamily
  2306.16968}}].

\bibitem{Mangano:2006rw}
M.~L.~Mangano, M.~Moretti, F.~Piccinini and M.~Treccani,
\emph{Matching matrix elements and shower evolution for top-quark production in hadronic collisions},
\href{https://doi.org/10.1088/1126-6708/2007/01/013}{JHEP {\bfseries 01} (2007) 013}
[\href{https://arxiv.org/abs/hep-ph/0611129}{\ttfamily hep-ph/0611129}].

\bibitem{Alwall:2007fs}
J.~Alwall, S.~Hoche, F.~Krauss, N.~Lavesson, L.~Lonnblad, F.~Maltoni, M.~L.~Mangano, M.~Moretti, C.~G.~Papadopoulos and F.~Piccinini, \textit{et al.}
\emph{Comparative study of various algorithms for the merging of parton showers and matrix elements in hadronic collisions},
\href{https://doi.org/10.1140/epjc/s10052-007-0490-5}{\emph{Eur. Phys. J. C} {\bfseries 53} (2008) 473} [\href{https://arxiv.org/abs/0706.2569}{\ttfamily 0706.2569}].

\bibitem{Catani:2001cc}
S.~Catani, F.~Krauss, R.~Kuhn and B.~R.~Webber,
\emph{QCD matrix elements + parton showers},
\href{https://doi.org/10.1088/1126-6708/2001/11/063}{\emph{JHEP} {\bfseries 11} (2001) 063} [\href{https://arxiv.org/abs/hep-ph/0109231}{\ttfamily hep-ph/0109231}].

\bibitem{Krauss:2002up}
F.~Krauss,
\emph{Matrix elements and parton showers in hadronic interactions},
\href{https://doi.org/10.1088/1126-6708/2002/08/015}{JHEP {\bfseries 08} (2002) 015}
[\href{https://arxiv.org/abs/hep-ph/0205283}{\ttfamily hep-ph/0205283}].
\bibitem{Hamilton:2012np}
  K.~Hamilton, P.~Nason and G.~Zanderighi,
  \emph{MINLO: Multi-Scale Improved NLO}, 
\href{https://doi.org/10.1007/JHEP10(2012)155}{\emph{JHEP} {\bfseries 10} (2012) 155}
[\href{https://arxiv.org/abs/1206.3572}{\ttfamily 1206.3572}].

\bibitem{Denner:2023eti}
A.~Denner, D.~Lombardi and G.~Pelliccioli,
\emph{Complete NLO corrections to off-shell $t\bar{t}Z$ production at the LHC},
\href{https://doi.org/10.1007/JHEP09(2023)072}{\emph{JHEP} {\bfseries 09} (2023) 072}
[\href{https://arxiv.org/abs/2306.13535}{\ttfamily 2306.13535}].

\bibitem{Stremmer:2024ecl}
D.~Stremmer and M.~Worek,
\emph{Complete NLO corrections to top-quark pair production with isolated photons},
 \href{https://doi.org/10.1007/JHEP07(2024)091}{\emph{JHEP}  {\bfseries 07} (2024) 091}
[\href{https://arxiv.org/abs/2403.03796}{\ttfamily 2403.03796}].


\end{thebibliography}\endgroup

\bibliographystyle{JHEP}

\providecommand{\href}[2]{#2}\begingroup\raggedright\endgroup

\end{document}